\documentclass[a4paper,12pt,twoside]{book}
\usepackage{enteteProjet}
\usepackage{newpxtext,newpxmath}
\defbibheading{bibliography}{\chapter*{References}} 
\usepackage{bigints}

\renewcommand{\lstlistingname}{\texttt{PARCS} input}

\def\subtitle#1{\gdef\@subtitle{#1}}
\usepackage{makecell}
\usepackage{nicematrix}
\usepackage{outlines}
\usepackage[version=4]{mhchem}

\hypersetup{
    colorlinks=true,      
    linkcolor=black,     
    citecolor=black,     
    urlcolor=black,      
    pdfborder={0 0 0}    
}

\usepackage{fancyhdr}
\usepackage{lastpage}
\usepackage[a4paper, left=2.54cm, right=2.54cm,bottom=2.7cm]{geometry}

\begin{document}

\newgeometry{bottom=2.5cm, left=2cm, right=2cm}
\begin{titlepage}

\begin{center}

\hspace*{-2cm} 
\begin{tabular}[H]{ p{0.4\textwidth} p{0.32\textwidth} p{0.2\textwidth}}
\vspace{15pt} \includegraphics[height=0.06\textwidth]{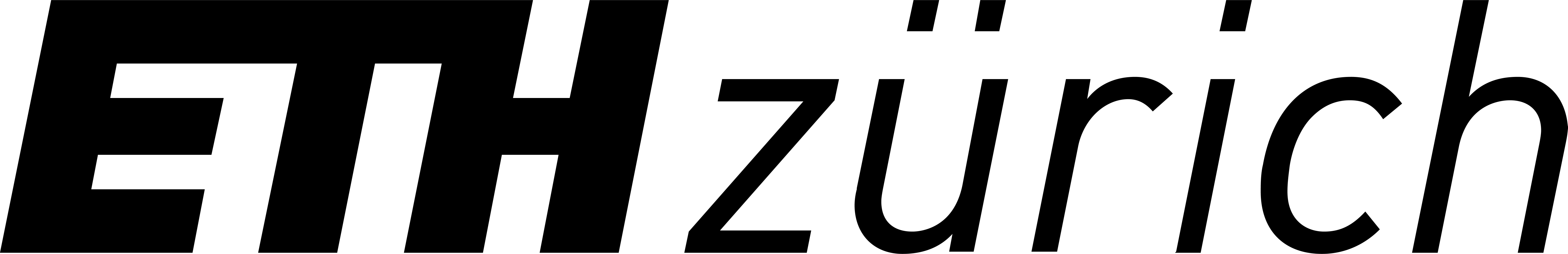} &
\vspace{0pt} \includegraphics[height=0.13\textwidth]{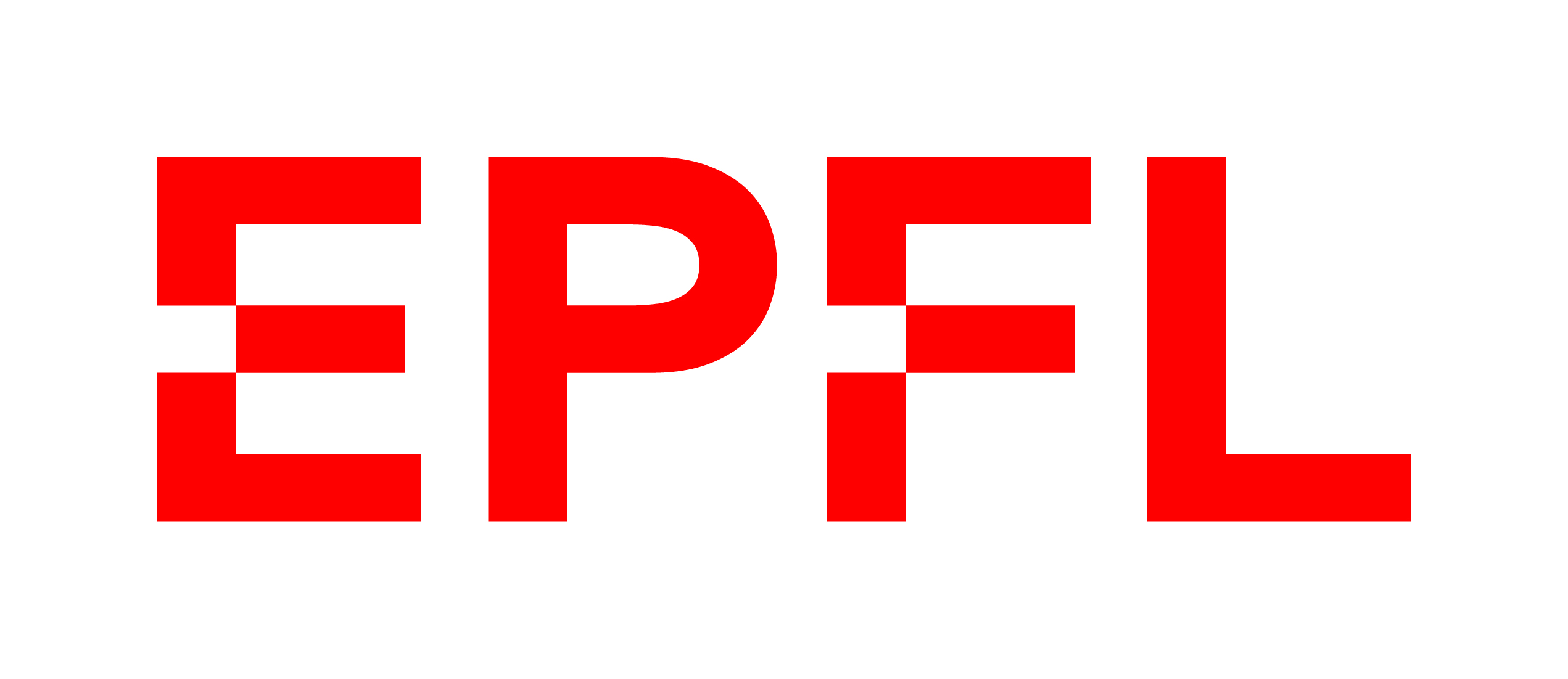} &
\vspace{0pt} \includegraphics[height=0.13\textwidth]{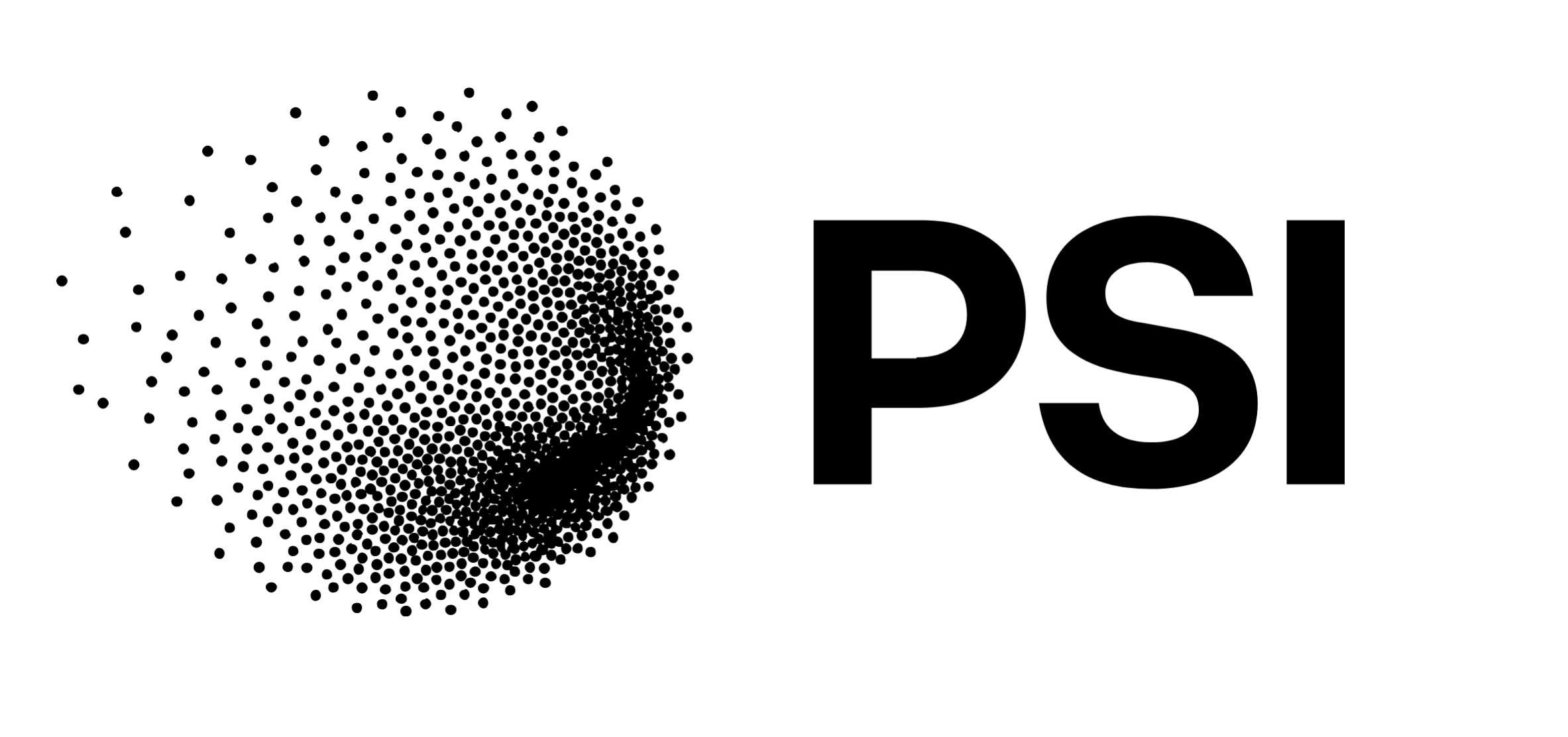}
\end{tabular}

\vspace{2cm}

{\huge \textbf{Multi-Cycle PWR Modeling\\ with \texttt{DRAGON} and \texttt{PARCS}: Implications for \\Nuclear Data Assessment}}

\vspace{0cm}
\definecolor{darkblue}{rgb}{0.0, 0.0, 0.5} 
 \textcolor{darkblue}{\rule{1\textwidth}{1pt}}

\vspace{1cm}

{\Large Master's Thesis\\[1cm]
EPFL/ETHZ MSc. in Nuclear Engineering\\
Laboratory for Reactor Physics and Systems Behaviour (LRS)}

\vfill

By\\[0.5cm]
{\Large \textbf{Benjamin Arthur Hugo MEUNIER}}

\vfill

\begin{center}
    \begin{tabular}{rl}
        Thesis Supervisor: & Dr.~MER~Mathieu HURSIN \\
        Expert: & Dr.~Vivian SALINO \\
    \end{tabular}
\end{center}
\vspace{0.5cm}
         \large
         August 16, 2024

\end{center}

\end{titlepage}
\restoregeometry

\pagenumbering{Roman}

\newpage
\section*{Acknowledgements}
I would like to extend my heartfelt gratitude to several individuals who have significantly contributed to the success of this project.\\[1\baselineskip]
First and foremost, I thank Mathieu Hursin, my supervisor. Our instructive discussions were invaluable, especially when new challenges arose. His guidance in characterising precisely these issues and identifying minimal examples to work with, as well as connecting me with experts in the field, has been instrumental in bringing this project to fruition. In addition, we have continued to work on the subject together and have updated some parts of this thesis for publication in a peer-reviewed paper~\cite{MEUNIER2025} and a conference~\cite{confParisNuc2024}. In particular, the modelling of Turkey-Point-3 was significantly improved.\\[0.5\baselineskip]
I am also deeply grateful to Vivian Salino from IRSN, whose PhD work served as a primary source of inspiration for my Thesis. His recent dissertation laid the groundwork for my project, and I built upon his contributions to the \texttt{DRAGON} procedures. After the submission of my Thesis, he assisted me in refining it for online publication in open access on arXiv.\\[0.5\baselineskip]
My appreciation extends to Andrew Ward from the University of Michigan, who introduced me to Git repository from Oak Ridge National Laboratory and assisted in debugging \texttt{PARCS}. His suggestions significantly advanced my work.\\[0.5\baselineskip]
I would like to acknowledge Maxime Roux, my research colleague in fluid mechanics for many years, for our insightful discussions and his assistance with the graphical design of my project.\\[0.5\baselineskip]
Finally, I wish to express my sincere thanks to my proofreaders: Mathieu Hursin, Maxime Roux, Timour Jestin, Barbara Ebert and Philippe Griveaux. Their meticulous attention to detail and invaluable feedback greatly enhanced the clarity and quality of my Thesis. I truly appreciate the time and effort they dedicated to reviewing my work.

\setcounter{tocdepth}{5}
\setcounter{secnumdepth}{5}

\tableofcontents

\newpage

\renewcommand\thesection{\arabic{section}}
\pagenumbering{arabic}

\lhead{}
\chead{}
\rhead{Page\ \thepage\ of\ \protect\pageref{LastPage}}
\renewcommand{\sectionmark}[1]{\markboth{#1}{}}
\lfoot{}
\cfoot{Chapter \thesection:\quad\leftmark}
\rfoot{}
\section{Introduction}

\fancypagestyle{plain}{
  \fancyhf{}
  \fancyhead[R]{Page\ \thepage\ of\ \protect\pageref{LastPage}}
}

Nuclear data libraries serve as the foundation for all calculations in the nuclear field. Their quality directly affects the accuracy of computations~\cite{Aliberti,hursinIntegralParamCASMO}. When new nuclear data libraries are released, they must undergo validation through the use of integral experimental data. This process can be achieved through criticality experiments (see~ICSBEP~\cite{ICSBEPI}), with Reactor Physics Experiments (see~IRPhE~\cite{IRPhE}) or with NPP (Nuclear Power Plant) data. However, the two former sources of experimental data fail to fully capture the complex behaviour of a nuclear reactor. The validation of new releases of nuclear data libraries focusing on these methods may not be able to detect a significant deterioration of the prediction of key physical quantities such as the reactivity loss~\cite{hursin_physor_2024, hursin_M_C_2023} during fuel evolution. A comprehensive assessment should integrate all these approaches.\\[0.5\baselineskip]
The aim of this Master's Thesis is to model multiple consecutive cycles of PWRs (Pressurised Water Reactors) and to gather publicly available data to build a pipeline to automate the assessment of the performance of novel nuclear data libraries using open source experimental data. These data encompass measurements performed during depletion cycles and startup physics tests. The measurements considered in this Master's Thesis are the following:
\begin{outline}
    \1 Local power measurements from detectors located in the instrumentation tubes of fuel assemblies;
    \1 Control Rod Worth assessment, which is defined as the difference in reactivity of the core measured between the insertion and the extraction of a specific set of control rods;
    \1 Axial offset, defined by equation~\eqref{eq:axialOffset}. It encapsulates the ratio of power produced in the upper half of the core compared to the power produced in the bottom half.
\end{outline}
\begin{equation}
 \textrm{Axial offset } (\%)=\frac{P_{\textrm{top}}-P_{\textrm{bot}}}{P_{\textrm{top}}+P_{\textrm{bot}}}\cdot 100
\label{eq:axialOffset}
\end{equation}
where $P_{\textrm{top/bot}}$ are the power above and below the core midplane.

\subsection{Thesis Objectives}
\begin{figure}
    \begin{subfigure}{1\textwidth}
        \centering
        \includegraphics[width=0.75\linewidth]{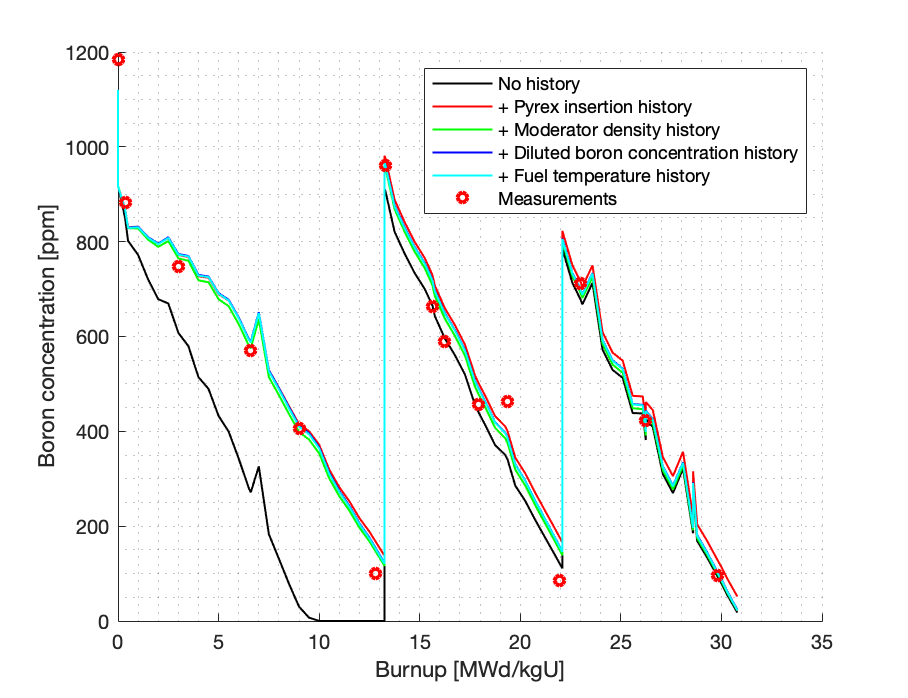}
        \caption{}
        \label{fig:TP3ImpactHistoriesBoron}
    \end{subfigure}
    \begin{subfigure}{1\textwidth}
        \centering
        \includegraphics[width=0.75\linewidth]{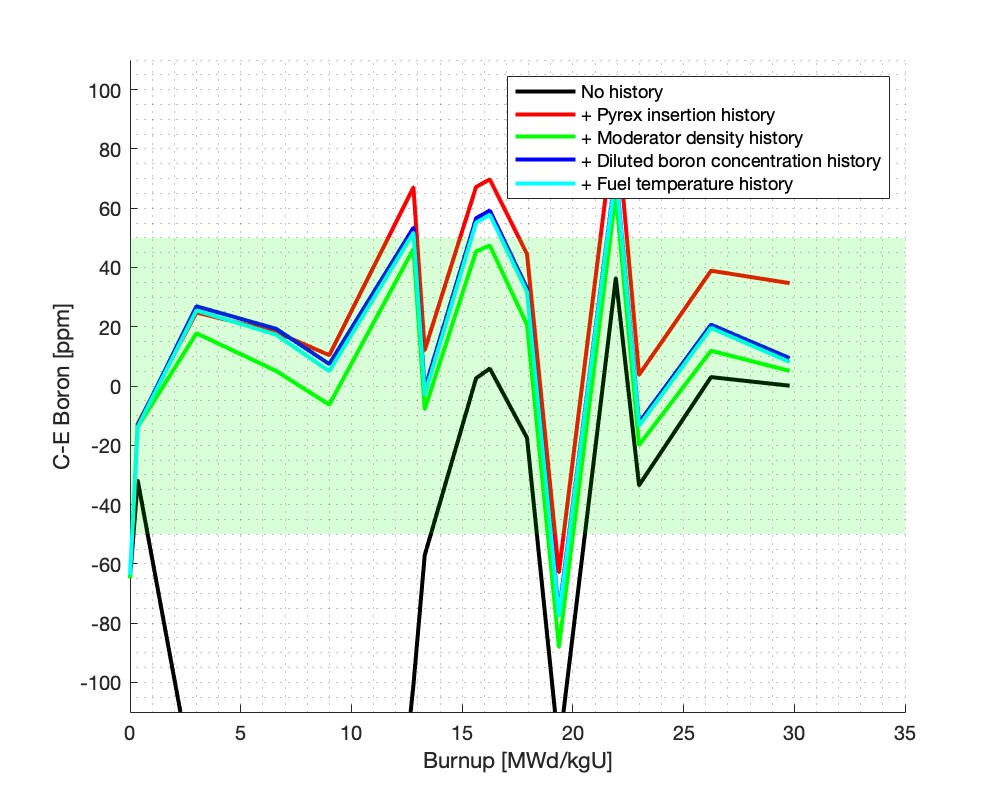}
        \caption{}
        \label{fig:TP3ImpactHistoriesDeltaBoron}
    \end{subfigure}
    \caption{Analysis of the significance of considering various histories with \texttt{PARCS}. The green shaded region represents the uncertainty range~\cite{JEFF_UncertaintyXS}. Each history is incrementally included in the simulation, starting from a scenario without any history and progressing to a case that includes all four histories. The macroscopic cross sections were obtained with \texttt{POLARIS} by the Oak Ridge National Laboratory~\cite{gitORNL}. (a) Evolution of the boron concentration in water during the first three cycles of Turkey-Point-3. (b) Difference with measurements.}
\label{fig:TP3ImpactHistories}
\end{figure}

Developing an automated simulation pipeline that transforms raw nuclear data into predictions for various reactors would significantly enhance validation processes, allowing for the assessment of potential biases in the prediction of integral measurements in nuclear power plants prior to the release of a nuclear data library. A key feature of this initiative is its accessibility and open-source nature, enabling integration into automated validation procedures utilised by various communities, such as the \texttt{JEFF} community~\cite{JEFF}. This consideration influenced the choice of \texttt{DRAGON}~\cite{Dragon} (an open-source tool) for lattice calculations and \texttt{PARCS}~\cite{PARCS_1_Inputs} (a university-developed code) for full core diffusion calculations. Additionally, \texttt{PyNJOY2016}~\cite{PyNJOY} (based on \texttt{NJOY2016}~\cite{NJOY2016} and also open source) will be employed to generate the microscopic nuclear data library for use by \texttt{DRAGON}. An important aspect of the open-source nature of this project is that the community can contribute to enriching the measurement database and improving the quality of the modeling.\\[0.5\baselineskip]
An important goal of this Master's Thesis is to develop models using \texttt{DRAGON} and \texttt{PARCS} to simulate three reactors, encompassing seven depletion cycles, and to validate these models against publicly available data from nuclear power plants.\\[0.5\baselineskip]
Another significant objective of the project is to establish this automated procedure.\\[0.5\baselineskip]
In addition, \texttt{DRAGON} model used in this Master's Thesis does not inherently support the generation of macroscopic cross sections under varying depletion conditions, known as histories. Figure~\ref{fig:TP3ImpactHistories} presents the evolution of the boron concentration in water during the first three cycles of Turkey-Point-3 with different history considerations. Each history is incrementally included in the simulation, starting from a scenario without any history and progressing to a case that includes all four histories. The macroscopic cross sections were obtained with \texttt{POLARIS} by the Oak Ridge National Laboratory~\cite{gitORNL}. It is important to note that the reference Pyrex insertion history (applied in the scenario without histories) reflects depletion conditions absent of Pyrex. Considering that Pyrex is predominantly present in the first cycle, the results are primarily affected during this cycle. When a reference history including Pyrex is considered, discrepancies become apparent in the second and third cycles.\\[0.5\baselineskip]
This analysis indicates that the Pyrex insertion history is crucial for predicting the boron let-down curve, especially for the first cycle. The moderator density history appears to be the second most important, especially as burnup increases (see~Fig.~\ref{fig:TP3ImpactHistoriesDeltaBoron}). The diluted boron history effect is notable but much less significant, while the fuel temperature history is barely noticeable in this analysis. Since the histories associated with the insertion of burnable poison rods and changes in moderator density have a significant impact on the quality of the modeling, they will be considered in the present work.\\[0.5\baselineskip]
A key objective of this Master's Thesis is to implement and verify a procedure for managing the generation of macroscopic cross sections with different histories using \texttt{DRAGON}.

\subsection{Organisation of Thesis}
Chapter~\ref{section:ComputationalScheme} presents the codes along with the computational scheme. The simulations discussed in the present Master's Thesis follow the two-step approach.\\[0.5\baselineskip]
Chapter~\ref{section:history} details the theoretical development requirements for modeling PWRs using \texttt{DRAGON} and \texttt{PARCS}. In particular, history effects are analysed, implemented in \texttt{DRAGON} and \texttt{PARCS} and verified code-to-code with a modelisation from the Oak Ridge National Lab~\cite{gitORNL} using \texttt{POLARIS} and \texttt{PARCS}.\\[0.5\baselineskip]
Chapter~\ref{section:modellingCycles} showcases the gathering of publicly available data for the initial cycles of PWRs, as well as the implementation and verification of the computational approach for seven PWR cycles.\\[0.5\baselineskip]
Chapter~\ref{section:nuclearDataValidation} provides a methodology to assess the performance of nuclear data libraries on the prediction PWR integral measurements.\\[0.5\baselineskip]
Finally, Chapter~\ref{section:conclusion} provides a concise conclusion that summarises the key findings and outlines potential directions for future researches.

\newpage
\section{Modeling of Pressurised Water Reactors with a Conventional Two-Steps Approach}
\label{section:ComputationalScheme}

\subsection{Overview of the Computational Scheme}
\label{subsection:schemeOverview}
The complexity of nuclear reactors prevents the use of analytical models that produce a comprehensive exact description of the system. They involve numerous interactions between neutrons, gamma rays and matter. Approximations have to be made to simulate the system's parameters and its evolution. To model complex systems, a common approach is the use of the Monte-Carlo method, modeling the fundamental physics of interactions between radiations and matter. Significant advantages of these stochastic methods are an improved geometry fidelity and avoiding to make approximations to get to equations that can be solved deterministically. However, this method comes with a large computational cost~\cite{whyNotMC}. A large amount of neutron histories need to be simulated. Approximations need to be make to reduce the variance, achieve faster convergence and reduce the computational cost~\cite{Kaiwen_MC_complex}. Monte-Carlo full core models suitable to perform core-follow calculations, e.g. model the evolution of a nuclear reactor over a cycle are few and difficult to use for routine application. They are not yet a standard approach in reactor analysis. Nonetheless, given the potential for reduced bias compared to deterministic methods (which rely on approximations like diffusion approximations to achieve results) heterogeneous continuous energy full-core Monte Carlo codes are currently being developed and validated, as they are crucial for safety and efficiency considerations~\cite{SerpentFullCoreValidation1,SerpentFullCoreValidation2,hursinTowardMC_FullCore}. \\[0.5\baselineskip]

The deterministic codes require the solving of equations governing the neutronic behaviour in a nuclear reactor and its isotopic content. These equations are the neutron transport equation (Eq.~\eqref{eq:transport}) and the Bateman equations (Eq.~\eqref{eq:Bateman}~\cite{PNR2}).
\begin{equation}
\begin{aligned}
    \frac{1}{v} \frac{d\phi}{dt} = & - \boldsymbol{\Omega} \cdot \boldsymbol{\nabla} \phi + \frac{1}{4\pi} \left[ \chi_P(E) \int_0^\infty dE' \int_{4\pi} d\Omega' \, \bar{\nu} \Sigma_f(t, r, E', T) \phi' + \sum_{i=1}^I \chi_{D,i}(E) \lambda_i N_i \right] \\
    & + \int_0^\infty dE' \int_{4\pi} d\Omega' \, \Sigma_S(t, r, \Omega' \to \Omega, E' \to E, T) \phi' \\
    & - \left[ \int_0^\infty dE' \int_{4\pi} d\Omega' \, \Sigma_S(t, r, \Omega \to \Omega', E \to E', T) + \Sigma_A(t, r, E, T) \right] \phi
\end{aligned}
\label{eq:transport}
\end{equation}
with $\phi:=\phi(t,\boldsymbol{r},E,\Omega)$, $\phi':=\phi(t,\boldsymbol{r},E',\Omega')$, $\Sigma_X=\displaystyle{\sum_{i=1}^{I}N_i\sigma_{X,i}}$, $v=v(E)$ and $N_i:=N_i(t,\boldsymbol{r},T)$. $\phi$ stands for the angular neutron flux, $v$ stands for the neutron scalar velocity, $\boldsymbol{\Omega}$ stands for the solid angle vector in direction of motion, $\Omega$ stands for the solid angle, $E$ stands for the neutron's energy, $T$ stands for the medium's temperature, $\lambda_i$ stands for the decay constant of delayed neutron precursor group $i$, $N_i$ stands for the number density of isotopes in the precursor group $i$, $I$ stands for the number of precursor groups, $\Sigma_S$ stands for the macroscopic scattering cross section, $\Sigma_f$ stands for the macroscopic fission cross section, $\Sigma_A$ stands for the macroscopic absorption cross section, $\bar{\nu}$ stands for the average number of neutrons emitted per fission, $\chi_P(E)$ stands for the probability density function of the generation energy of prompt neutrons, $\chi_{D,i}(E)$ stands for the probability density function of the generation energy of delayed neutrons from precursor group $i$.

\begin{equation}
\begin{aligned}
    \frac{dN_i}{dt} = & \sum_{j \neq i}^I \left[ \lambda_{j \to i} N_j + \int_0^\infty dE' \int_{4\pi} d\Omega' \, \sigma_{j \to i}(E', T) \phi' \right] \\
    & - \left[ \int_0^\infty dE' \int_{4\pi} d\Omega' \, \sigma_{i \to j}(E', T) \phi' \right] - \lambda_i N_i
\end{aligned}
\label{eq:Bateman}
\end{equation}
where $N_i$ stands for the number density of nucleus $i$, $\sigma_{j\to i}$ (resp.~$\lambda_{j\to i}$) stands for the microscopic cross section of nucleus $j$ to transmute to nucleus $i$ due to a neutron interaction (resp.~ due to spontaneous decay).

While they are direct whole core transport calculations codes (refer to \texttt{MPACT}~\cite{MPACTdirectTransport} or \texttt{nTRACER}~]\cite{nTRACERdirectTransport}), they demand substantial computational resources. Due to the high computational burden associated with these methods, this approach will not be used in this Master's Thesis.\\[0.5\baselineskip]
The work presented in this Master's Thesis uses a conventional two-step approach to solve this set of coupled equations. The general idea of the this computational scheme is to simplify the whole core simulation by splitting the task in two consecutive simpler steps. The spatial scale is gradually increased throughout the steps. The description of the physics is also simplified, with careful attention to preserve the local reaction rates. Complicated physics (e.g.~Doppler broadening and corrections for self-shielding in both space and energy) is incorporated into spatially homogenised few-group cross sections. This spatial homogenisation enables the execution of full core calculations at a manageable computational cost and makes it possible to use diffusion theory for flux solutions. \\[0.5\baselineskip]
First, the physics of lattice cells with reflexive boundary conditions is simulated. In this master's project, the lattice cells consist of two-dimensional representations of the different types of fresh fuel assemblies, based on their isotopic composition. The goal of this step is to generate a case matrix, i.e. a set of energy condensed spatially homogeneous lattice cell for a set of state variables to fully characterise fuel design for \texttt{PARCS}. These homogenised cells are meant to be physically equivalent to the heterogeneous description when it comes to predicting the global parameters of the reactor (the effective neutron multiplication factor $k_\textrm{eff}$, diluted boron concentration in water, smooth representation of power distributions, \ldots).
Although this step can be done with a Monte-Carlo code (for example with \texttt{Serpent}~\cite{Serpent_LatticeCalculations}) the deterministic code \texttt{DRAGON}~\cite{Dragon} is used as the lattice code for this Master's Thesis. This decision is driven by the entirely open-source nature of the project and by the availability of open-source models of the fuel assemblies of Fessenheim-2 and Tihange-1~\cite{gitSalino}. Moreover, \texttt{PyNJOY}\cite{PyNJOY} (the code used to process nuclear data libraries for the input of \texttt{DRAGON}) is also open-source and uses \texttt{Python}, which is also open-source.\\
The second stage of the two-step approach consists in the simulation of the full core by solving the diffusion form of the neutron transport equation and the Bateman equation. Diffusion theory is based on the following approximations:
\begin{itemize}
    \item Linear anisotropy of the angular flux;
    \item The neutron current density linearly depends on the scalar flux gradient (Fick's first law, see~Eq.~\eqref{eq:1stFickLaw});
    \begin{equation}
    J_g(r,t) = -D_g \nabla \phi_g(r,t)
    \label{eq:1stFickLaw}
    \end{equation}
    \item The medium is infinite and homogeneous in space;
    \item Scattering prevails over absorption and is isotropic within the laboratory frame of reference;
    \item Broad energy groups are used in multi-group condensation.
\end{itemize}

These approximations enable the formulation of the group neutron diffusion equation~\eqref{eq:diffusionNeutron}.
\begin{equation}
\frac{1}{v_g} \frac{\partial}{\partial t} \phi_g(r,t) - D_g \nabla^2 \phi_g(r,t) + \Sigma_{r,g} \phi_g(r,t) = \sum_{g' \neq g} \Sigma_{s,g' \rightarrow g} \phi_{g'}(r,t) + \frac{\chi_g}{k} \sum_{g'} \nu \Sigma_{f,g'} \phi_{g'}(r,t)
\label{eq:diffusionNeutron}
\end{equation}
where $v_g $ stands for the group velocity of neutrons in the energy group $g$, $\phi_g(r,t)$ stands for the neutron flux in the energy group $g$ at position $r$ and time $t$, $D_g$ stands for the diffusion coefficient for neutrons in the energy group $g$, $\Sigma_{r,g}$ stands for the macroscopic removal cross-section for neutrons in the energy group $g$, $\Sigma_{s,g' \rightarrow g}$ stands for the macroscopic scattering cross-section from energy group $g'$ to energy group $g$, $\chi_g$ stands for the fraction of neutrons produced from fission events that are emitted in the energy group $g$, $k$ stands for the effective multiplication factor of the system, $\nu$ stands for the average number of neutrons produced per fission event, $\Sigma_{f,g'}$ stands for the macroscopic fission cross-section for neutrons in the energy group $g'$.

As the diffusion equation solved in the second step assumes spatial homogenisation it is not valid near localised neutron absorbers (e.g.~control rods, fuel rods) and large moderator regions (e.g. reflectors). Hence, the lattice cells to homogenised is chosen to be fuel assemblies ($\sim$~20~cm).

\begin{figure}
    \centering
    \includegraphics[scale=0.7]{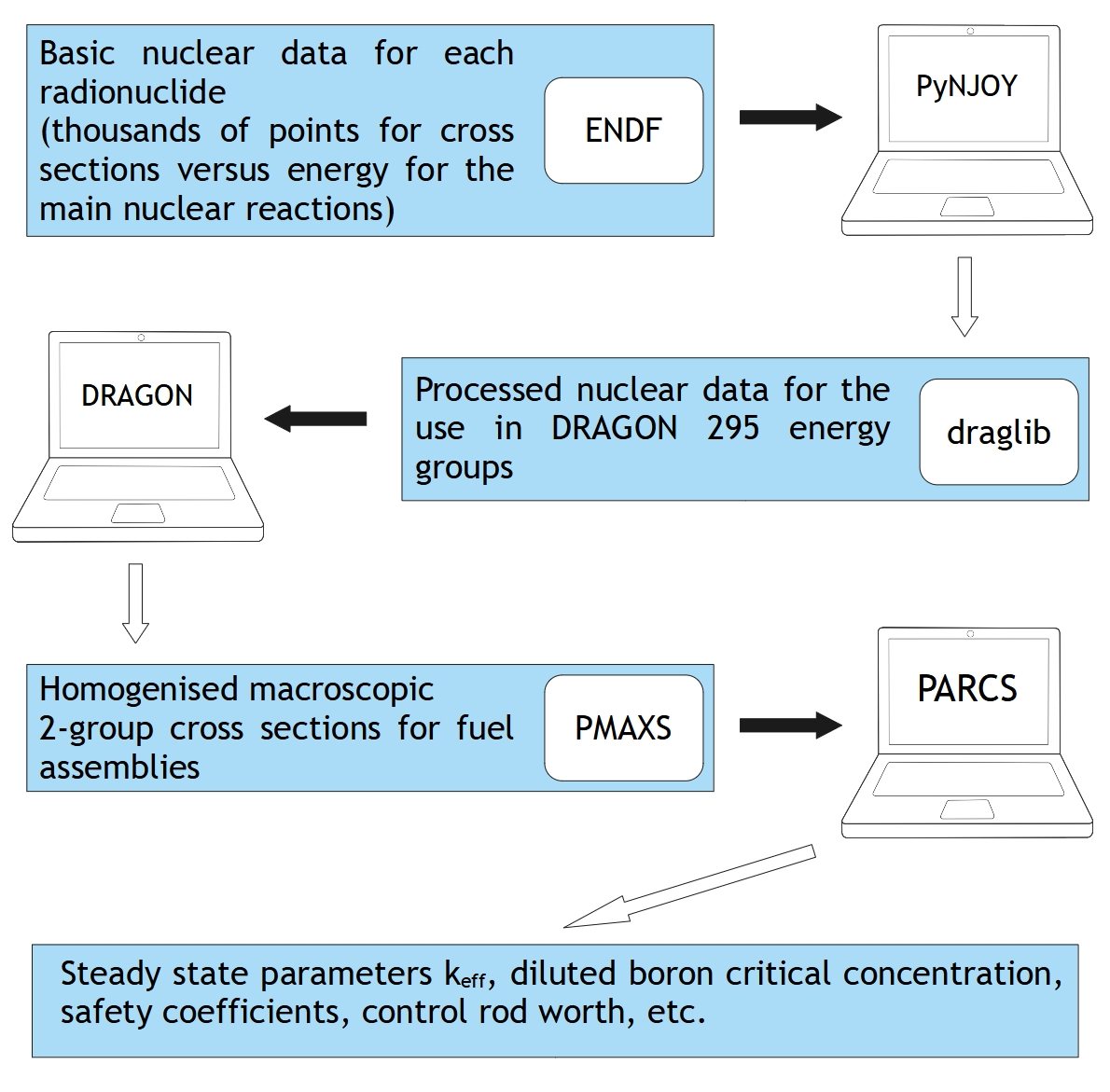}
    \caption{Schematics of the computational method used to model nuclear reactors. The outline of this schematics is inspired by a schematics of the ETHZ's course \textit{Physics of Nuclear Reactors II}~\cite{PNR2}.}
    \label{fig:computationScheme}
\end{figure}

The computational approach is schematically represented in figure~\ref{fig:computationScheme}. The specifics of each step will be examined in the following subsections.\\

\subsection{Generation of the \texttt{DRAGON} Nuclear Data Library with \texttt{PyNJOY}}
\begin{figure}[h]
    \centering
    \includegraphics[width=\linewidth]{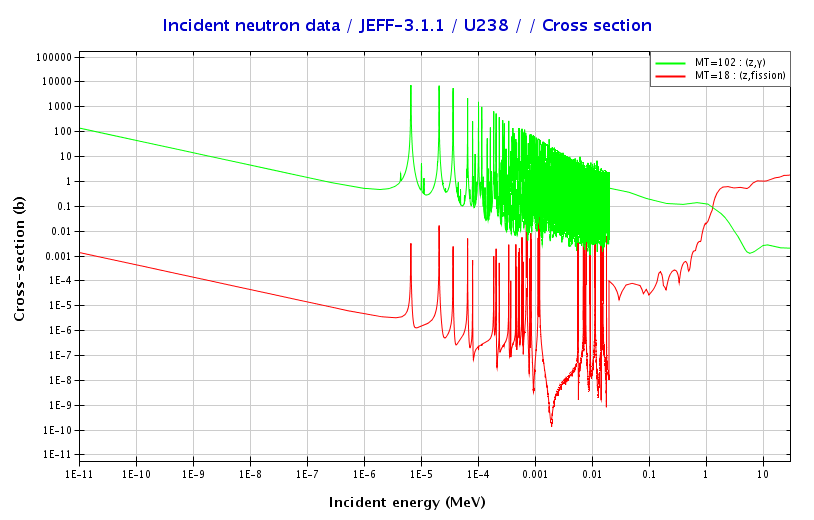}
    \caption{Point wise cross section for neutron capture and fission for \ce{^{238}U} in \texttt{JEFF--3.1.1}~\cite{JANIS,JEFF}.}
    \label{fig:XS_U238}
\end{figure}

At first, the nuclear data processing code \texttt{PyNJOY2016}~\cite{PyNJOY} is used to reduce the raw nuclear data libraries to a few hundreds of energy groups for downstream use in Dragon. This is done once for a nuclear data library as it is not specific to any particular physical system to model. The raw nuclear data libraries, encompassing thousands of data points for cross sections as a function of energy across various nuclear reactions, such as neutron capture, production, absorption, (n,\,2n), scattering, fission spectrum, total and gamma transport. This data spans an energy range of approximately 10~$\mu$eV to 20~MeV. In this Master's Thesis, the \texttt{JEFF--3.1.1}~\cite{JEFF}, \texttt{JEFF--4T3}~\cite{JEFF}, and \texttt{ENDF/B--VIII.0}~\cite{ENDF} libraries are utilised. These libraries are formatted as \texttt{ENDF} (Evaluated Nuclear Data File) with a point-wise structure. An example of these data is presented for \ce{^{238}U} on figure~\ref{fig:XS_U238}. A collection of nuclear data libraries are publicly made available by the IAEA Nuclear Data Services~\cite{dataIAEA}.

\texttt{PyNJOY2016}~\cite{PyNJOY} is used to convert nuclear data from the \texttt{ENDF} format to a \texttt{draglib} file, which features a group-wise structure with 295 energy groups using the SHEM 295 group structure~\cite{SHEM295}, allowing for a fine energy group resolution. This \texttt{draglib} file can be used as an input for \texttt{DRAGON}. The objective is to reduce the size of the database by partitioning the entire energy range into distinct groups and calculating averaged values for each group. The main feature of this energy condensation is to preserve the integrated reaction rate inside the energy range of each group.

\subsection{Lattice Cell Calculations with \texttt{DRAGON}}
This first stage of the two-step approach is performed with a slightly modified version of  \texttt{DRAGON} \cite{gitSalino}, based on version 5.0.8 revision 2417, to get macroscopic 2-group cross sections for chosen assembly models.

\texttt{DRAGON} is a deterministic lattice code developed in \texttt{FORTRAN-2003} at Polytechnique Montréal. It comprises a collection of modules designed to simulate the neutronic behaviour of a lattice cell~\cite{DragonUserGuide}. These modules include, but are not limited to, multigroup multidimensional neutron flux calculations, self-shielding of microscopic cross sections, and isotopic depletion calculations.
The objective of \texttt{DRAGON} is to match an heterogeneous lattice cell to a spatially homogenised one with a coarse energy group structure. The output of this calculation is a set of two-group cross-sections describing spatially homogenised fuel assemblies, which can be used to model nodes by \texttt{PARCS}. It accomplishes this by ensuring that the homogeneous cell maintains the same cell-integrated reaction rates as its heterogeneous counterpart.
The \texttt{DRAGON} code calculates the detailed flux distribution, which is necessary for condensing and homogenising cross sections, for various depletion condition histories (so-called histories) and different instantaneous core parameters (so-called branch calculations) to account for the different conditions fuel assemblies will encounter in the reactor. This topic will be explored in details in Chapters~\ref{section:history} and \ref{section:modellingCycles}.

In the context of this research, it is essential to highlight that IRSN's modified version of \texttt{DRAGON}, dedicated to the production of \texttt{PMAXS} files, is not based on the latest iteration of \texttt{DRAGON5}~\cite{Taforeau}. This distinction leads to specific dependencies, including the requirement for the library \texttt{libgfortran.so.3}, which is incompatible with the most recent versions of the Ubuntu operating system. To effectively work with \texttt{DRAGON}, a Docker container can be used to create a suitable environment~\cite{docker}. A working configuration includes \texttt{Ubuntu 18.04} along with the following compiler versions: \texttt{gcc 6}, \texttt{g++ 6}, and \texttt{gfortran 6} (see the docker configuration in Appendix A.1). This setup ensures that all necessary dependencies are met, allowing for the successful execution of the \texttt{DRAGON} code in a controlled and reproducible manner. To integrate the \texttt{DRAGON} modules into a unified computational framework, procedures written in the supervisory language \texttt{CLE-2000}~\cite{CLE2000UserGuide} are employed, specifically designed for this purpose.

Then, \texttt{GENPMAXS} is used for the interface from the output of \texttt{DRAGON} to a file format readable by the full core simulator \texttt{PARCS}~\cite{GenPMAXS_manual}. The resulting files, known as \texttt{PMAXS} (\texttt{PARCS} Macroscopic Cross-Section set), contain a series of cross-section tables (case matrices) tailored to specific types of fuel assemblies or reflectors. These tables provide 2-group macroscopic cross-sections as a function of various state parameters. The state parameters considered are burnup, diluted boron concentration, fuel temperatures, moderator density, and control rod insertion.

\subsection{Full Core Simulations with \texttt{PARCS}}
In the second stage of the two-step approach, \texttt{PARCS} (Purdue Advanced Reactor Core Simulator)~\cite{PARCS_1_Inputs} is used to obtain the steady state parameters, such as the effective neutron multiplication factor $k_{\textrm{eff}}$, the diluted boron critical concentration, feedback coefficients and control rod worths. \texttt{PARCS} is a deterministic three-dimensional reactor core simulator which solves the steady-state and time-dependent multi-group neutron diffusion and low-order neutron transport equations in Cartesian or hexagonal fuel geometries. It is a stand-alone code which includes a simple Thermal-Hydraulics solvers suitable for PWR and BWR analysis. To solve the neutron diffusion equation, it uses the \texttt{PARCS} hybrid solver, which uses a combination of the nodal expansion method (NEM) and the analytic nodal method (ANM). It has been created and developed at the University of Purdue and is currently under development at the University of Michigan. The U.S. Nuclear Regulatory Commission (NRC) uses this code as part of their regulatory framework and safety assessments~\cite{NRC_PARCS}. It has been validated a for few systems including Pressurised Water Reactors (PWRs)~\cite{NRC_PARCS, ORNL_TP3}.

To satisfy the assumptions of the diffusion equation, \texttt{PARCS} decomposes the nuclear reactor in so-called nodes. These nodes are homogeneous cells where the diffusion equation is solved. The cross sections generated with \texttt{DRAGON} are used to represent the physics in these cells. A radial decomposition of the core as a set of assemblies is chosen. Axially, each assembly is decomposed into dozens of nodes to capture the axial dependencies.

\texttt{PARCS} allow to simulate multiple cycle with fuel assembly shuffle between them since \texttt{PARCS v3.4.2}. The parameters that can change during each cycle are the size of the depletion steps, the power level during each step, the water flow rate, the core inlet water temperature, the core exit water pressure and the control banks insertion. During the intervals between cycles, the reactor enter a shutdown cooling phase, which is represented by a power level of -1 parameter of the \texttt{POWER\_LEV} \texttt{PARCS} card~\cite{PARCS_1_Inputs}. However, \texttt{PARCS v3.4.2} struggles to handle this phase effectively when multiple cycle depletions are involved. It appears necessary to implement depletion steps immediately before and after the cooling phase to prevent an unexpected termination of the simulation. A functional implementation of a shutdown cooling cycle in \texttt{PARCS} is demonstrated in \texttt{PARCS}~input~\ref{lst:shutdown_cooling}, using minimal depletion steps of $10^{-5}$~MW/kgU to minimise the impact on the simulation.

\begin{minipage}{\linewidth}
\begin{lstlisting}[caption={30 days shutdown cooling at $100^{\textrm{o}}$C.}, label={lst:shutdown_cooling}]
CYCLE_DEF 2
DEPL_STEP -0.00001 30 -0.00001
POWER_LEV   3 -1 3 3
BANK_SEQ 4*1
INLET_ENT -562 -372 2*-562
FLOW_RATE 4*82
\end{lstlisting}
\end{minipage}

Between cycles, it is possible to shuffle fuel assemblies, which involves changing the locations of some fuel assemblies and loading fresh fuel assemblies. If the spent fuel assemblies are directly from the previous cycle, multiple cycles can be modelled using a single \texttt{PARCS} input file. In this scenario, the shuffle consists of assigning each spent fuel assembly ID (defined with the \texttt{LOCATION} card) to its new location using the \texttt{SHUF\_MAP} card. However, if a spent fuel assembly does not come from the spent fuel pool (such as the central fuel assembly of the third cycle of Turkey-Point-3, which was only burned during the first cycle) a different \texttt{PARCS} input file is required for that cycle.

To ensure accurate modeling, the output options for all relevant history variables must be set to true in the \texttt{OUT\_OPT} card for all previous cycles that include a fuel assembly that will be used in the cycle of interest (in this case, the third cycle of Turkey-Point-3). The \texttt{PMAXS} output file, which has a name ending in \verb|parcs_cyc-*| (or the file ending in \verb|parcs_dep| if depletion was performed without the multiple cycle option), along with the file containing the fuel assembly locations from previous cycles, will serve as inputs for the \texttt{PREV\_CYC} card. The rest of the process functions the same manner as with the previous method.

Pyrex rods being modelled in \texttt{DRAGON} as control rods with special composition and locations, they are treated as such in \texttt{PARCS}, allowing for their removal at any time during and between cycles.

\newpage
\section{Modeling History Effects with \texttt{DRAGON} and \texttt{PARCS}}
\label{section:history}

\subsection{Handling of Instantaneous and History Variables in \texttt{PARCS}}
\label{subsection:understandingBranchesHistories}

The concepts of histories and branches can exhibit variability depending on the nuclear codes used. The aim of this section is to provide a comprehensive understanding of how histories and branches are defined and implemented in \texttt{PARCS}.

A preliminary depletion calculation, commonly referred to as a history calculation, is conducted under specific conditions (characterised by fuel temperature, moderator density, diluted boron concentration, and control rod insertion) for a particular type of fuel assembly using \texttt{DRAGON}. Following this initial computation, a static analysis is performed by introducing a perturbation to the reference conditions at a specified burnup level; this analysis is termed a branch. In this perturbed system, the macroscopic cross sections for each energy group (encompassing transport, absorption, and fission) along with the average energy per fission and the average number of neutrons emitted per fission, are calculated.\\[0.5\baselineskip]
The branches are classified as follows~\cite{GenPMAXS_manual}:
\begin{outline}
    \1 Start by choosing a reference branch, which will be referred to as "RE". The cross sections for this branch are stored as is in the \texttt{PMAXS} file.
    \1 Then change one variable $A$ by an amount $\Delta A$, which can be positive or negative. Store the differential cross sections $\left(\frac{\partial \Sigma}{\partial A}\right)_1$, as computed with equation~\eqref{eq:branchA1}, in the \texttt{PMAXS} file. This will be the first branch of type $A$.

\begin{equation}
\left(\frac{\partial \Sigma}{\partial A}\right)_1=\frac{\Sigma(\textrm{RE})-\Sigma(\textrm{RE}+\Delta A)}{\Delta A}
\label{eq:branchA1}
\end{equation}
where $\textrm{RE}+\Delta A$ refers to the conditions of the reference branch where the variable $A$ is perturbed by an amount $\Delta A$.

    \1 Then change an other variable $B$ by an amount $\Delta B$. Variable $B$ will be associated with two branches. Store the differential cross sections ${\left(\frac{\partial \Sigma}{\partial B}\right)}_1$, as computed with equation~\eqref{eq:branchB1}, in the \texttt{PMAXS} file for the first branch of type $B$ and store  the differential cross sections ${\left(\frac{\partial \Sigma}{\partial B}\right)}_2$, as computed with equation~\eqref{eq:branchB2}, in the \texttt{PMAXS} file for the second branch of type $B$.
    
\begin{equation}
{\left(\frac{\partial \Sigma}{\partial B}\right)}_1=\frac{\Sigma(\textrm{RE})-\Sigma(\textrm{RE}+\Delta B)}{\Delta B}
\label{eq:branchB1}
\end{equation}

\begin{equation}
{\left(\frac{\partial \Sigma}{\partial B}\right)}_2=\frac{\Sigma(\textrm{RE}+\Delta A)-\Sigma(\textrm{RE}+\Delta A+\Delta B)}{\Delta B}
\label{eq:branchB2}
\end{equation}
    
\end{outline}

\noindent Then repeat for all histories in other depletion conditions with the same branches.

Consider a scenario with two control positions, denoted as $c$ (out and in, corresponding to $c=0$ and $c=1$, respectively), and three coolant densities, $DC^i$ ($i\in\{1,2,3\}$). The cross section is obtained through interpolation, using equation~\eqref{eq:interpolationXS_DC} in conjunction with equations~\eqref{eq:differentialXS_DC} and \eqref{eq:weight_DC}. The interpolation process is conducted in a manner analogous to that used for branches.
\begin{equation}
\Sigma(c, DC) = \Sigma^r + c \frac{\partial \Sigma}{\partial Cr} \Bigg|_{(Cr^1 / 2)} + (DC - DC^r) \frac{\partial \Sigma}{\partial DC} \Bigg|_{(c, DC^m)}
\label{eq:interpolationXS_DC}
\end{equation}

\begin{equation}
    \frac{\partial \Sigma}{\partial DC} \bigg|_{(c, DC^m)} = w_1 \frac{\partial \Sigma}{\partial DC} \bigg|_{(0, DC^{1m})} + w_2 \frac{\partial \Sigma}{\partial DC} \bigg|_{(0, DC^{2m})} + w_3 \frac{\partial \Sigma}{\partial DC} \bigg|_{(1, DC^{3m})} + w_4 \frac{\partial \Sigma}{\partial DC} \bigg|_{(1, DC^{4m})}
\label{eq:differentialXS_DC}
\end{equation}

\begin{equation}
    \begin{aligned}
        w_1 &= (1 - c) \frac{DC - DC^2}{DC^1 - DC^2} \\
        w_2 &= (1 - c) \left( 1 - \frac{DC - DC^2}{DC^1 - DC^2} \right) \\
        w_3 &= c \frac{DC - DC^4}{DC^3 - DC^4} \\
        w_4 &= c \left( 1 - \frac{DC - DC^4}{DC^3 - DC^4} \right)
    \end{aligned}
\label{eq:weight_DC}
\end{equation}

\noindent In \texttt{PARCS}, history variables are computed using equation~\eqref{eq:historyFormula}
\begin{equation*}
\begin{aligned}
    H(B_0 + \Delta B) = \frac{H(B_0) \cdot B_0 + \omega\cdot V_1 \Delta B}{B_0 + \omega\cdot\Delta B}
\end{aligned}
\label{eq:historyFormula}
\end{equation*}

where:

\begin{itemize}
    \item $H(B)$: History value at burnup $B$,
    \item $B_0$: Burnup at beginning of this depletion step,
    \item $\Delta B$: Burnup increment during this step,
    \item $V_1$: Value of the parameter at the end of the depletion step.
    \item $\omega$: Nonlinear weighting factor. The \texttt{PARCS} manual recommends selecting values between 2 and 2.5 for all history treatments~\cite{PARCS_1_Inputs}. This factor is influenced by the observation that complex feedback effects in a reactor tend to moderate the impact of histories over time. A value of 2.25 was chosen for the Pyrex insertion history, while a value of 1.0 was used for the moderator density history. However, there appears to be no theoretical justification for these specific values.
\end{itemize}

\begin{figure}
    \centering
    \includegraphics[width=\linewidth]{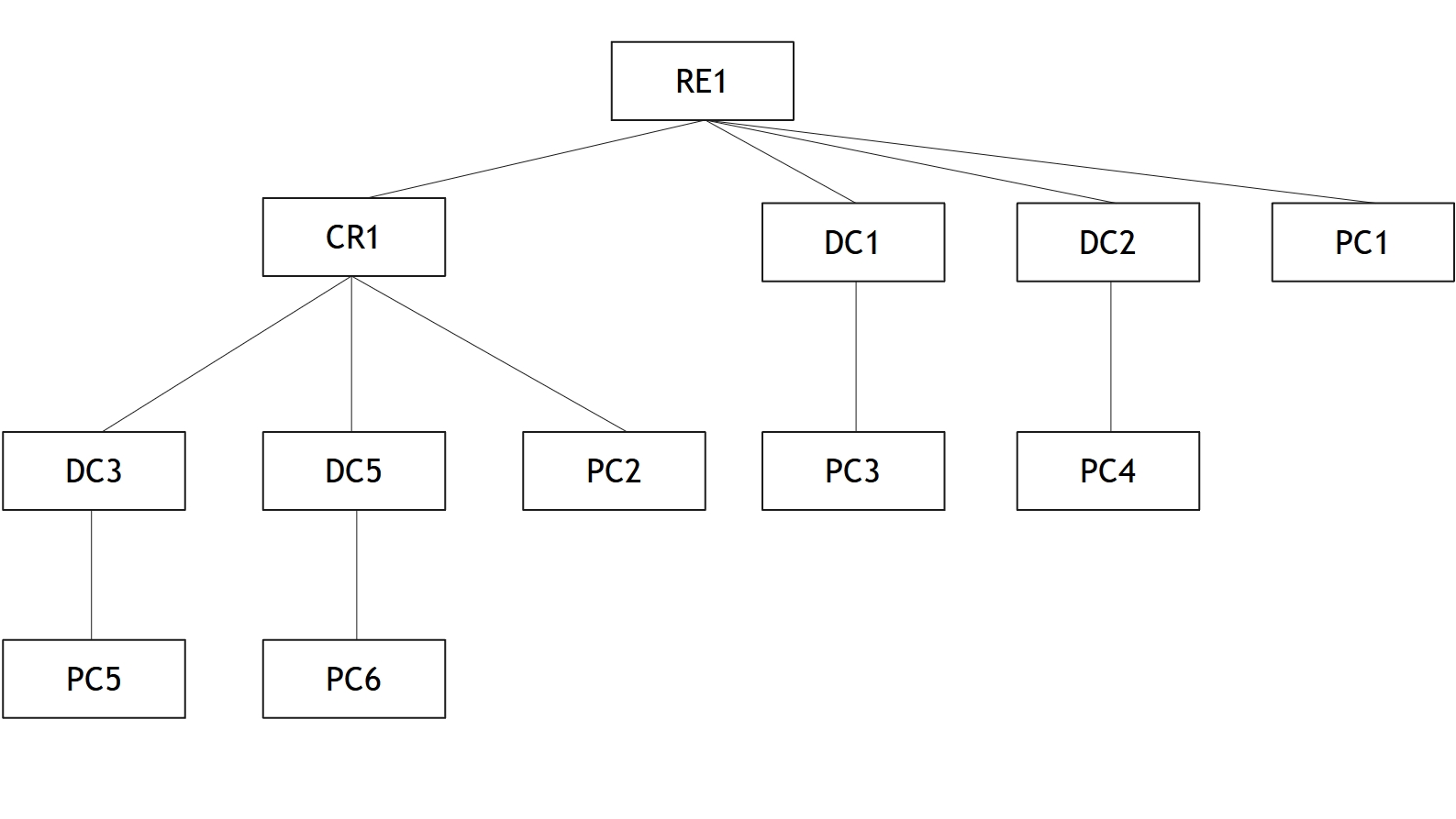}
    \caption{Graph representing the connections between branches in a basic example. Two branches are connected if and only the branch above is taken as a reference to compute the differential cross section of the branch below (see~Eq.~\eqref{eq:branchA1}-\eqref{eq:branchB2}).}
    \label{fig:branchOperation}
\end{figure}

To illustrate the recovery of the cross sections from the differential cross sections, let's consider an example. Consider the branches given in table~\ref{tab:branchOperation}. Using the branch construction process described before in this section, we can represent schematically the situation (see~Fig.~\ref{fig:branchOperation}).

The cross section in the conditions of PC5 is obtained using equation~\eqref{eq:computeBranch_PC5}.

\begin{equation}
\begin{aligned}
    \Sigma(\textrm{PC5})&=\Sigma(\textrm{DC3})+\Delta c_{\textrm{bor}}\textcolor{blue}{\frac{\partial \Sigma}{\partial c_{\textrm{bor}}}\bigg|_{5}}\\
    \Sigma(\textrm{DC3})&=\Sigma(\textrm{CR1})+\Delta d_{\textrm{mod}}\textcolor{blue}{\frac{\partial \Sigma}{\partial d_{\textrm{mod}}}\bigg|_{3}}\\
    \Sigma(\textrm{CR1})&=\textcolor{blue}{\Sigma(\textrm{RE1})}+\Delta c\textcolor{blue}{\frac{\partial \Sigma}{\partial c}\bigg|_{1}}\\
\end{aligned}
\label{eq:computeBranch_PC5}
\end{equation}
where $\Sigma(\textrm{A})$ stands for a macroscopic cross section for branch A, $\Delta c_{\textrm{bor}}=-1500$ is the difference of soluble boron concentration between branch DC3 and branch PC5, $\Delta d_{\textrm{mod}}=0.04$ is the difference of moderator density between branch CR1 and branch DC1 and $\Delta c=1$ stands for the insertion of control rods to go from branch RE1 to branch CR1. The quantities in blue are given in the \texttt{PMAXS} file. Each equation consists in going up one layer in figure~\ref{fig:branchOperation}, until the RE1 branch is reached.

\begin{table}
\resizebox{\linewidth}{!}{
\begin{NiceTabular}{|c|c|c|c|c|c|}
  \hline
  \cellcolor{blue!5}\makecell{Branch\\ type} & \cellcolor{blue!5} Label & \cellcolor{blue!5}\makecell{Control rod\\ insertion} & \cellcolor{blue!5}\makecell{Moderator\\ density} & \cellcolor{blue!5}\makecell{Soluble boron\\ concentration (ppm)} & \cellcolor{blue!5}\makecell{Fuel\\ temperature (K)}\\
  \hline
  RE & 1 & 0 & 0.71 & 1500 & 923.15 \\
  \hline
  CR & 1 & 1 & 0.71 & 1500 & 923.15 \\
  \hline
  DC & 1 & 0 & 0.75 & 1500 & 923.15 \\
  DC & 2 & 0 & 0.69 & 1500 & 923.15 \\
  DC & 3 & 1 & 0.75 & 1500 & 923.15 \\
  DC & 4 & 1 & 0.69 & 1500 & 923.15 \\
  \hline
  PC & 1 & 0 & 0.71 &    0 & 923.15 \\
  PC & 2 & 1 & 0.71 &    0 & 923.15 \\
  PC & 3 & 0 & 0.75 &    0 & 923.15 \\
  PC & 4 & 0 & 0.69 &    0 & 923.15 \\
  PC & 5 & 1 & 0.75 &    0 & 923.15 \\
  PC & 6 & 1 & 0.69 &    0 & 923.15 \\
  \hline
\end{NiceTabular}
}
\caption{Branch description in a basic example analysed in section~\ref{subsection:understandingBranchesHistories}}
\label{tab:branchOperation}
\end{table}

\begin{figure}
    \centering
    \includegraphics[width=\linewidth]{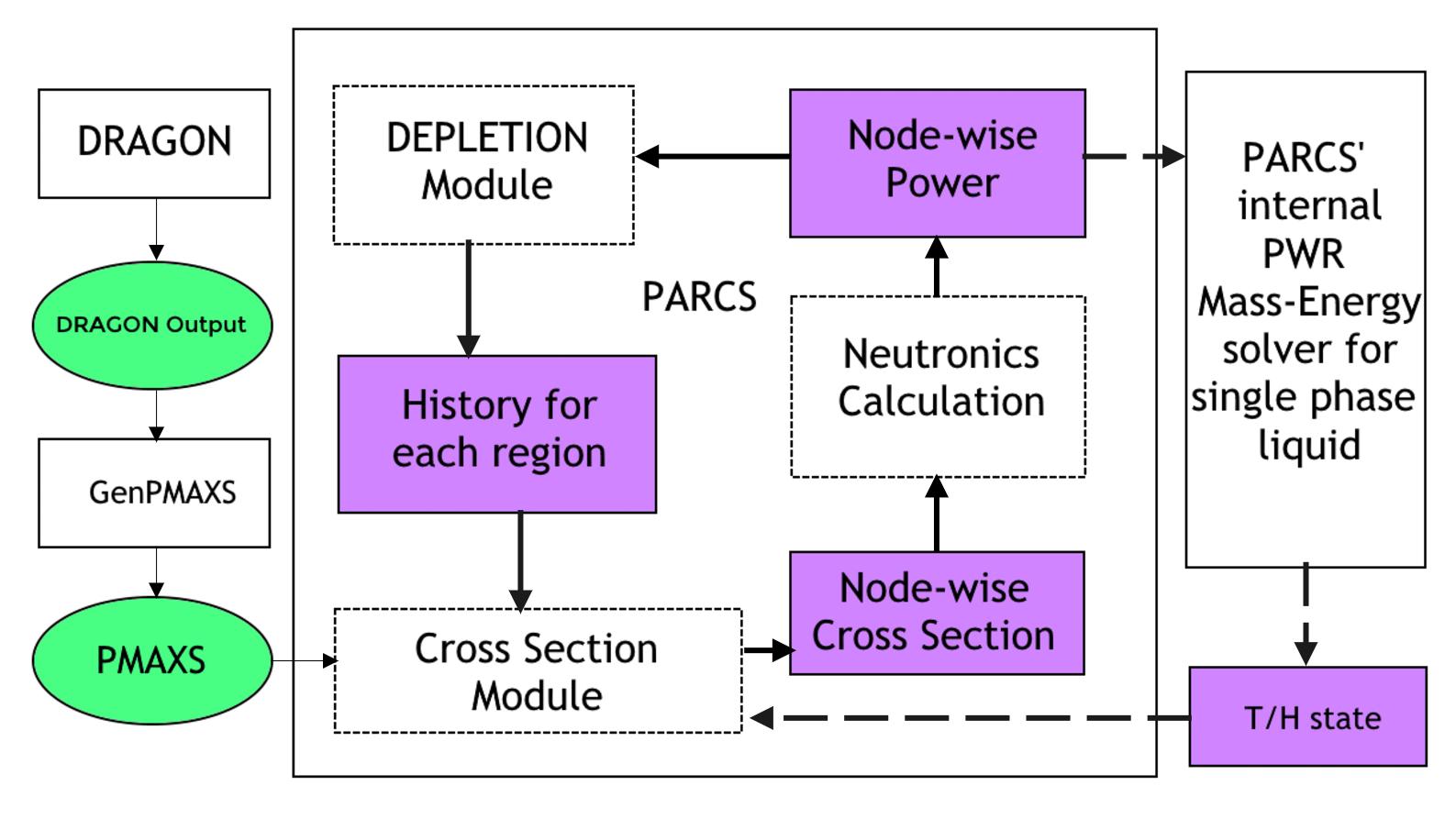}
    \caption{Outlines of \texttt{PARCS}' computational scheme for depletion. This schematics is inspired by a figure in the \texttt{GenPMAXS} manual~\cite{GenPMAXS_manual}.}
    \label{fig:PARCSloopTH}
\end{figure}

A high-level schematics of the place that have histories and cross section interpolation in \texttt{PARCS} is depicted on figure~\ref{fig:PARCSloopTH}.

\subsection{Motivations to use Histories}
\label{subsection:motivationHistory}

As outlined in the Introduction, the histories of Pyrex insertion and moderator density hold particular significance in the modeling of PWRs (see Fig.~\ref{fig:TP3ImpactHistoriesBoron}). This section aims to present a more formal theoretical justification for the use of these histories.\\[0.5\baselineskip]
In nominal conditions of PWRs, the density of water exhibits significant variation, ranging from ca.~0.75 to 0.65, as one moves from the bottom to the top of a fuel assembly. These different densities imply differences in moderation, which affect the neutron energy spectrum. The differences in the neutron spectrum subsequently influence the Bateman equations~\eqref{eq:Bateman}. Reduced moderation efficiency contributes to increased absorption in U-238, thereby facilitating greater plutonium buildup, and vice versa. Given that plutonium radionuclides are fissile, these changes in isotopic composition have a substantial effect on local reactivity. Such effects cannot be overlooked over extended operational periods when assessing axial power distributions and boron letdown curves. Consequently, long-term reactivity effects are modelled using histories of moderation density (depletion calculations at densities of 0.65, 0.71, and 0.77), as a single depletion condition proves inadequate.

\begin{figure}
    \centering
    \includegraphics[width=0.5\linewidth]{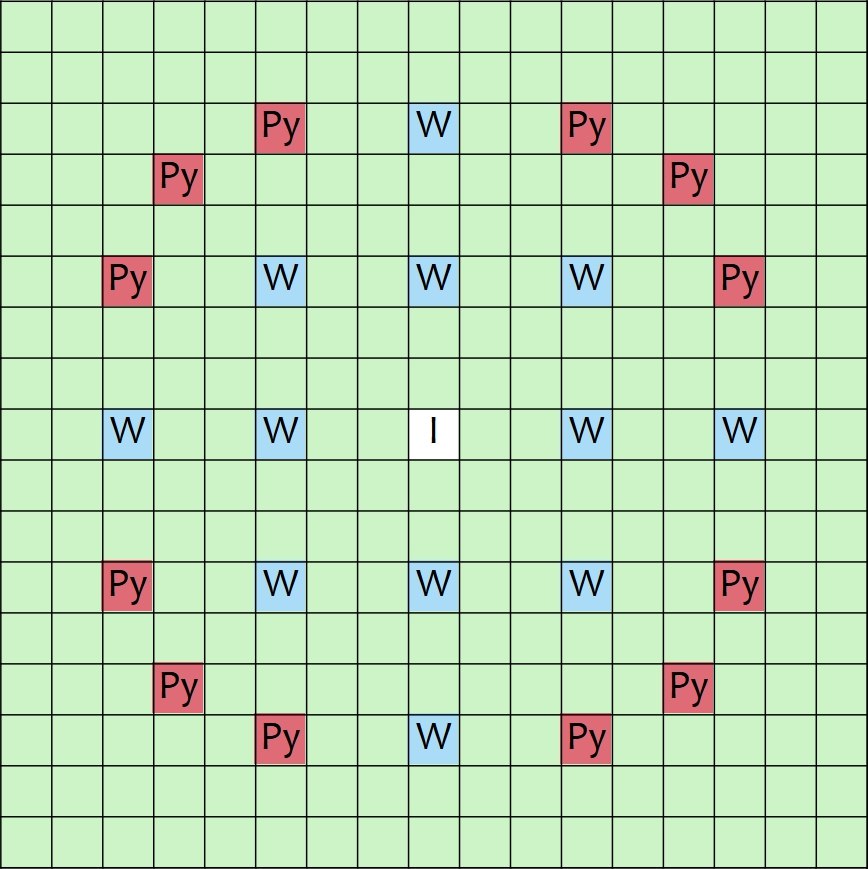}
    \caption{Schematics representation of a 15x15 fuel assembly of Turkey-Point-3, with 12 burnable absorber rods, used during its first cycles of operation. \texttt{I} stands for the central guide tube for measurement instrument insertion, \texttt{W} stands for water holes, \texttt{Py} stands for Pyrex rods and the green not-labelled cases represent fuel rods. In fuel assemblies without burnable poison rods, control rod banks can be inserted in place of the \texttt{W} and \texttt{Py} locations.}
    \label{fig:TP3_PyrexConfig}
\end{figure}

Pyrex serves as a burnable absorber (BA), also referred to as a burnable poison (BP). This material is a type of borosilicate glass containing 12.9 wt\% of \ce{B_2O_3}, with natural boron (comprising 80\,\% \ce{^{11}B} and 20\,\% \ce{^{10}B}). Notably, \ce{^{10}B} acts as a thermal neutron absorber. During the initial cycles of reactor operation, burnable absorbers can be employed to extend cycle lengths and mitigate excess reactivity at the beginning of each cycle. This Master's Thesis investigates designs incorporating Pyrex rods within specific fuel assemblies, in places where control rods are typically located (see~Fig~\ref{fig:TP3_PyrexConfig}). After one or two cycles, these rods are removed, leaving water holes and creating space for potential control rod insertion. As a thermal absorber, Pyrex influences the neutron energy spectrum in its vicinity, particularly within the fuel assembly. Therefore, similar to the considerations for water density, the history of burnable absorber usage must be accounted for to accurately predict power distribution and the soluble boron critical concentration in water to offset excess reactivity.

\subsection{Generate \texttt{PMAXS} Cross Sections with Histories using \texttt{DRAGON}}
\label{subsection:concatenateHiatories}
To the author's knowledge, \texttt{DRAGON} does not supply tools to create \texttt{PMAXS} files with histories with \texttt{GenPMAXS}. The following process is used to deal with this matter. It is to be repeated for each fuel assembly type.

\begin{figure}
    \centering
    \includegraphics[scale=0.7]{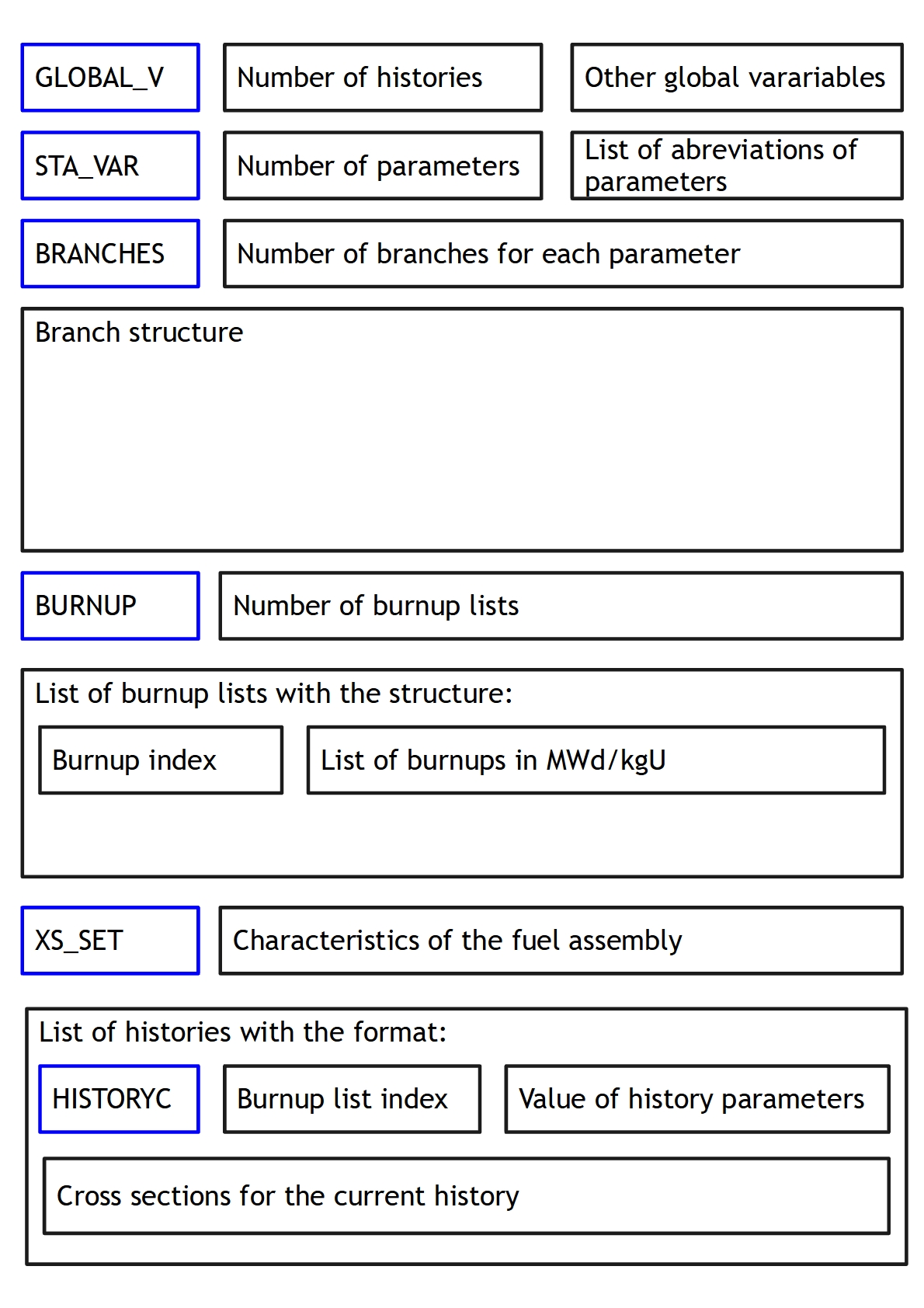}
    \caption{Structure of \texttt{PMAXS} files. Understanding this structure makes it possible to construct \texttt{PMAXS} files with histories for cross sections generated with \texttt{DRAGON} in various depletion conditions.}
    \label{fig:PMAXS_hst}
\end{figure}

A first depletion calculation in nominal core average conditions is performed with \texttt{DRAGON}, with 25 burnup steps from 0~MWd/kgU to 41~MWd/kgU. Then for each of these burnup steps, the state of the core is perturbed (by changing the instantaneous control rod insertion, moderator density, soluble boron concentration or the fuel temperature) and the macroscopic 2-group cross sections are computed. These form the branch calculations. A \texttt{PMAXS} file is generated with \texttt{GenPMAXS} following the algorithm presented in section~\ref{subsection:understandingBranchesHistories}.\\[0.5\baselineskip]
This process is repeated with various depletion conditions. Contrary to branches, for each history calculation, only one parameter differs from the conditions of the reference histories. These constitute the history calculations. Each history is associated with a \texttt{PMAXS} file. However, to use histories in \texttt{PARCS}, they have to be in a single \texttt{PMAXS} file. The concatenation of all histories into one \texttt{PMAXS} file is performed with a bash script with the following instructions, based on the structure of a \texttt{PMAXS} file containing one or more histories depicted in figure~\ref{fig:PMAXS_hst}. \\[0.5\baselineskip]

First, change the number of histories on the first line of the \texttt{PMAXS} constituting the reference history to be the total number of histories considered. Then concatenate at the end of this file the part after "HISTORYC" of the \texttt{PMAXS} corresponding to other histories, one after another in no particular order. This home-made concatenation of histories being not native, its performance will be analysed and discussed in section~\ref{subsection:history_validation}.

\subsection{Verification of History Implementation: Code-to-Code Verification between \texttt{DRAGON} and \texttt{POLARIS}}
\label{subsection:history_validation}

This section will focus on the verification of the implementation of history effect with the concatenation of \texttt{PMAXS} files. The histories implemented in \texttt{DRAGON} and verified are the Pyrex insertion history and the moderator density history. The \texttt{PARCS} prediction with and without history consideration will be analysed. The \texttt{PMAXS} cross sections used as input for \texttt{PARCS} come from the \texttt{DRAGON} model described in the former sections and from a \texttt{POLARIS} model developed by the Oak Ridge National Laboratory~\cite{gitORNL,POLARIS}. The branch parameters and the histories for Pyrex insertion and moderator density are the same in both cases (see~Tab.~\ref{tab:BranchesValuesVerifHistories}).

\begin{table}[H]
    \centering
    \begin{NiceTabular}{|c|c|c|c|c|}
        \hline
         & \makecell{Pyrex\\insertion} & \makecell{Moderator\\density} & \makecell{Boron concentration [ppm]} & \makecell{Fuel\\temperature [$^{\textrm{o}}C$]} \\
        \hline
        Value & \makecell{0 (Out)\\1 (In)} & \makecell{0.60832\\0.70081\\0.76999} & 500.0 & 800.0 \\
        \hline
    \end{NiceTabular}
    \caption{Values of branch and history value considered for the analysis in section~\ref{subsection:history_validation}.}
    \label{tab:BranchesValuesVerifHistories}
\end{table}

The lattice cell used to generate the \texttt{PMAXS} cross sections used in this section corresponds to a fuel assembly used in the first cycles of Turkey-Point-3~\cite{TurkeyPoint}. Its \ce{^235U} enrichment is 2.56\,\% and Pyrex can be inserted according to the map shown on figure~\ref{fig:TP3_PyrexConfig}."

\begin{table}[H]
    \centering
    \begin{tabular}{|c|c|c|}
        \hline
         & \texttt{DRAGON} & \texttt{POLARIS} \\
        \hline
        $k_{\textrm{inf}}$ & 1.062022 & 1.05125 \\
        \hline
    \end{tabular}
    \caption{Infinite neutron neutron multiplication factors in for lattice cells discussed in section~\ref{subsection:history_validation}, obtained with \texttt{DRAGON} (with diluted spacer grids) and with \texttt{POLARIS} (without spacer grids)}
    \label{tab:kinfBurn0_POLvsDRAG}
\end{table}

Table~\ref{tab:kinfBurn0_POLvsDRAG} shows that the $k_\infty$ obtained in this case with \texttt{DRAGON} and \texttt{POLARIS} are very different, with more than 1000~pcm of discrepancy. In Chapter 4, this discrepancy will be discussed again, but several sources of discrepancies can already be pinpointed. Firstly, the deterministic methods are different, which implies that the deterministic biases are different. Secondly, the nuclear data used with these different codes are different: \texttt{JEFF-3.1.1} with \texttt{DRAGON}, \texttt{ENDF/B-VII.1} with \texttt{POLARIS}. A third difference between both models is how they deal with spacer grids. Spacer grids are structures made of Inconel-718~\cite{TurkeyPoint}. They offer axial structural support, minimise rod vibration and bowing, and, in certain instances, facilitate coolant flow mixing~\cite{gridSpacerImpact}. The \texttt{DRAGON} model consider them homogeneously diluted in the lattice cell. On the other hand, the \texttt{POLARIS} model consider two different lattices cells for every fuel assembly type, thereby two \texttt{PMAXS} cross section sets. The first lattice cell corresponds to a radial section of a fuel assembly without spacer grid, the second one considers a radial section of a fuel assembly where spacer grids are located. To avoid the consideration of axial effects, the \texttt{PMAXS} cross sections corresponding to a lattice cell without fuel spacers has been considered here. This increase in moderation (attributed to the presence of water in space of spacer grids) will lead to a higher reactivity for the \texttt{PMAXS} obtained with \texttt{POLARIS}. Finally, it is difficult at this stage to rule out an error in the use of one of these two codes. A more definite conclusion will be reached in Chapter 4.

\begin{figure}[h]
    \centering
    \includegraphics[width=0.9\linewidth]{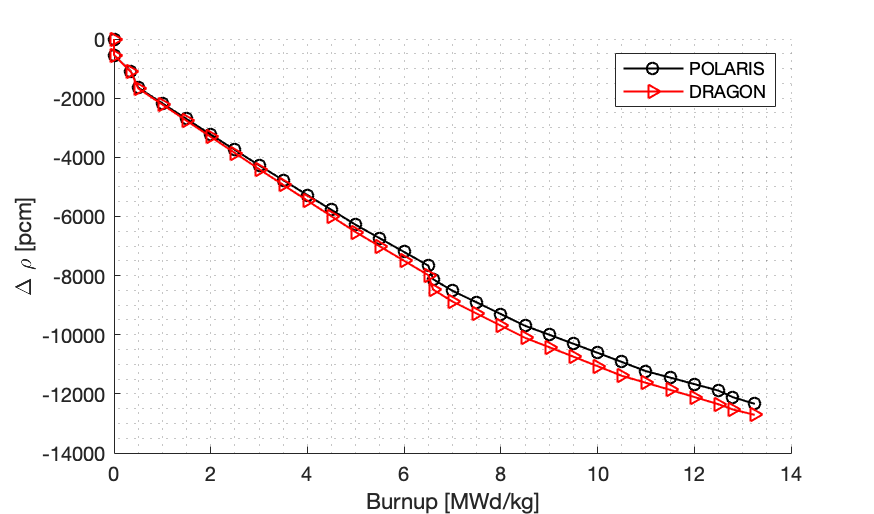}
    \caption{Difference in reactivity with and without the control rod history option for \texttt{PMAXS} cross sections generated with \texttt{DRAGON} and \texttt{POLARIS}.}
    \label{fig:keff_hstCR}
\end{figure}

To study the effect of Pyrex insertion history, a fuel assembly with Pyrex fully inserted is modelled with \texttt{PARCS}. No thermal feedback is considered. As Pyrex is modelled as a control rod in \texttt{PARCS}, the control rod history option is utilised. When this option is set to false, \texttt{PARCS} uses the reference history (the first one in the \texttt{PMAXS} file). When this option is activated, \texttt{PARCS} interpolates between histories using the procedure described in section \ref{subsection:understandingBranchesHistories}. The reference history in \texttt{PMAXS} files corresponds to a lattice cell with Pyrex extracted during depletion. Figure~\ref{fig:keff_hstCR} shows the difference in reactivity with and without the control rod history option for both \texttt{PMAXS} cross section sets. The trends of both models are the same, i.e. a strong reactivity reduction is observed when the Pyrex history effects are considered. Therefore, the procedure developed in section~\ref{subsection:concatenateHiatories} is considered verified for the Pyrex insertion history.

\begin{figure}[H]
    \centering
    \includegraphics[width=0.9\linewidth]{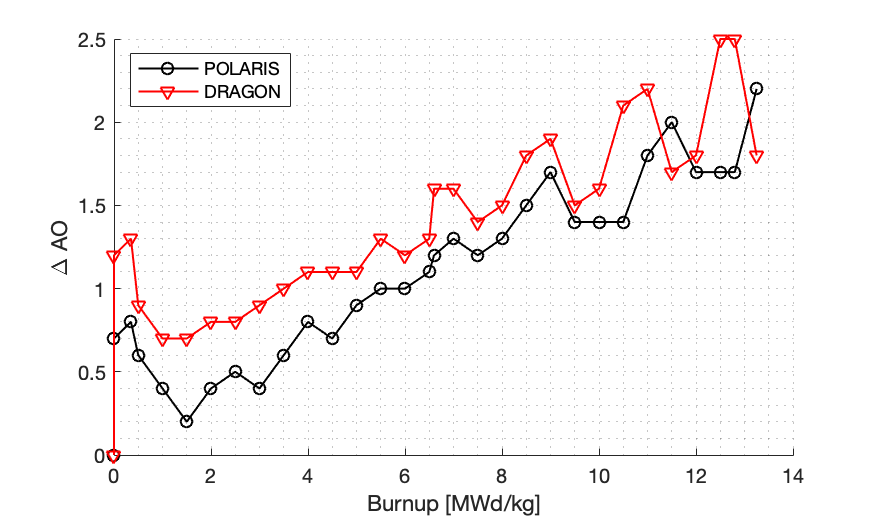}
    \caption{Difference in axial offset with and without moderator density history for \texttt{PMAXS} cross sections generated with \texttt{DRAGON} and \texttt{POLARIS}.}
    \label{fig:AO1D_hstmod}
\end{figure}

To study the effect of moderation density history, a fuel assembly without Pyrex but with thermal feedback is considered. Reflective boundary conditions are used on every face of this single assembly. The evolution of the axial offset, defined by equation~\eqref{eq:axialOffset}, is analysed during depletion. As coolant is injected from the bottom of the fuel assembly and is heated up as it travels to the the top of the assembly, the water is colder and denser on the bottom than on the top. As discussed in section~\ref{subsection:motivationHistory}, the reactivity is expected to evolve to be higher at the top of the core. This would lead to an axial offset increasing during depletion, starting from 0 for fresh fuel. Figure~\ref{fig:AO1D_hstmod} shows the difference in axial offset with and without moderator density history for cross sections generated with the \texttt{DRAGON} and the \texttt{POLARIS} models. Similarly as for the Pyrex insertion history, the two curves show clear evidence of following the same trend. The non-overlap is expected to come from the small difference in isotopic composition between both assemblies. Therefore, the procedure developed in section~\ref{subsection:concatenateHiatories} is considered verified for the moderator density history.

Given that this procedure is independent of the type of history variable and has demonstrated effectiveness with two different history variables, it can be regarded as verified. This result is the first major outcome of this Master's Thesis.

\newpage
\section{Modeling of Seven Pressurised Water Reactors Cycles}
\label{section:modellingCycles}

Measurements performed in commercial nuclear reactors are crucial for validating computational models and evaluating nuclear data. However, this data is often sensitive and subject to strict regulatory protections, making it challenging to access as it is not readily available in the public domain. Despite these challenges, measurements conducted in European and American PWRs during the 1970s are accessible. A substantial amount of information is required for meaningful validation, which makes any publicly available data extremely valuable. A robust and extensive collection of data points is essential to ensure the accuracy and credibility of the validation process. The following section will focus on identifying such measurements and compare them with the results of the numerical modeling with \texttt{DRAGON} and \texttt{PARCS}.

\subsection{Review of Publicly Available Pressurised Water Reactor Measurements}
\label{subsection:ReviewPublicData}
The first part of this section will focus on identifying PWRs with available measurement data. As the first cycle of the Fessenheim-2 reactor is already modelled with \texttt{DRAGON} and \texttt{PARCS} \cite{Salino,gitSalino}, the goal is to modify only slightly the modellisation to model multiple cycles for various reactors, to minimise the risk of modeling errors and to be able to test specifically the performances of different libraries on these cores. The suggestions offered by Dr.~Vivian~Salino in his PhD Thesis \cite{Salino} are the primary source of inspiration. Additionally, Dr.~Andrew Ward from the University of Michigan suggested exploring the \textit{SCALE Reactor Physics Validation} public Git repository~\cite{gitORNL}. This repository revealed itself to be a gold mine for the modeling of multiple PWR depletion cycle calculations with PARCS and for the analysis of NPP operation reports. As this was discovered later in the course of conducting the present Master's Thesis, it was not thoroughly explored or utilised. It contains detailed digitalised raw measurements, modellisation with \texttt{POLARIS}~\cite{POLARIS} (American lattice code developed by ORNL for the U.S.~NRC) and \texttt{PARCS}, as well as analyses for the first depletion cycles of BEAVRS, Watts-Bar-1, Surry-1, Turkey-Point-3, Monticello, Peach-Bottom-2, Hatch-1 and Cities-1. These may serve as documentation in a future study to continue the work started during  the present Master's Thesis and include more experimental data for the validation of nuclear data libraries.

While the burnable absorber is referred to as "Pyrex" for these reactors, a note in the Zion-2 EPRI report~\cite{Zion} points out that the makeup of the borosilicate glass tubing suggests that it may be "Tempax", an other borosilicate glass close to Pyrex but with a higher \ce{B_2O_3} weight fraction (17.5~wt\% instead of 12.9~wt\%)~\cite{morey_glassProperties}. As this is the poison, it would significantly impact the modeling. This may be the case but not noticed for other reactors as well. In the rest of this Master's Thesis, the burnable absorber is assumed to be Pyrex, but this fact is to be kept in mind when it comes to the gathering of modeling data.

An exhaustive description of the geometry and characteristic of the nuclear nuclear reactor and as well as operating conditions is necessary to model the cores. A wrap-up of the main characteristics of eight nuclear reactors with publicly available data is presented in table~\ref{tab:LitReviewMODELLING}.

\begin{table}[H]
\resizebox{\linewidth}{!}{
\begin{NiceTabular}{|c|c|c|c|c|c|c|}
  \hline
   \cellcolor{pink!20} Reactor & \cellcolor{blue!5}\makecell{Fessenheim-1,\\ Fessenheim-2, \\ and Bugey-2 \cite{Enaam}} & \cellcolor{blue!5}\makecell{Almaraz-2 \cite{IAEA}} & \cellcolor{blue!5}\makecell{Three-Mile-\\Island-1\\(TMI-1)\cite{TMI}} & \cellcolor{blue!5}\makecell{Zion-2 \cite{Zion}} & \cellcolor{blue!5}\makecell{Turkey-\\ Point-3 \cite{TurkeyPoint}} & \cellcolor{blue!5}\makecell{Surry-1 \cite{Surry,ORNL_Surry}} \\
  \hline
  \cellcolor{blue!10}\makecell{Number of\\ assemblies} & 157 & 157 & 157 & 193 & 157 & 157\\
  \hline
  \cellcolor{blue!10}\makecell{Assembly\\Lattice} & 17x17 & 17x17 & 15x15 & 15x15 & 15x15 & 15x15\\
  \hline
  \cellcolor{blue!10}\makecell{BA material} & Pyrex & Pyrex & \ce{Al_2O_3-B_4C} & Pyrex & Pyrex & Pyrex \\
  \hline
  \cellcolor{blue!10}\makecell{Number of\\ BA rods per\\ BA assembly} & 12, 16, 20 & 12, 16, 20 & 16 & \makecell{8, 9, 12,\\ 16, 19, 20} & 12 & 8, 12, 20 \\
  \hline
  \cellcolor{blue!10}\makecell{Cycle 1\\U-235\\ enrichment [\%]} & \makecell{2.1,\\ 2.6 (BA),\\ 3.1,\\ 3.1 (BA)} & \makecell{2.1,\\ 2.6, 2.6 (BA),\\ 3.1, 3.1 (BA)} & \makecell{2.06,\\ 3.05\\2.75 (BA)} & \makecell{2.248, 2.248 (BA),\\ 2.789, 2.789 (BA),\\ 3.292, 3.292 (BA)} & \makecell{1.86,\\ 2.56, 2.56 (BA),\\ 3.10, 3.10 (BA)} & \makecell{1.868,\\ 2.573, 2.573 (BA),\\ 3.117, 3.117 (BA)}\\
  \hline
  \cellcolor{blue!10}\makecell{Cycle 2\\U-235\\ enrichment [\%]} & \makecell{2.1,\\ 2.9,\\ 3.1,\\ 3.25} & \makecell{2.1,\\ 2.6,\\3.1,\\ 3.15} & \makecell{2.75,\\ 3.05\\ 2.64} & \makecell{2.248, 2.248 (BA),\\ 2.789, 2.789 (BA),\\ 3.063, 3.063 (BA), \\ 3.292, 3.292 (BA)} & \makecell{1.86,\\2.56, 2.56 (BA),\\3.10} & \makecell{1.860, 1.860 (BA),\\1.868,\\ 2.573,\\3.117,\\2.610 (BA),\\3.330}\\
  \hline
  \cellcolor{blue!10}\makecell{Cycle 3\\U-235\\ enrichment [\%]} & \makecell{N/A} & \makecell{N/A} & \makecell{N/A} & \makecell{N/A} & \makecell{1.86,\\2.56,\\2.61,\\2.90,\\3.10} & \makecell{1.860,\\1.868,\\2.100,\\2.573,\\2.610,\\3.117,\\3.330}\\
  \hline
  \cellcolor{blue!10}\makecell{Comments} & \makecell{N/A} & \makecell{N/A} & \makecell{N/A} & \makecell{The core and\\ BA assembly\\ layouts are \\ asymmetric\\$\Rightarrow$ can't be\\ modelled with\\ PARCS V3.4.2} & \makecell{N/A} & \makecell{To be taken\\with careful precautions.\\\\ Some unreadable data\\(e.g. loading patterns)\\ and errors in\\ the operational\\ EPRI report,\\corrected to be\\ realistic by ORNL~\cite{ORNL_Surry}.}\\
  \hline
\end{NiceTabular}
}
\caption{Comparison of the main characteristics of nuclear reactors similar to \text{Fessenheim-2} with publicly available measurements for consective depletion cycles. "(BA)" stands for fuel assemblies with Burnable Absorber rods inserted.}
\label{tab:LitReviewMODELLING}
\end{table}

The nuclear reactors showcased in table~\ref{tab:LitReviewMODELLING} exhibit similar properties. Considering that TMI-1 fuel assembly design uses a different material for the burnable absorber and that Zion-2 can not be modelled with the version of \texttt{PARCS} used for this report, these reactors are not analysed in this Master's Thesis. Nevertheless, it is important to note that they may be utilised in future research endeavours. Table~\ref{tab:LitReviewDATA} presents the available measurement data for the other reactors.

In view of table~\ref{tab:LitReviewMODELLING} Almaraz-2 appears to the most straightforward reactor to model due to its close similarity with Fessenheim-2. Their very similar conditions of operation allow to use the same assembly-homogenised macroscopic cross sections in the full-core computations for most of the assemblies. Almaraz-2 also makes it possible to compare the simulation results with measurements of the boron letdown curve for two cycles (see~Tab.~\ref{tab:LitReviewDATA}). This additional information is precious to be able to discriminate small variations in nuclear data libraries and analyse their performance at higher fuel exposure.

The modeling of Turkey-Point-3 and Surry-1 requires the use of a 15x15 fuel assembly type in \texttt{DRAGON} (see~Tab.~\ref{tab:LitReviewMODELLING}) and new Pyrex configurations in fuel assemblies compared to the existing \texttt{DRAGON} 15x15 model of Tihange-1~\cite{gitSalino}. However, these reactors make it possible to compare data for 3 consecutive cycles and have a lot of measurement data available (see~Tab.~\ref{tab:LitReviewDATA}). The modeling of Turkey-Point-3 will be performed and analysed in this Master's Thesis thanks to the reduced number of lattice cells to model with \texttt{DRAGON} and due to the poor quality of the Surry-1 operational report (see comments in Tab.~\ref{tab:LitReviewMODELLING}). It is important to highlight that the recent updates to this report by ORNL~\cite{ORNL_Surry} enable the modeling of Surry-1; however, these corrections a posteriori suggest that caution should be exercised in their interpretation.

\begin{table}[H]
\centering
\resizebox{1.0\linewidth}{!}{
\begin{NiceTabular}{|c|c|c|c|c|}
  \hline
   \cellcolor{pink!20} Reactor & \cellcolor{blue!5}\makecell{Fessenheim-1,\\ Fessenheim-2, \\ and Bugey-2 \cite{Enaam}} & \cellcolor{blue!5}\makecell{Almaraz-2 \cite{IAEA}} & \cellcolor{blue!5}\makecell{Turkey-\\ Point-3 \cite{TurkeyPoint}} & \cellcolor{blue!5}\makecell{Surry-1 \cite{Surry,ORNL_Surry}} \\
  \hline
  \cellcolor{blue!10}\makecell{Data detectors \\ cycle 1} & \cellcolor{green!25}\makecell{ 2D: 9x50=450} & \cellcolor{green!25}\makecell{1D:~3x58=174\\2D:~9x26=234} & \cellcolor{green!25}\makecell{3D: 5x50x29=7250} & \cellcolor{green!25}\makecell{ 3D: 7921} \\
  \hline
  \cellcolor{blue!10}\makecell{Data boron \\ cycle 1} & \cellcolor{green!25} 13 & \cellcolor{green!25} 14 & \cellcolor{green!25} 6 & \cellcolor{green!25} 125 \\
  \hline
  \cellcolor{blue!10}\makecell{HZP SPT\\cycle 1} & \cellcolor{red!25}N & \cellcolor{green!25}\makecell{CRW (6)\\ITC (3)\\PPF (2)\\ARO MC (1)\\ ARO DC (1)} & \cellcolor{green!25}\makecell{CRW(4)} & \cellcolor{red!25}\makecell{N}\\
  \hline
  \cellcolor{blue!10}\makecell{Other data\\cycle 1} & \cellcolor{red!25}N & \cellcolor{green!25}\makecell{AO (9)\\PPF (9)} & \cellcolor{red!25}N & \cellcolor{red!25}\makecell{N}\\
  \hline
  \cellcolor{blue!10}\makecell{Data detectors \\ cycle 2} & \cellcolor{green!25}\makecell{ 2D:1x50=50} & \cellcolor{green!25}\makecell{1D:3x58\\2D:~6x26=156} & \cellcolor{green!25}\makecell{3D: 7x50x29=10150} & \cellcolor{green!25}\makecell{3D: 117x61=7137} \\
  \hline
  \cellcolor{blue!10}\makecell{Data boron \\ cycle 2} & \cellcolor{red!25}N & \cellcolor{green!25} 13 & \cellcolor{green!25}7 & \cellcolor{green!25} 75 \\
  \hline
  \cellcolor{blue!10}\makecell{HZP SPT\\cycle 2} & \cellcolor{red!25}N & \cellcolor{green!25}\makecell{CRW (7)\\ITC (2)\\PPF (1)\\ARO MC (1)\\ ARO DC (1)} & \cellcolor{green!25}\makecell{CRW (4)} & \cellcolor{red!25}\makecell{N}\\
  \hline
  \cellcolor{blue!10}\makecell{Other data\\cycle 2} & \cellcolor{red!25}N & \cellcolor{green!25}\makecell{AO (6)\\PPF (6)} & \cellcolor{red!25}N & \cellcolor{red!25}\makecell{N}\\
  \hline
  \cellcolor{blue!10}\makecell{Data detectors \\ cycle 3} & \cellcolor{red!25}N & \cellcolor{red!25}N & \cellcolor{green!25}\makecell{3D: 4x50x29=5800} & \cellcolor{green!25}\makecell{3D: 73x61=4453} \\
  \hline
  \cellcolor{blue!10}\makecell{Data boron \\ cycle 3} & \cellcolor{red!25}N & \cellcolor{red!25}N & \cellcolor{green!25} 5 & \cellcolor{red!25}N \\
  \hline
  \cellcolor{blue!10}\makecell{HZP SPT\\cycle 3} & \cellcolor{red!25}\makecell{N} & \cellcolor{red!25}\makecell{N} & \cellcolor{green!25}\makecell{CRW (5)} & \cellcolor{red!25}N \\
  \hline
  \cellcolor{blue!10}\makecell{Other data\\cycle 3} & \cellcolor{red!25}\makecell{N} & \cellcolor{red!25}\makecell{N} & \cellcolor{red!25}N & \cellcolor{red!25}N \\
  \hline
  \cellcolor{blue!10}\makecell{Total number\\ of data points} & \makecell{513} & \makecell{704} & \makecell{23231} & \makecell{19641}\\
  \hline
  \cellcolor{blue!10}\makecell{Comments} & \makecell{Data of the\\ three reactors\\ are often merged\\together\\\\ Data can also be\\ found in the\\ PhD theses of\\ Dr.~Henri Panek \cite{Panek}\\ and\\ Dr.~Ahmed Hassini \cite{Hassini}.} & \makecell{2D detector data\\ averaged for an\\ octant core} & \makecell{CRW are measured\\ with a so-called \\ \textit{reactivity computer},\\ without any \\ indication of\\ what it is.\\ Not clear if\\ it is measurement \\ or calculations.} & \makecell{Only data\\ clearly readable\\ were counted.\\ \\ Be careful: \\the last 2\\ pages of the\\ EPRI report\\ are inverted.}\\
  \hline
\end{NiceTabular}
}
\caption{Synthesis of publicly available measurement data for the first three cycles of nuclear reactors similar to \text{Fessenheim-2}. \texttt{N} stands for unavailable data. The figures refer to the number of measurement points available for each physical quantity. Acronyms are detailed at the end of section~\ref{subsection:ReviewPublicData}.}
\label{tab:LitReviewDATA}
\end{table}

\noindent For enhance clarity, acronyms were used in table~\ref{tab:LitReviewDATA}:
\begin{outline}
    \1 HZP: Hot Zero Power. It refers to a state where the reactor is critical with negligible power compared to nominal conditions.
    \1 SPT: Startup Physics Tests. They consist of tests performed  before the beginning of a depletion cycle to verify that the characteristics of the core are compliant with safety standards with the core design.
    \1 "1D: $n_1$x$n_2$": $n_1$ measurements of radially averaged power at various burnups where performed, each at $n_2$ different axial locations.
    \1 "2D: $n_1$x$n_2"$: $n_1$ measurements of axially averaged power at various burnups where performed, each in $n_2$ different fuel assemblies.
    \1 "3D: $n_1$x$n_2$x$n_3$": $n_1$ measurements of power at various burnups where performed, each in $n_2$ different fuel assemblies at $n_3$ different axial locations.
    \1 CRW: Control Rod Worth. Difference of reactivity of the core measured between the insertion and the extraction of a specific set of control rods.
    \1 ITC: Isothermal Temperature Coefficient. Difference in reactivity measured when the core temperature is raised by 1~K.
    \1 MC: Moderator coefficient. Difference of reactivity measured when the moderator temperature is raised by 1~K.
    \1 DC: Doppler coefficient, also known as fuel temperature coefficient. Difference of reactivity measured when the fuel temperature is raised by 1~K.
    \1 PPF: Ratio of the maximal power in the core versus the average power.
    \1 ARO: All rods out. Denotes a state where all control rods are extracted from the core.
    \1 AO: Axial offset. Defined by equation~\eqref{eq:axialOffset}. It encapsulates the ratio of power produced in the upper half of the core compared to the power produced in the bottom half.
\end{outline}

\subsection{Strategies for Digitising Old Operational Reports: Comparative Analysis of Various Methods}
A challenge encountered when analysing extensive datasets within scanned documents is their digitisation. The validation of the modeling of nuclear reactors (performed in this Master's Thesis in section~\ref{subsection:validation7Cycles}) requires a large amount of experimental data. Additionally, the documents being examined date back to the 1970s and 1980s. Their varying original quality, the preservation of the ink, and the quality of the scanning present challenges when working with them. Several techniques were evaluated and deliberated upon:

\begin{enumerate}
    \item \textbf{Hand Copying}
    \begin{itemize}
        \item \textbf{Advantages}: This method is straightforward and yields a reasonable level of accuracy.
        \item \textbf{Disadvantages}: It is time-consuming and can be physically exhausting. Additionally, a secondary verification process is necessary, particularly for lengthy copying sessions.
    \end{itemize}
    
    \item \textbf{Python with img2table~\cite{img2table} and Pandas Libraries~\cite{pandasPython}, and Optical Character Recognition (OCR) Capabilities via PaddleOCR~\cite{PaddleOCR}}
    \begin{itemize}
        \item \textbf{Advantages}: This approach provides sufficient quality for digitising the IAEA Almaraz-2 benchmark document~\cite{IAEA}. It requires minimal secondary verification and performs effectively with screenshots of tables.
        \item \textbf{Disadvantages}: The method struggles with handwriting, poorly scanned documents (e.g., Turkey-Point-3), and tables that exhibit non-standard geometries (e.g., octant cores).
    \end{itemize}
    
    \item \textbf{ChatGPT-4o~\cite{chatgpt4o} with Image Input, Enhanced by Prompt Engineering}
    \begin{itemize}
        \item \textbf{Advantages}: This technique demonstrates significantly improved performance compared to Python, particularly for poorly scanned documents. It offers a user-friendly interface that does not require programming skills, although a secondary verification is still necessary due to the model's propensity to generate inaccurate information with confidence~\cite{bullshit}.
        \item \textbf{Disadvantages}: As of August 2024, access to this tool is limited, and it is restricted to public documents due to confidentiality concerns.
    \end{itemize}
\end{enumerate}

The following prompt can be used to extract data and automatically sort them in a database with a bash script (tested with the Turkey-Point-3 2D detector measurements on the $2^{\textrm{nd}}$ August, 2024):
\begin{verbatim}
Analyze the table shown in this screenshot and extract the text content.
Then, write a bash script to perform the following tasks:

    1. Create a new directory named assembly_power_cycleX_XXXBurn.
    2. In this directory, create one file for each column of the table.
       Name each file according to the column header.
    3. Populate each file with the respective column's data, with each
       row's value on a new line.
\end{verbatim}

\noindent The scripting part saves time but increases the computational cost, thereby restricting the access of the service and raising its cost. The first line of the prompt is sufficient to extract the data in plain text. This method is anticipated to become obsolete in the coming months or years due to the continuous advancement and democratisation of large language models (LLMs), along with the emergence of open-source alternatives and smaller versions that can operate on local machines.

A combination of these three techniques was used for this Master's Thesis work. The result of this digitisation was saved in text files to be used for validation in section~\ref{subsection:validation7Cycles}~(see Appendix B). A directory for each cycle was created, with a sub-directory for each depletion cycle inside. Boron letdown curves and axial offsets were saved with two columns: the first one for the physical quantity of interest, the second one for the burnup corresponding to these measurements. The control rod worths were stored with two columns as well: the first one for the name of the control bank inserted (e.g. \textit{ABC}), the second one for the measured reactivity worth. The power maps were stored directly as matrices, the name of the file referring to the burnup value. This systematic organisation of the experimental measurement database allow for their automatic processing. This aspect is an essential step toward an automated nuclear data library assessment using these data.

\subsection{Modeling of Pressurised Water Reactor Fuel Assembly Lattices with \texttt{DRAGON}}
\label{subsection:ImplementationDRAGON}
The work presented in this Master's Thesis builds upon the procedures developed by A.~Bruneau, M.~Cordier, V.~Salino, G.~Tixier, G.~Drouard, L.~Liponi, and R.~Nguyen Van Ho~\cite{gitSalino}. These procedures were design to model the 17x17 and 15x15 fuel assemblies corresponding to the first cycles of Fessenheim-2 and Tihange-1 reactors, respectively.\\
The method presented in section~\ref{subsection:concatenateHiatories} is followed to generate a \texttt{PMAXS} file with various branches and histories. Considering the fact that the branch calculations at different burnup steps are independent from each other, this step is parallelised. The computation of \texttt{PMAXS} cross-sections for Turkey-Point-3, which involves 25 burnup steps, as well as moderator density and Pyrex insertion histories, requires approximately one day of processing time on a computer equipped with 50 gigabytes of RAM and 25 processing cores. Each instantaneous variable is evaluated at three distinct values during this process. To enhance execution speed, further parallelisation could be implemented, leveraging the independence of each fuel assembly calculation. This approach would be beneficial for future research in this area and could facilitate the acquisition of results within an overnight time frame.

\subsection{Modeling of Multiple Cycle Depletion with \texttt{PARCS}}
\label{subsection:multipleCyclesPARCS}

The implementation of multiple cycle depletion discussed in this Master's Thesis requires \texttt{PARCS v3.4.2} or a later version. Specifically, it is not compatible with \texttt{PARCS v3.1.1}. The modeling of multiple cycles for Fessenheim-2 and Almaraz-2 with \texttt{PARCS} is based on a depletion model of the first cycle of Fessenheim-2 developped jointly EPFL, PSI and IRSN~\cite{hursinSalinoFessenheimCycle1}. The multiple cycle modeling of Turkey-Point-3 with \texttt{PARCS} is based on a model developed by the Oak Ridge National Lab available through a git repository~\cite{gitORNL}.\\[0.5\baselineskip]

\noindent For the depletion calculations, the following \texttt{PARCS} models and options were used.

\begin{itemize}
    \item Usage of the internal simple PWR mass-energy solver, specifically designed for single-phase liquid analysis.
    \item Transient treatments for xenon and samarium.
    \item Division of each fuel assembly into 22 nodes, comprising 20 fuel nodes, one reflector node at the top, and one reflector node at the bottom.
    \item Usage of a 2x2 radial sub-meshing approach for enhanced 2D power distributions for Fessenheim-2 and Almaraz-2. This was not used for Turkey-Point-3.
    \item Incorporation of flux shape corrections to account for the control rod cusping effect, thereby considering the implications of partial rod insertion within the nodes.
    \item Manual tuning of the fuel-pin gap conductance to obtain, at full power conditions, the fuel temperature indicated in the operational reports~\cite{TurkeyPoint,TMI,Enaam,IAEA,Zion,Surry}.
\end{itemize}

\subsection{Validation of the Pressurised Water Reactor Modeling with \texttt{DRAGON} and \texttt{PARCS}}
\label{subsection:validation7Cycles}

The results presented in this section are obtained with the nuclear data library \texttt{JEFF--3.1.1} \cite{JEFF}.

\subsubsection{Validation of the Modeling of the First Two Cycles of Almaraz-2}
\begin{figure}
    \centering
    \includegraphics[width=\linewidth]{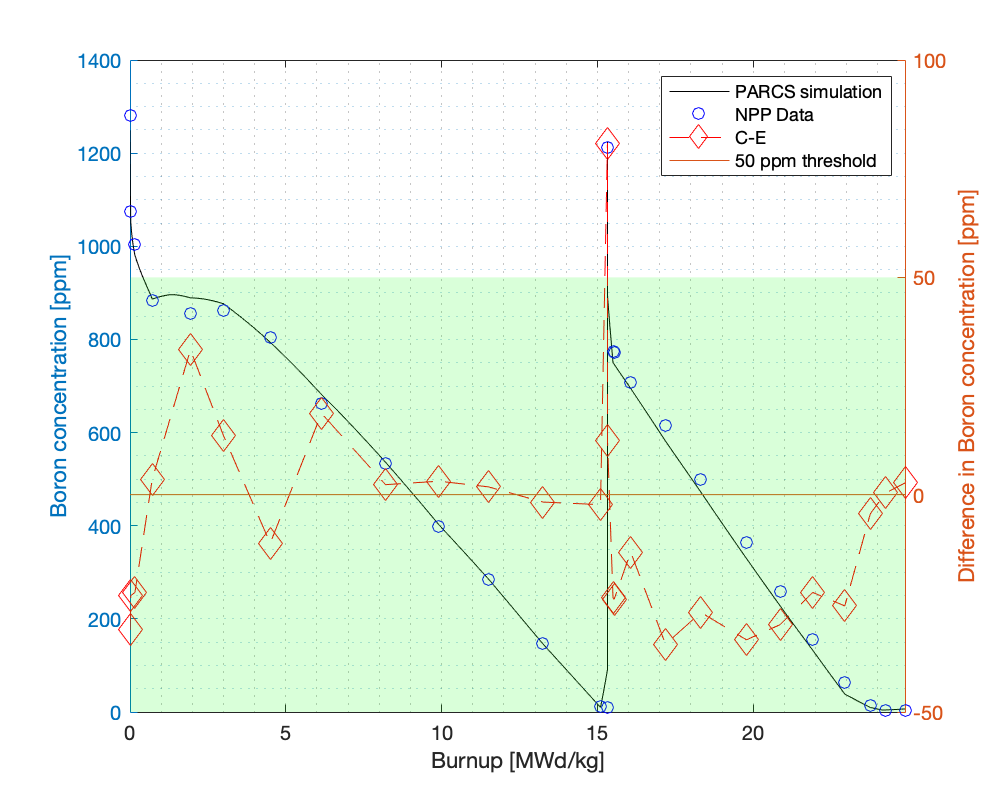}
    \caption{Evolution of the boron concentration in water during the first two cycles of Almaraz-2: comparison between the simulation of \texttt{PARCS} (with macroscopic cross sections generated with \texttt{DRAGON}) and experimental measurements. The green shaded region represents the uncertainty range attributed to variations in nuclear data~\cite{JEFF_UncertaintyXS}.}
    \label{fig:Almaraz2CyclesBoron}
\end{figure}

Figure~\ref{fig:Almaraz2CyclesBoron} illustrates the comparison between the predicted boron letdown curve and the experimental measurements during the first two depletion cycles of Almaraz-2. The predictions align with the overall trend of the measurements, remaining within a margin of $\pm\, 40$~ppm.

\begin{figure}
    \centering
    \includegraphics[width=\linewidth]{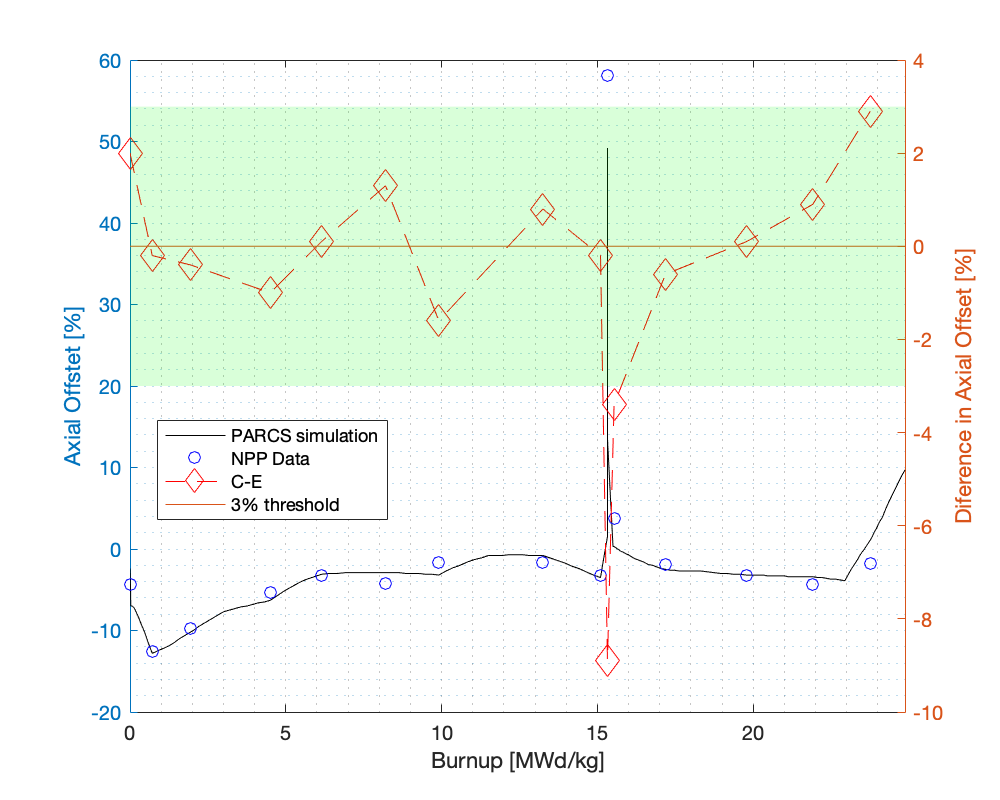}
    \caption{Evolution of the axial offset during the first two cycles of Almaraz-2: comparison between the simulation of \texttt{PARCS} (with macroscopic cross sections generated with \texttt{DRAGON}) and experimental measurements. The green shaded region represents the uncertainty range~\cite{JEFF_UncertaintyXS}.}
    \label{fig:Almaraz2CyclesAO}
\end{figure}

Figure~\ref{fig:Almaraz2CyclesAO} shows the comparison between the predicted axial offset and the experimental measurements. The predictions are consistent with the overall trend of the measurements, staying approximately within an error margin of $\pm\, 3$\,\%. Although explicitly modeling the spacer grid could lead to more precise results, all measurements are within the uncertainty range assessed at  $\pm\,3$\,\% in a previous analysis~\cite{JEFF_UncertaintyXS}).

However, there is one notable exception to these observations: a measurement taken at the conclusion of the first cycle during stretch-out operations. The uniqueness of this data point indicates that there may be insufficient information to model the reactor accurately at that stage. Another potential issue could be errors in the IAEA report~\cite{IAEA} used for modeling and measurement extraction. This concern is underscored by the fact that Oak Ridge National Laboratory identified similar errors in their report on Surry-1~\cite{ORNL_Surry}, indicating a precedent for such error.

The control rod worths measured during the startup physics tests before the first cycle of Almaraz-2 are presented in table~\ref{tab:CRW_Almaraz_cyc1}. The relative error of the predictions all fall in the uncertainty range, assessed to be $\pm$\,10\,\% in a previous analysis~\cite{JEFF_UncertaintyXS}. Therefore, the modeling of these control rod worth measurements can be considered validated.

\begin{table}[H]
    \centering
    \resizebox{1.0\linewidth}{!}{
    \begin{NiceTabular}{|c|c|c|c|}
        \hline 
         \makecell{Control Rod Bank \\ Label} & \makecell{Reactivity Worth\\ Measured [pcm]} & \makecell{Reactivity Worth\\ Simulated [pcm]} & \makecell{(Simulated - Measured)/Measured\\ [\%]} \\
        \hline
        D & 1394 & 1471.0 & $5.5$ \\
        \hline
        C (D in) & 1192 & 1239.4 & $4.0$ \\
        \hline
        B (DC in) & 1964 & 1941.5 & $-\,1.1$ \\
        \hline
        A (DCB in) & 1253 & 1338.0 & $6.7$ \\
        \hline
        SB (DCBA in) & 1022 & 921.4 & $-\,9.8$ \\
        \hline
    \end{NiceTabular}
    }
    \caption{Control Rod Worth assessment before the first cycle of Almaraz-2. Each label refers to a distinct group of control rods.}
    \label{tab:CRW_Almaraz_cyc1}
\end{table}

Aside from the consideration at the end of the first cycle, the Almaraz-2 model developed using \texttt{DRAGON} and \texttt{PARCS} to predict the boron letdown curve, the axial offset and the control rod worths is deemed validated for the first two depletion cycles and is suitable for evaluating the nuclear data library.

\subsubsection{Validation of the Modeling of the First Two Cycles of Fessenheim-2}

\begin{figure}
    \centering
    \includegraphics[width=\linewidth]{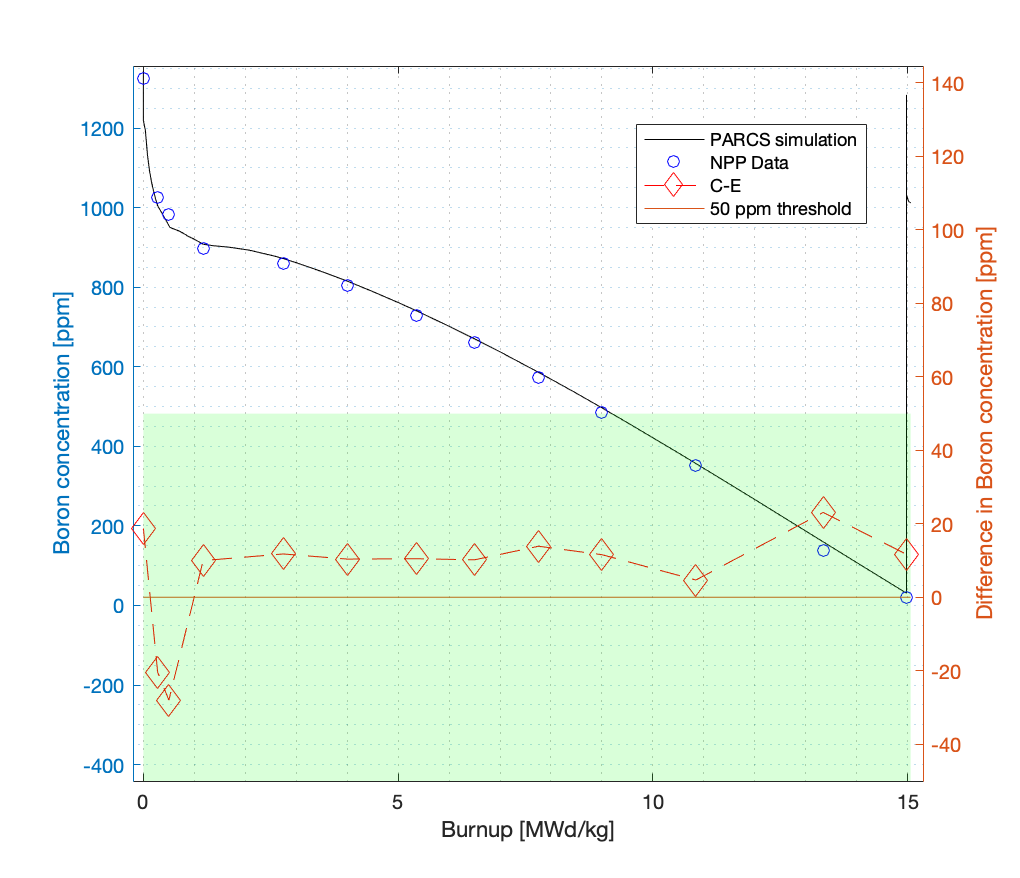}
    \caption{Evolution of the boron concentration in water during the first two cycles of Fessenheim-2: comparison between the simulation of \texttt{PARCS} (with macroscopic cross sections generated with \texttt{DRAGON}) and experimental measurements. The green shaded region represents the uncertainty range~\cite{JEFF_UncertaintyXS}.}
    \label{fig:Fessenheim2CyclesBoron}
\end{figure}

\begin{figure}
    \centering
    \includegraphics[width=0.8\linewidth]{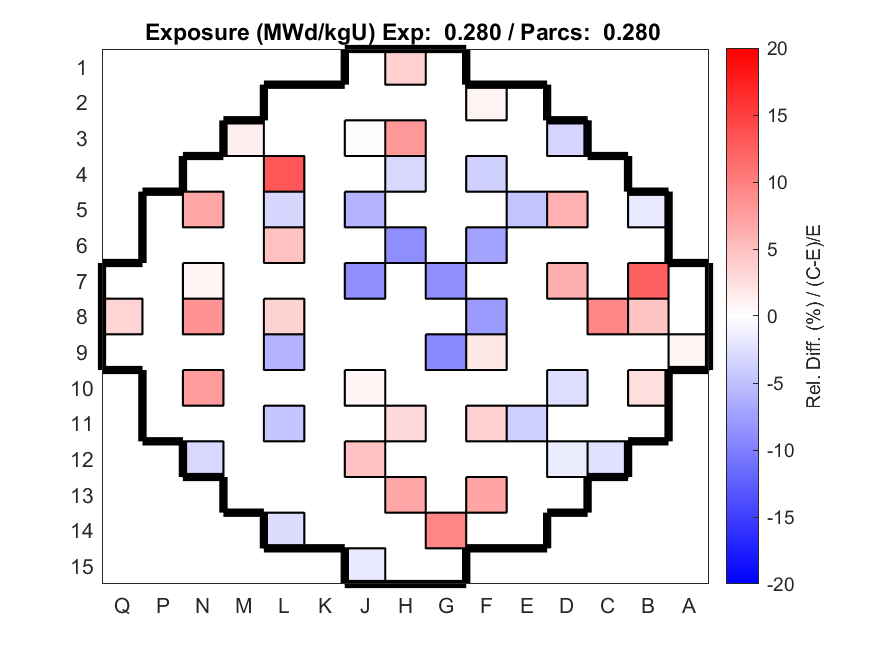}
    \caption{Response of fission chamber detectors during the first cycle of the Fessenheim-2 reactor at a burnup of $0.280~\mathrm{MWd/kgU}$. Standard deviation: 5.80\,\%.}
    \label{fig:STEP_002_Detectors}
\end{figure}

\begin{figure}
    \centering
    \includegraphics[width=0.8\linewidth]{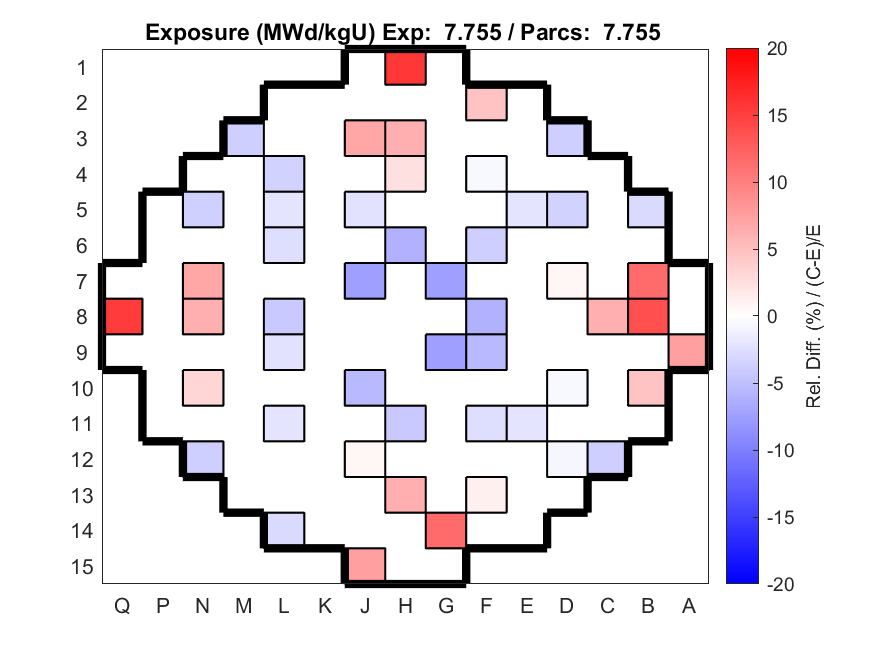}
    \caption{Response of fission chamber detectors during the first cycle of the Fessenheim-2 reactor at a burnup of $7.755~\mathrm{MWd/kgU}$. Standard deviation: 6.09\,\%.}
    \label{fig:STEP_008_Detectors}
\end{figure}

\begin{figure}
    \centering
    \includegraphics[width=0.8\linewidth]{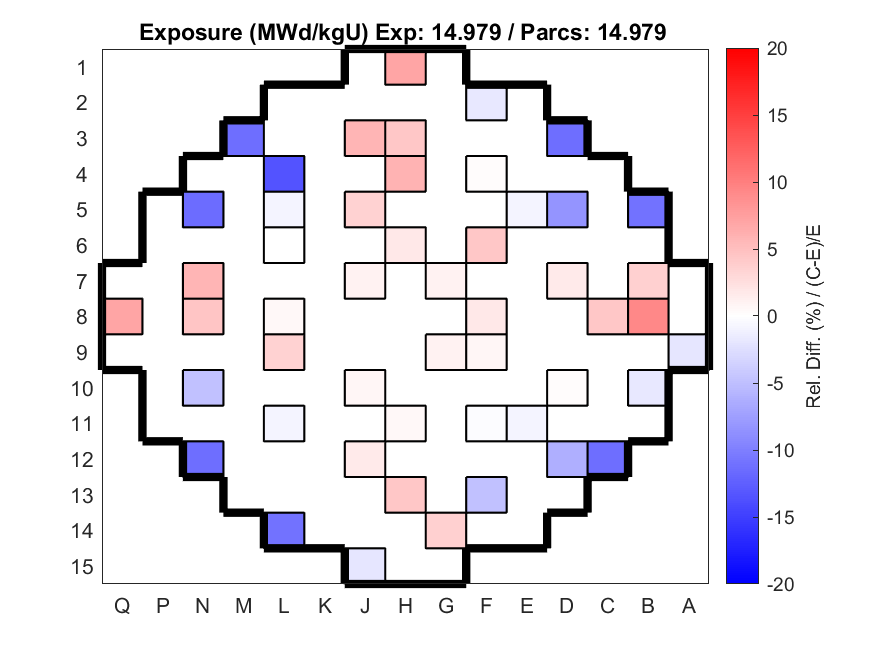}
    \caption{Response of fission chamber detectors during the first cycle of the Fessenheim-2 reactor at a burnup of $14.979~\mathrm{MWd/kgU}$. Standard deviation: 5.79\,\%.}
    \label{fig:STEP_010_Detectors}
\end{figure}

\begin{figure}
    \centering
    \includegraphics[width=0.8\linewidth]{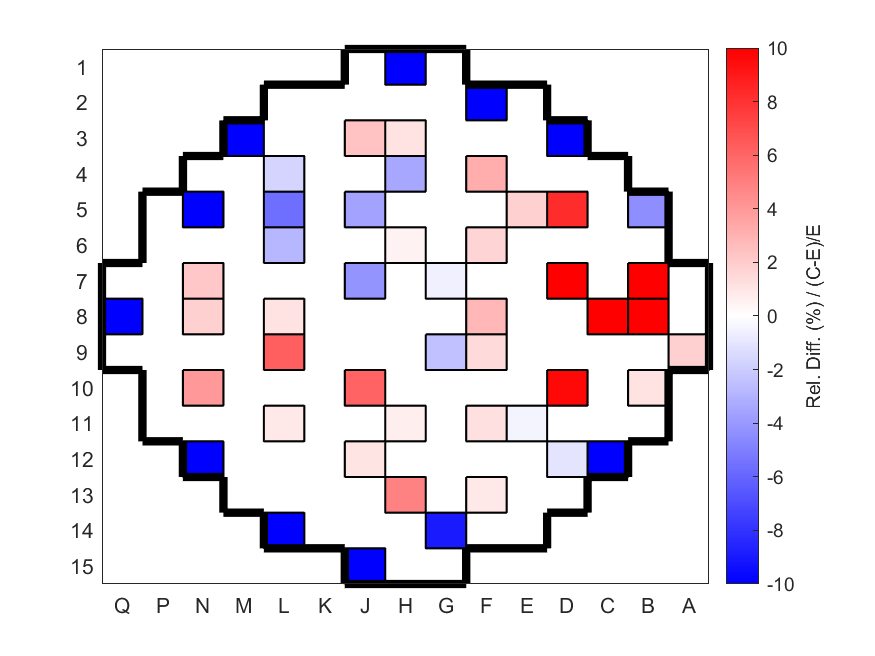}
    \caption{Response of fission chamber detectors at hot zero power at the beginning of the second cycle of Fessenheim-2. Standard deviation: 9.58\,\%.}
    \label{fig:Fessenheim_C2_HZP_Detectors}
\end{figure}

Figure~\ref{fig:Fessenheim2CyclesBoron} illustrates the comparison between the predicted boron letdown curve and the experimental measurements during the first depletion cycles of Fessenheim-2 and the beginning of the second cycle. The predictions align with the overall trend of the measurements, remaining within an error margin of $\pm\, 25$~ppm. Additionally, the predicted responses of the detectors used for power measurement during the first cycle of Fessenheim-2 are depicted in figures~\ref{fig:STEP_002_Detectors}, \ref{fig:STEP_008_Detectors}, and \ref{fig:STEP_010_Detectors}. The observed discrepancies, which approximate 6\,\% of standard deviation, indicate a need for further refinement in the predictive model.\\
Furthermore, figure~\ref{fig:Fessenheim_C2_HZP_Detectors} illustrates the predictions of the fission chamber detectors' response at hot zero power at the onset of the second cycle of Fessenheim-2. The results reveal substantial discrepancies, with relative errors reaching as high as 35\,\%. This highlights the necessity for improved accuracy in the predictive modeling of detector responses in future analyses.\\
The power response reconstruction was carried out using a script developed by Dr.~Mathieu Hursin prior to this Master's Thesis. The interface between \texttt{DRAGON} and \texttt{PARCS} for the detector response remains in a preliminary stage and requires further enhancement in future research.\\[0.5\baselineskip]

The Fessenheim-2 model developed using \texttt{DRAGON} and \texttt{PARCS} to predict the boron letdown curve is deemed validated for the first depletion cycle and is suitable for evaluating the nuclear data library. The prediction of the detector's response during the first cycle and at the beginning of the second cycle failed to be validated.


\subsubsection{Validation of the Modeling of the First Three Cycles of Turkey-Point-3}

\begin{figure}
    \centering
    \includegraphics[width=\linewidth]{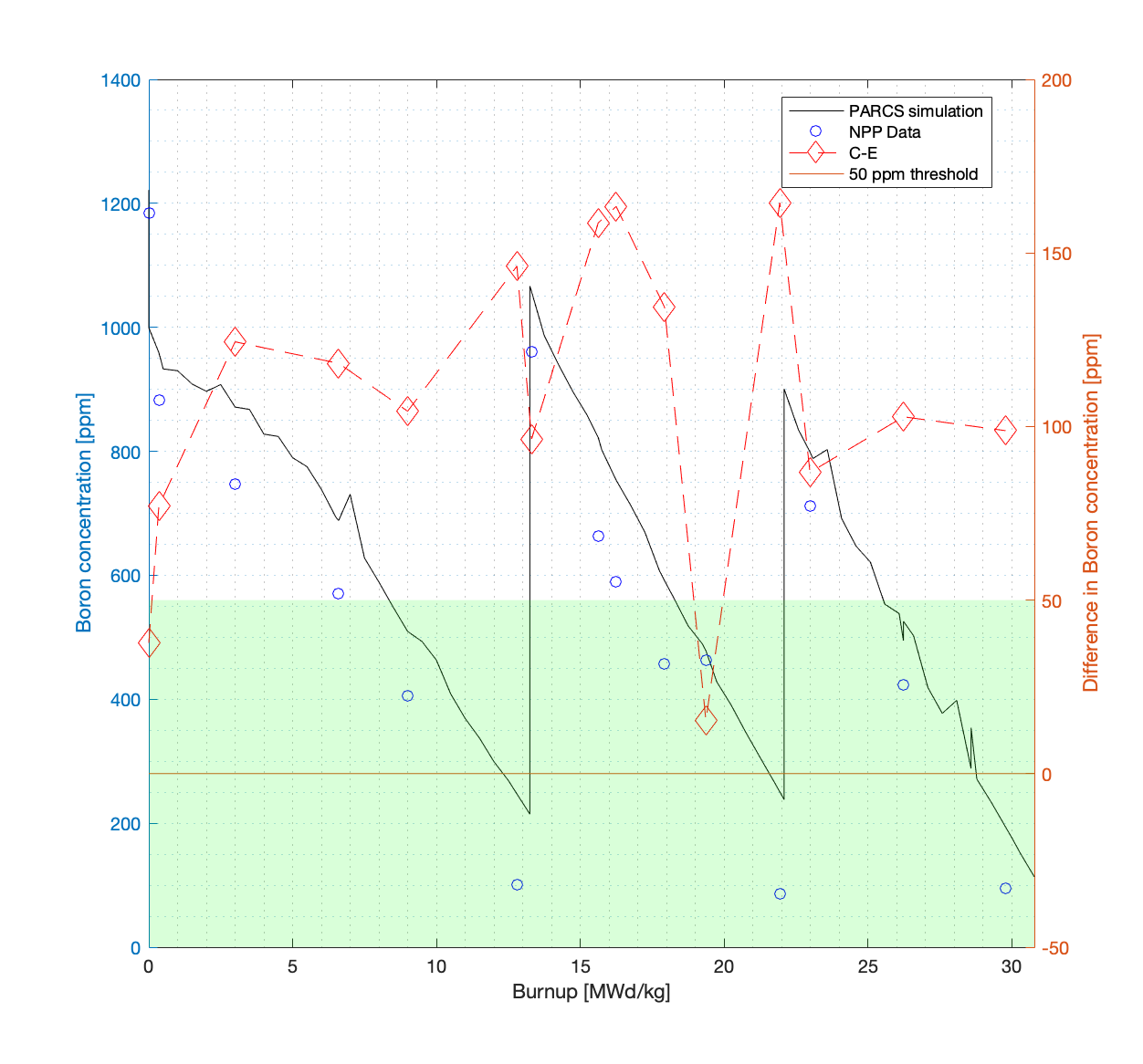}
    \caption{Evolution of the boron concentration in water during the first three cycles of Turkey-Point-3: comparison between the simulation of \texttt{PARCS} (with macroscopic cross sections generated with \texttt{DRAGON}) and experimental measurements. The green shaded region represents the uncertainty range attributed to variations in nuclear data~\cite{JEFF_UncertaintyXS}.}
    \label{fig:TP3CyclesBoron}
\end{figure}

Figure~\ref{fig:TP3CyclesBoron} illustrates the comparison between the predicted boron let-down curve and the experimental measurements during the first two depletion cycles of Turkey-Point-3. The predictions consistently exceed the actual measurements, with most values falling outside the acceptable error range of $\pm\, 50$~ppm. This overestimation is attributed to an inflated prediction of the core reactivity that was present from the start of the first cycle. The modeling carried out using the \texttt{POLARIS} cross section, as presented in figure~\ref{fig:TP3ImpactHistoriesBoron}, does not display this reactivity tilt, despite employing the same \texttt{PARCS} modeling framework.

Consequently, the issue can be identified as a modeling concern associated with our use of \texttt{DRAGON}. Moreover, this excess reactivity had already been observed in Chapter 3, with DRAGON results 1000~pcm above POLARIS, on this specific $15 \times 15$ Turkey-Point-3 case. Also, an order of magnitude of the differential boron worth is 10 pcm/ppm. Therefore, it seems consistent to find 100~ppm excess in boron concentration. As a perspective, this problem could be investigated with a comparison of \texttt{POLARIS}, \texttt{DRAGON} and possibly a Monte-Carlo code (such as \texttt{SERPENT}) with identical nuclear data, by removing all grids, on this single $15 \times 15$ assembly.

\noindent\textit{Note that this discrepancy was traced back to an error in the chemical composition of the lattice cell. The corrected analysis was later published by Meunier and Hursin~\cite{MEUNIER2025}}.

\newpage
\section{Nuclear Data Validation}
\label{section:nuclearDataValidation}
This Chapter focuses on presenting a method to assess the performance of novel nuclear data libraries in predicting integral measurements in PWRs.
\subsection{Introduction to the Methodology}
The proposed method involves a systematic analysis of nuclear data libraries, aiming to identify their respective biases in predicting reactor integral measurements introduced in section~\ref{subsection:ReviewPublicData}.\\[0.5\baselineskip]
A key component of this analysis is the relative error, defined by 
$(C - E)/E$, where $C$ represents the computed value and $E$ the experimental value. This metric is centred around zero, allowing for a straightforward interpretation of biases: positive values indicate an overestimation, while negative values suggest an underestimation.
This metric is visualised through histograms, providing a clear representation of the distribution of discrepancies between predictions and measurements.\\[0.5\baselineskip]
Weighted averages, standard deviations and $\chi^2$ statistics can be calculated to summarise the performance of each library across various physical quantities of interest (including the diluted boron critical concentration and fission chambers' responses) and multiple reactors.\\[0.5\baselineskip]
The analysis is conducted by splitting the data according to the library used. The results can be categorised according to the physical quantity of interest or the reactor considered, facilitating a thorough investigation of biases. It is important to ensure that the dataset of each analysis is sufficiently large to draw reliable conclusions.\\[0.5\baselineskip]
Scripts were developed with the objective of automating the execution and post-processing of all calculations associated with each reactor and each physical quantity to be analysed. In particular, three distinct scripts were developed: one for the DRAGON calculations, another for the PARCS simulations, and a third for the post-processing tasks. The post-processing script encompasses a range of functions, including the extraction of physical quantities of interest, the computation of control rod worths, the generation of graphical figures, and the execution of statistical analyses. The automation not only enhances the efficiency of the computational workflow but also ensures consistency and accuracy in the analysis of reactor behaviour. Additionally, a meta-script that sequentially calls all these scripts would simplify their usage and make them suitable for routine application. An enhancement could also involve expanding these scripts to evaluate more physical quantities, thereby encompassing all the quantities listed in table~\ref{tab:LitReviewDATA}.

\subsection{Preliminary Analysis on the First Two Cycles of Almaraz-2 and the First Cycle of Fessenheim-2}
\begin{figure}
    \begin{subfigure}{1\textwidth}
        \centering
        \includegraphics[width=0.8\linewidth]{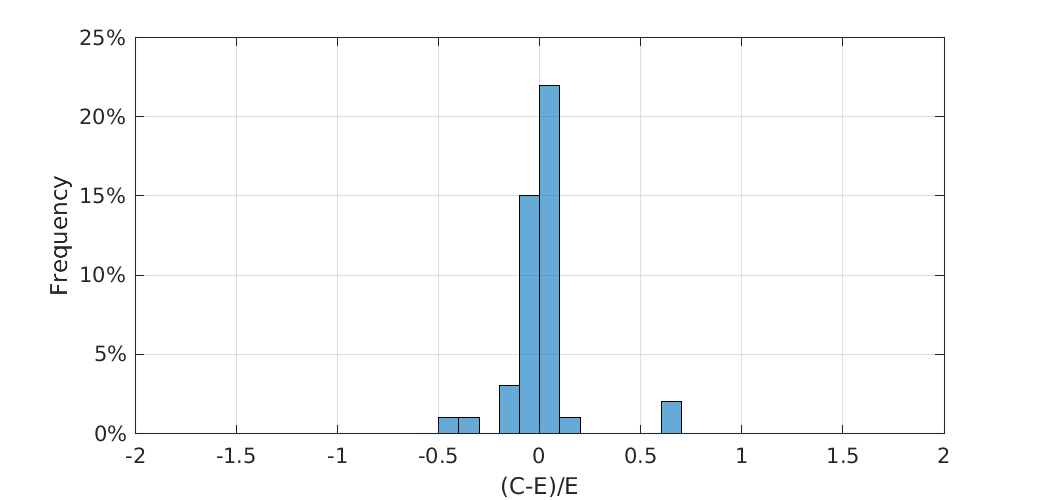}
        \caption{}
        \label{fig:boronCE_J3p1p1}
    \end{subfigure}
    \begin{subfigure}{1\textwidth}
        \centering
        \includegraphics[width=0.8\linewidth]{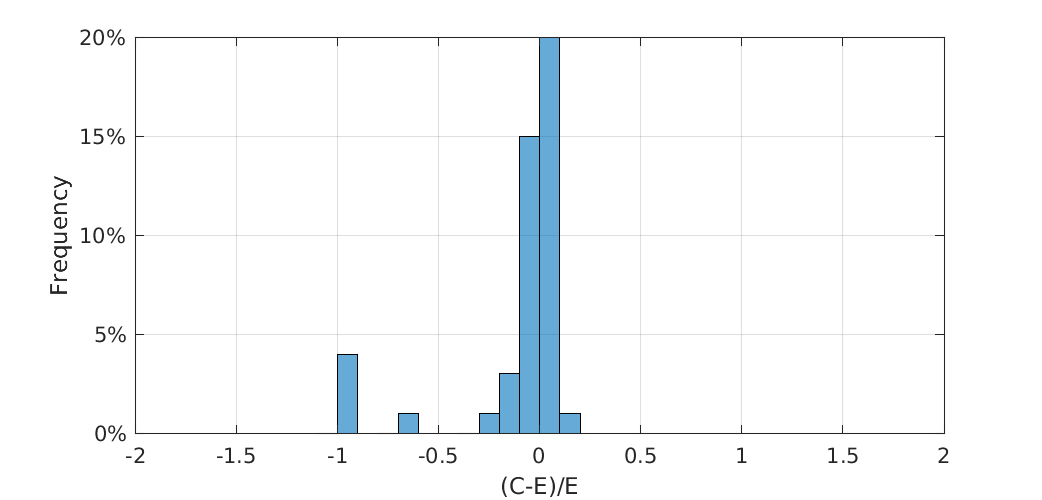}
        \caption{}
        \label{fig:boronCE_J4t30}
    \end{subfigure}
    \begin{subfigure}{1\textwidth}
        \centering
        \includegraphics[width=0.8\linewidth]{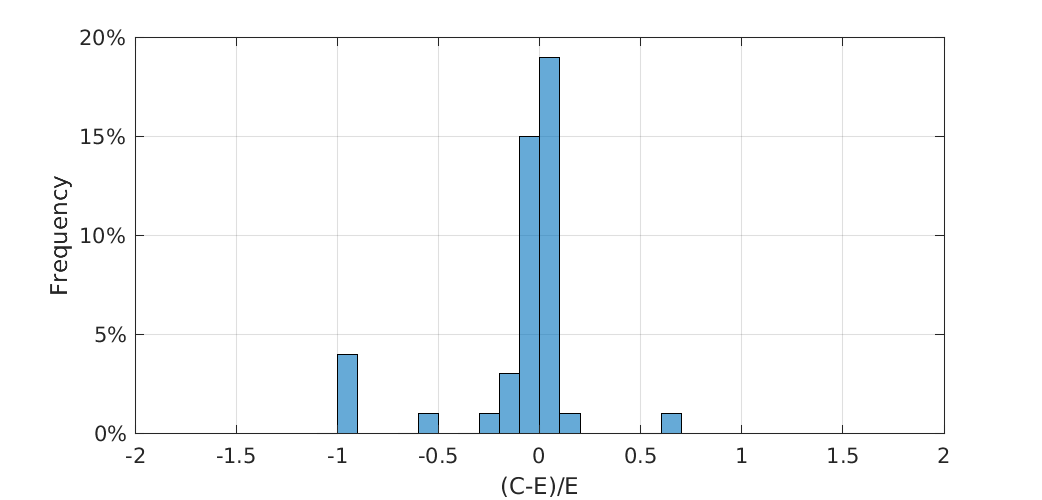}
        \caption{}
        \label{fig:boronCE_E81b1}
    \end{subfigure}
    \caption{Comparison of the performances of different nuclear data libraries to predict the control rod worths at the beginning of the first cycle of Almaraz-2 and the boron concentration in water during the first two cycles of Almaraz-2 and the first cycle of Fessenheim-2 using (a) \texttt{JEFF--3.1.1} (average $=-0.57\,\%$, $\sigma=16.77$\,\%) (b) \texttt{JEFF--4T3} (average $=-11.33\,\%$, $\sigma=29.72$\,\%) (c) \texttt{ENDF/B--VIII.0} (average $=-10.16\,\%$, $\sigma=31.25$\,\%).}
\label{fig:boronCE}
\end{figure}

A preliminary analysis to examine if this procedure is viable is conducted. The performances of the nuclear data libraries \texttt{JEFF--3.1.1}~\cite{JEFF}, \texttt{JEFF--4T3}~\cite{JEFF}, and \texttt{ENDF/B--VIII.0}~\cite{ENDF} to predict the control rod worths at the beginning of cycle 1 of Almaraz-2, the boron concentration in water during the first two cycles of Almaraz-2 and the boron concentration in water during the first cycle of Fessenheim-2 are assessed. As only 45 data points are available for these quantities, no conclusion can be drawn. The goal is to determine whether clear distinct behaviours can be observed between nuclear data libraries. The relative error of the predictions compared to measurements $(C-E)/E$ are illustrated on figure~\ref{fig:boronCE_J3p1p1}, \ref{fig:boronCE_J4t30} and \ref{fig:boronCE_E81b1}.\\[0.5\baselineskip]
According to table~\ref{tab:boronCE}, the library \texttt{JEFF--3.1.1} appears to be more accurate ($(C-E)/E$ average closer to zero) and precise (smaller standard deviation) than the other libraries considered. These preliminary results suggest that nuclear data library biases can be identified using this method. However, a more extensive study is required to draw conclusions about the performance of the libraries examined.

\begin{table}[H]
    \centering
    \begin{NiceTabular}{|c|c|c|c|}
        \hline 
         \makecell{} & \makecell{\texttt{JEFF--3.1.1}} & \makecell{\texttt{JEFF--4T3}} & \makecell{\texttt{ENDF/B--VIII.0}} \\
        \hline
        \makecell{Average of\\$(C-E)/E$ [\%]} & -\,0.57 & -\,11.33 & -\,10.16 \\
        \hline
        \makecell{Standard Deviation\\ $\sigma$ [\%]} & 16.77 & 29.72 & 31.25 \\
        \hline
    \end{NiceTabular}
    \caption{Comparison of the performances of different nuclear data libraries to predict the control rod worths at the beginning of the first cycle of Almaraz-2 and the boron concentration in water during the first two cycles of Almaraz-2 and the first cycle of Fessenheim-2.}
    \label{tab:boronCE}
\end{table}

\newpage
\section{Conclusion}
\label{section:conclusion}

The objective of the present Master's Thesis was to model multiple consecutive cycles of PWRs and to gather publicly available data, with the ultimate goal being to establish a foundation for building a pipeline to automate the assessment of the performance of novel nuclear data libraries using open source experimental data. This work represents a significant step towards the development of such an automated assessment framework.
The first step involved developing a scheme to manage history variables through the two-step approach using \texttt{DRAGON} and \texttt{PARCS}. The implementation was subsequently verified through a code-to-code comparison with their modeling, using a simple core featuring a single assembly and reflective boundary conditions. The verification demonstrated success in predicting the effects of historical data on the boron letdown curve for the Pyrex insertion history, as well as the axial offset for the moderator density histories. \\[0.5\baselineskip]
The second step consisted in the modeling of seven PWR depletion cycles. A literature review of publicly available PWR measurements was carried out. The reactors Fessenheim-1, Fessenheim-2, Bugey-2, Almaraz-2, Turkey-Point-3 were identified as reactors with a substantial amount of data for multiple depletion cycles. Subsequently, strategies to digitise old operational reports were presented and compared. These strategies were used to obtain the experimental data in this Master's Thesis, and can be used to extend the database of experimental data. Subsequently, the modeling of two cycles of Fessenheim-2, two cycles of Almaraz-2 and three cycles of Turkey-Point-3 was validated against these measurements.

The modeling of Almaraz-2 was successfully validated for the prediction of the boron letdown curve and the prediction of axial offset for both cycles, with the exception of the stretch out operation at the end of the first cycle.

The modeling of Fessenheim-2 was successfully validated for the prediction of the boron letdown curve for the first cycle. However, it failed to predict the detector's response for both cycles. The modeling of both Fessenheim-2 and the detector's response has to be improved.

The modeling of Turkey-Point-3 failed to be validated for the boron letdown curve. The current modeling of reactor's fuel assemblies in \texttt{DRAGON} was identified as the problem\\[0.5\baselineskip]
Finally, the last part consisted in presenting a methodology for the establishment of an automated procedure to assess to quality of novel nuclear data libraries against measurements performed in PWRs. The method proved to have potential based on the analysis of the diluted boron concentration in the first two cycles of Almaraz-2 and the first cycle of Fessenheim-1, as well as the control rod worths before the first cycle of Almaraz-2. \\[0.5\baselineskip]
In conclusion, the main achievements of this Master's Thesis can be summarised as follows:
\begin{outline}
    \1 Implementation and verification of the handling of history variables in \texttt{DRAGON} to determine the macroscopic cross sections of the nodal code \texttt{PARCS}.
    \1 Development of models to simulate seven depletion cycles across three reactors using \texttt{DRAGON} and \texttt{PARCS} and validation against publicly available data.
    \1 Establishment of a methodology to assess the quality of novel nuclear data libraries against experimental measurements in PWRs.
\end{outline}

\noindent Further efforts are necessary for the establishment of an automated procedure using the methodology introduced in the present Master's Thesis, regarding both reactor modeling and code development:
\begin{outline}
    \1 The modeling of Turkey-Point-3 and the second depletion cycle of Fessenheim-2 must be enhanced before they can be integrated into this procedure.
    \1 The interface between \texttt{DRAGON} and \texttt{PARCS} for the reconstruction of fission chambers' response used for the power measurements remains in a preliminary stage and requires further enhancement to be used in a nuclear data library validation procedure.
    \1 A substantial amount of experimental data points and reactor information is essential for the proper validation of nuclear data libraries. Therefore, a digitised database of open data needs to be developed.
    \1 The current procedure consists of several separate scripts. Creating a meta-script to consolidate these would facilitate easier use for routine applications.
    \1 Some components of the procedure are currently implemented in \texttt{MATLAB}. Transitioning these to an open-source language like \texttt{Python} is necessary.
\end{outline}
\newpage
\part*{Appendix}
\appendix
\addcontentsline{toc}{section}{Appendix}
\setcounter{section}{1}
\renewcommand{\thesection}{\Alph{section}}

\cfoot{Appendix}
\label{Appendix}

\subsection{Docker configuration for the use of \texttt{DRAGON}}

\renewcommand{\lstlistingname}{Dockerfile}
\setcounter{lstlisting}{0}

\lstset { %
    language=Matlab,
    backgroundcolor=\color{pink!8}, 
    basicstyle=\footnotesize,
}

\lstdefinestyle{customc}{
  language=Bash,
  showstringspaces=false,
  basicstyle=\footnotesize\ttfamily,
  keywordstyle=\bfseries\color{black},
  commentstyle=\itshape\color{red},
  identifierstyle=\color{black},
  stringstyle=\color{orange},
}
\lstset{escapechar=@,style=customc}

\begin{minipage}{\linewidth}
\begin{lstlisting}[caption={Docker configuration for the use of \texttt{DRAGON}}, label={lst:Dockerfile}]
# Configurations with lighter Operating Systems (such as Alpine) may work as well
FROM ubuntu:18.04

RUN apt-get update && \
    apt-get install -y software-properties-common && \
    add-apt-repository ppa:ubuntu-toolchain-r/test && \
    apt-get update && \
    apt-get -y install python && \
    apt-get -y install vim  && \
    apt-get -y install nano  && \
    apt-get -y install make && \
    apt-get -y install sudo && \
    apt-get install -y gcc-6 g++-6 gfortran-6 && \
    update-alternatives --install /usr/bin/gcc gcc /usr/bin/gcc-6 50 && \
    update-alternatives --install /usr/bin/g++ g++ /usr/bin/g++-6 50 && \
    update-alternatives --install /usr/bin/gfortran gfortran /usr/bin/gfortran-6 50 && \
    apt-get -y install g++ && \
    apt-get -y update && \
    apt-get clean && \
    rm -rf /var/lib/apt/lists/*

# Displays the version of gcc, gfortran and python in the Docker image
RUN gcc --version && \
    gfortran --version && \
    python --version
\end{lstlisting}
\end{minipage}
\newpage
\subsection{Measurement Database used in this Master's Thesis}

\renewcommand{\lstlistingname}{Database file}
\setcounter{lstlisting}{0}
\lstset{
    basicstyle=\tiny\ttfamily,
}

\lstset{numbers=none}
\subsubsection{Almaraz-2 Cycle 1}
\begin{minipage}{\linewidth}
\begin{lstlisting}[caption={almaraz/cycle1/boron.txt}]
0       1280
25      1074
159     1003
715     883
1940    856
3000    863
4500    805
6146    663
8200    534
9912    399
11500   284
13250   148
15100   12
15323   10   
\end{lstlisting}
\end{minipage}
\begin{minipage}{\linewidth}
\begin{lstlisting}[caption={almaraz/cycle1/axOff.txt}]
0        -4.4
715      -12.6
1940     -9.8
4500     -5.3
6146     -3.2
8200     -4.2
9912     -1.6
13250    -1.6
15100    -3.3
\end{lstlisting}
\end{minipage}
\begin{minipage}{\linewidth}
\begin{lstlisting}[caption={almaraz/cycle1/power1D\_Burn0.txt}]
0.        0.181
1.7544    0.203
3.5088    0.295
5.2631    0.386
7.0175    0.472
8.7719    0.55
10.5263   0.631
12.2806   0.705
14.035    0.772
15.7894   0.823
17.5438   0.836
19.2982   0.962
21.0525   1.048
22.8069   1.113
24.5613   1.171
26.3157   1.222
28.07     1.26
29.8244   1.279
31.5788   1.227
33.3332   1.353
35.0876   1.419
36.842    1.46
38.5964   1.491
40.3507   1.508
42.1051   1.514
43.8595   1.496
45.6139   1.403
47.3682   1.5
49.1226   1.543
50.877    1.55
52.6314   1.547
54.3857   1.531
56.1401   1.504
57.8945   1.454
59.6489   1.33
61.4033   1.383
63.1577   1.391
64.912    1.363
66.6664   1.325
68.4208   1.279
70.1752   1.223
71.9296   1.15
73.6839   1.015
75.4383   1.015
77.1927   0.984
78.9471   0.927
80.7015   0.863
82.4558   0.792
84.2102   0.713
85.9646   0.626
87.719    0.512
89.4734   0.459
91.2277   0.395
92.9821   0.319
94.7365   0.248
96.4909   0.182
98.2453   0.135
100.      0.1
\end{lstlisting}
\end{minipage}
\begin{minipage}{\linewidth}
\begin{lstlisting}[caption={almaraz/cycle1/power1D\_Burn715.txt}]
0.        0.305
1.7544    0.356
3.5088    0.49
5.2631    0.614
7.0175    0.725
8.7719    0.822
10.5263   0.906
12.2806   0.978
14.035    1.036
15.7894   1.07
17.5438   1.052
19.2982   1.178
21.0525   1.242
22.8069   1.283
24.5613   1.313
26.3157   1.334
28.07     1.343
29.8244   1.333
31.5788   1.245
33.3332   1.35
35.0876   1.386
36.842    1.396
38.5964   1.4
40.3507   1.395
42.1051   1.383
43.8595   1.355
45.6139   1.247
47.3682   1.321
49.1226   1.347
50.877    1.343
52.6314   1.331
54.3857   1.312
56.1401   1.286
57.8945   1.248
59.6489   1.144
61.4033   1.18
63.1577   1.193
64.912    1.175
66.6664   1.15
68.4208   1.119
70.1752   1.083
71.9296   1.034
73.6839   0.938
75.4383   0.927
77.1927   0.924
78.9471   0.885
80.7015   0.839
82.4558   0.787
84.2102   0.727
85.9646   0.656
87.719    0.558
89.4734   0.498
91.2277   0.449
92.9821   0.38
94.7365   0.307
96.4909   0.235
98.2453   0.176
100.      0.127
\end{lstlisting}
\end{minipage}
\begin{minipage}{\linewidth}
\begin{lstlisting}[caption={almaraz/cycle1/power1D\_Burn13250.txt}]
0.        0.488
1.7544    0.578
3.5088    0.759
5.2631    0.871
7.0175    0.975
8.7719    1.054
10.5263   1.097
12.2806   1.125
14.035    1.134
15.7894   1.106
17.5438   1.029
19.2982   1.116
21.0525   1.125
22.8069   1.117
24.5613   1.116
26.3157   1.105
28.07     1.09
29.8244   1.057
31.5788   0.967
33.3332   1.037
35.0876   1.055
36.842    1.048
38.5964   1.052
40.3507   1.048
42.1051   1.04
43.8595   1.019
45.6139   0.938
47.3682   1.006
49.1226   1.035
50.877    1.036
52.6314   1.049
54.3857   1.051
56.1401   1.05
57.8945   1.039
59.6489   0.966
61.4033   1.018
63.1577   1.066
64.912    1.072
66.6664   1.091
68.4208   1.098
70.1752   1.102
71.9296   1.095
73.6839   1.028
75.4383   1.061
77.1927   1.117
78.9471   1.121
80.7015   1.126
82.4558   1.117
84.2102   1.098
85.9646   1.061
87.719    0.966
89.4734   0.913
91.2277   0.898
92.9821   0.812
94.7365   0.709
96.4909   0.588
98.2453   0.489
100.      0.442
\end{lstlisting}
\end{minipage}
\begin{minipage}{\linewidth}
\begin{lstlisting}[caption={almaraz/cycle1/power2D\_Burn0.txt}]
0.000  0.000  0.000  0.000  0.000  0.000  0.633  0.843  0.633  0.000  0.000  0.000  0.000  0.000  0.000
0.000  0.000  0.000  0.000  0.593  0.911  1.022  1.082  1.022  0.911  0.593  0.000  0.000  0.000  0.000
0.000  0.000  0.000  0.630  0.913  0.989  1.143  1.189  1.143  0.989  0.913  0.630  0.000  0.000  0.000
0.000  0.000  0.630  0.855  0.977  1.123  1.126  1.192  1.126  1.123  0.977  0.855  0.630  0.000  0.000
0.000  0.593  0.913  0.977  1.120  1.127  1.178  1.078  1.178  1.127  1.120  0.977  0.913  0.593  0.000
0.000  0.911  0.989  1.123  1.127  1.200  1.142  1.176  1.142  1.200  1.127  1.123  0.989  0.911  0.000
0.633  1.022  1.143  1.126  1.178  1.142  1.157  1.061  1.157  1.142  1.178  1.126  1.143  1.022  0.633
0.843  1.082  1.189  1.192  1.078  1.176  1.061  1.138  1.061  1.176  1.078  1.192  1.189  1.082  0.843
0.633  1.022  1.143  1.126  1.178  1.142  1.157  1.061  1.157  1.142  1.178  1.126  1.143  1.022  0.633
0.000  0.911  0.989  1.123  1.127  1.200  1.142  1.176  1.142  1.200  1.127  1.123  0.989  0.911  0.000
0.000  0.593  0.913  0.977  1.120  1.127  1.178  1.078  1.178  1.127  1.120  0.977  0.913  0.593  0.000
0.000  0.000  0.630  0.855  0.977  1.123  1.126  1.192  1.126  1.123  0.977  0.855  0.630  0.000  0.000
0.000  0.000  0.000  0.630  0.913  0.989  1.143  1.189  1.143  0.989  0.913  0.630  0.000  0.000  0.000
0.000  0.000  0.000  0.000  0.593  0.911  1.022  1.082  1.022  0.911  0.593  0.000  0.000  0.000  0.000
0.000  0.000  0.000  0.000  0.000  0.000  0.633  0.843  0.633  0.000  0.000  0.000  0.000  0.000  0.000
\end{lstlisting}
\end{minipage}
\begin{minipage}{\linewidth}
\begin{lstlisting}[caption={almaraz/cycle1/power2D\_Burn715.txt}]
0.000  0.000  0.000  0.000  0.000  0.000  0.595  0.771  0.595  0.000  0.000  0.000  0.000  0.000  0.000
0.000  0.000  0.000  0.000  0.585  0.862  0.963  1.018  0.963  0.862  0.585  0.000  0.000  0.000  0.000
0.000  0.000  0.000  0.633  0.895  0.965  1.116  1.156  1.116  0.965  0.895  0.633  0.000  0.000  0.000
0.000  0.000  0.633  0.859  0.980  1.122  1.143  1.201  1.143  1.122  0.980  0.859  0.633  0.000  0.000
0.000  0.585  0.895  0.980  1.127  1.148  1.215  1.131  1.215  1.148  1.127  0.980  0.895  0.585  0.000
0.000  0.862  0.965  1.122  1.148  1.233  1.202  1.236  1.202  1.233  1.148  1.122  0.965  0.862  0.000
0.595  0.963  1.116  1.143  1.215  1.202  1.230  1.137  1.230  1.202  1.215  1.143  1.116  0.963  0.595
0.771  1.018  1.156  1.201  1.131  1.236  1.137  1.206  1.137  1.236  1.131  1.201  1.156  1.018  0.771
0.595  0.963  1.116  1.143  1.215  1.202  1.230  1.137  1.230  1.202  1.215  1.143  1.116  0.963  0.595
0.000  0.862  0.965  1.122  1.148  1.233  1.202  1.236  1.202  1.233  1.148  1.122  0.965  0.862  0.000
0.000  0.585  0.895  0.980  1.127  1.148  1.215  1.131  1.215  1.148  1.127  0.980  0.895  0.585  0.000
0.000  0.000  0.633  0.859  0.980  1.122  1.143  1.201  1.143  1.122  0.980  0.859  0.633  0.000  0.000
0.000  0.000  0.000  0.633  0.895  0.965  1.116  1.156  1.116  0.965  0.895  0.633  0.000  0.000  0.000
0.000  0.000  0.000  0.000  0.585  0.862  0.963  1.018  0.963  0.862  0.585  0.000  0.000  0.000  0.000
0.000  0.000  0.000  0.000  0.000  0.000  0.595  0.771  0.595  0.000  0.000  0.000  0.000  0.000  0.000
\end{lstlisting}
\end{minipage}
\begin{minipage}{\linewidth}
\begin{lstlisting}[caption={almaraz/cycle1/power2D\_Burn1940.txt}]
0.000  0.000  0.000  0.000  0.000  0.000  0.573  0.738  0.573  0.000  0.000  0.000  0.000  0.000  0.000
0.000  0.000  0.000  0.000  0.568  0.841  0.946  0.981  0.946  0.841  0.568  0.000  0.000  0.000  0.000
0.000  0.000  0.000  0.621  0.892  0.972  1.105  1.156  1.105  0.972  0.892  0.621  0.000  0.000  0.000
0.000  0.000  0.621  0.857  0.996  1.128  1.157  1.196  1.157  1.128  0.996  0.857  0.621  0.000  0.000
0.000  0.568  0.892  0.996  1.135  1.174  1.218  1.154  1.218  1.174  1.135  0.996  0.892  0.568  0.000
0.000  0.841  0.972  1.128  1.174  1.230  1.225  1.243  1.225  1.230  1.174  1.128  0.972  0.841  0.000
0.573  0.946  1.105  1.157  1.218  1.225  1.245  1.178  1.245  1.225  1.218  1.157  1.105  0.946  0.573
0.738  0.981  1.156  1.196  1.154  1.243  1.178  1.236  1.178  1.243  1.154  1.196  1.156  0.981  0.738
0.573  0.946  1.105  1.157  1.218  1.225  1.245  1.178  1.245  1.225  1.218  1.157  1.105  0.946  0.573
0.000  0.841  0.972  1.128  1.174  1.230  1.225  1.243  1.225  1.230  1.174  1.128  0.972  0.841  0.000
0.000  0.568  0.892  0.996  1.135  1.174  1.218  1.154  1.218  1.174  1.135  0.996  0.892  0.568  0.000
0.000  0.000  0.621  0.857  0.996  1.128  1.157  1.196  1.157  1.128  0.996  0.857  0.621  0.000  0.000
0.000  0.000  0.000  0.621  0.892  0.972  1.105  1.156  1.105  0.972  0.892  0.621  0.000  0.000  0.000
0.000  0.000  0.000  0.000  0.568  0.841  0.946  0.981  0.946  0.841  0.568  0.000  0.000  0.000  0.000
0.000  0.000  0.000  0.000  0.000  0.000  0.573  0.738  0.573  0.000  0.000  0.000  0.000  0.000  0.000
\end{lstlisting}
\end{minipage}
\begin{minipage}{\linewidth}
\begin{lstlisting}[caption={almaraz/cycle1/power2D\_Burn4500.txt}]
0.000  0.000  0.000  0.000  0.000  0.000  0.554  0.700  0.554  0.000  0.000  0.000  0.000  0.000  0.000
0.000  0.000  0.000  0.000  0.574  0.825  0.937  0.944  0.937  0.825  0.574  0.000  0.000  0.000  0.000
0.000  0.000  0.000  0.635  0.919  0.986  1.077  1.145  1.077  0.986  0.919  0.635  0.000  0.000  0.000
0.000  0.000  0.635  0.874  1.030  1.111  1.171  1.173  1.171  1.111  1.030  0.874  0.635  0.000  0.000
0.000  0.574  0.919  1.030  1.125  1.186  1.203  1.190  1.203  1.186  1.125  1.030  0.919  0.574  0.000
0.000  0.825  0.986  1.111  1.186  1.200  1.247  1.232  1.247  1.200  1.186  1.111  0.986  0.825  0.000
0.554  0.937  1.077  1.171  1.203  1.247  1.235  1.216  1.235  1.247  1.203  1.171  1.077  0.937  0.554
0.700  0.944  1.145  1.173  1.190  1.232  1.216  1.225  1.216  1.232  1.190  1.173  1.145  0.944  0.700
0.554  0.937  1.077  1.171  1.203  1.247  1.235  1.216  1.235  1.247  1.203  1.171  1.077  0.937  0.554
0.000  0.825  0.986  1.111  1.186  1.200  1.247  1.232  1.247  1.200  1.186  1.111  0.986  0.825  0.000
0.000  0.574  0.919  1.030  1.125  1.186  1.203  1.190  1.203  1.186  1.125  1.030  0.919  0.574  0.000
0.000  0.000  0.635  0.874  1.030  1.111  1.171  1.173  1.171  1.111  1.030  0.874  0.635  0.000  0.000
0.000  0.000  0.000  0.635  0.919  0.986  1.077  1.145  1.077  0.986  0.919  0.635  0.000  0.000  0.000
0.000  0.000  0.000  0.000  0.574  0.825  0.937  0.944  0.937  0.825  0.574  0.000  0.000  0.000  0.000
0.000  0.000  0.000  0.000  0.000  0.000  0.554  0.700  0.554  0.000  0.000  0.000  0.000  0.000  0.000
\end{lstlisting}
\end{minipage}
\begin{minipage}{\linewidth}
\begin{lstlisting}[caption={almaraz/cycle1/power2D\_Burn6146.txt}]
0.000  0.000  0.000  0.000  0.000  0.000  0.560  0.708  0.560  0.000  0.000  0.000  0.000  0.000  0.000
0.000  0.000  0.000  0.000  0.576  0.827  0.946  0.950  0.946  0.827  0.576  0.000  0.000  0.000  0.000
0.000  0.000  0.000  0.636  0.934  1.006  1.064  1.141  1.064  1.006  0.934  0.636  0.000  0.000  0.000
0.000  0.000  0.636  0.870  1.043  1.105  1.172  1.152  1.172  1.105  1.043  0.870  0.636  0.000  0.000
0.000  0.576  0.934  1.043  1.115  1.191  1.185  1.193  1.185  1.191  1.115  1.043  0.934  0.576  0.000
0.000  0.827  1.006  1.105  1.191  1.195  1.245  1.209  1.245  1.195  1.191  1.105  1.006  0.827  0.000
0.560  0.946  1.064  1.172  1.185  1.245  1.214  1.220  1.214  1.245  1.185  1.172  1.064  0.946  0.560
0.708  0.950  1.141  1.152  1.193  1.209  1.220  1.209  1.220  1.209  1.193  1.152  1.141  0.950  0.708
0.560  0.946  1.064  1.172  1.185  1.245  1.214  1.220  1.214  1.245  1.185  1.172  1.064  0.946  0.560
0.000  0.827  1.006  1.105  1.191  1.195  1.245  1.209  1.245  1.195  1.191  1.105  1.006  0.827  0.000
0.000  0.576  0.934  1.043  1.115  1.191  1.185  1.193  1.185  1.191  1.115  1.043  0.934  0.576  0.000
0.000  0.000  0.636  0.870  1.043  1.105  1.172  1.152  1.172  1.105  1.043  0.870  0.636  0.000  0.000
0.000  0.000  0.000  0.636  0.934  1.006  1.064  1.141  1.064  1.006  0.934  0.636  0.000  0.000  0.000
0.000  0.000  0.000  0.000  0.576  0.827  0.946  0.950  0.946  0.827  0.576  0.000  0.000  0.000  0.000
0.000  0.000  0.000  0.000  0.000  0.000  0.560  0.708  0.560  0.000  0.000  0.000  0.000  0.000  0.000
\end{lstlisting}
\end{minipage}
\begin{minipage}{\linewidth}
\begin{lstlisting}[caption={almaraz/cycle1/power2D\_Burn8200.txt}]
0.000  0.000  0.000  0.000  0.000  0.000  0.569  0.709  0.569  0.000  0.000  0.000  0.000  0.000  0.000
0.000  0.000  0.000  0.000  0.586  0.830  0.960  0.945  0.960  0.830  0.586  0.000  0.000  0.000  0.000
0.000  0.000  0.000  0.645  0.952  1.025  1.057  1.142  1.057  1.025  0.952  0.645  0.000  0.000  0.000
0.000  0.000  0.645  0.879  1.066  1.097  1.178  1.136  1.178  1.097  1.066  0.879  0.645  0.000  0.000
0.000  0.586  0.952  1.066  1.106  1.192  1.158  1.196  1.158  1.192  1.106  1.066  0.952  0.586  0.000
0.000  0.830  1.025  1.097  1.192  1.166  1.233  1.179  1.233  1.166  1.192  1.097  1.025  0.830  0.000
0.569  0.960  1.057  1.178  1.158  1.233  1.182  1.221  1.182  1.233  1.158  1.178  1.057  0.960  0.569
0.709  0.945  1.142  1.136  1.196  1.179  1.221  1.186  1.221  1.179  1.196  1.136  1.142  0.945  0.709
0.569  0.960  1.057  1.178  1.158  1.233  1.182  1.221  1.182  1.233  1.158  1.178  1.057  0.960  0.569
0.000  0.830  1.025  1.097  1.192  1.166  1.233  1.179  1.233  1.166  1.192  1.097  1.025  0.830  0.000
0.000  0.586  0.952  1.066  1.106  1.192  1.158  1.196  1.158  1.192  1.106  1.066  0.952  0.586  0.000
0.000  0.000  0.645  0.879  1.066  1.097  1.178  1.136  1.178  1.097  1.066  0.879  0.645  0.000  0.000
0.000  0.000  0.000  0.645  0.952  1.025  1.057  1.142  1.057  1.025  0.952  0.645  0.000  0.000  0.000
0.000  0.000  0.000  0.000  0.586  0.830  0.960  0.945  0.960  0.830  0.586  0.000  0.000  0.000  0.000
0.000  0.000  0.000  0.000  0.000  0.000  0.569  0.709  0.569  0.000  0.000  0.000  0.000  0.000  0.000
\end{lstlisting}
\end{minipage}
\begin{minipage}{\linewidth}
\begin{lstlisting}[caption={almaraz/cycle1/power2D\_Burn9912.txt}]
0.000  0.000  0.000  0.000  0.000  0.000  0.573  0.710  0.573  0.000  0.000  0.000  0.000  0.000  0.000
0.000  0.000  0.000  0.000  0.600  0.846  0.974  0.948  0.974  0.846  0.600  0.000  0.000  0.000  0.000
0.000  0.000  0.000  0.650  0.979  1.048  1.057  1.144  1.057  1.048  0.979  0.650  0.000  0.000  0.000
0.000  0.000  0.650  0.886  1.081  1.095  1.175  1.119  1.175  1.095  1.081  0.886  0.650  0.000  0.000
0.000  0.600  0.979  1.081  1.098  1.182  1.140  1.184  1.140  1.182  1.098  1.081  0.979  0.600  0.000
0.000  0.846  1.048  1.095  1.182  1.140  1.215  1.148  1.215  1.140  1.182  1.095  1.048  0.846  0.000
0.573  0.974  1.057  1.175  1.140  1.215  1.157  1.199  1.157  1.215  1.140  1.175  1.057  0.974  0.573
0.710  0.948  1.144  1.119  1.184  1.148  1.199  1.149  1.199  1.148  1.184  1.119  1.144  0.948  0.710
0.573  0.974  1.057  1.175  1.140  1.215  1.157  1.199  1.157  1.215  1.140  1.175  1.057  0.974  0.573
0.000  0.846  1.048  1.095  1.182  1.140  1.215  1.148  1.215  1.140  1.182  1.095  1.048  0.846  0.000
0.000  0.600  0.979  1.081  1.098  1.182  1.140  1.184  1.140  1.182  1.098  1.081  0.979  0.600  0.000
0.000  0.000  0.650  0.886  1.081  1.095  1.175  1.119  1.175  1.095  1.081  0.886  0.650  0.000  0.000
0.000  0.000  0.000  0.650  0.979  1.048  1.057  1.144  1.057  1.048  0.979  0.650  0.000  0.000  0.000
0.000  0.000  0.000  0.000  0.600  0.846  0.974  0.948  0.974  0.846  0.600  0.000  0.000  0.000  0.000
0.000  0.000  0.000  0.000  0.000  0.000  0.573  0.710  0.573  0.000  0.000  0.000  0.000  0.000  0.000
\end{lstlisting}
\end{minipage}
\begin{minipage}{\linewidth}
\begin{lstlisting}[caption={almaraz/cycle1/power2D\_Burn13250.txt}]
0.000  0.000  0.000  0.000  0.000  0.000  0.598  0.731  0.598  0.000  0.000  0.000  0.000  0.000  0.000
0.000  0.000  0.000  0.000  0.620  0.859  1.004  0.961  1.004  0.859  0.620  0.000  0.000  0.000  0.000
0.000  0.000  0.000  0.677  1.005  1.065  1.058  1.145  1.058  1.065  1.005  0.677  0.000  0.000  0.000
0.000  0.000  0.677  0.907  1.101  1.081  1.167  1.104  1.167  1.081  1.101  0.907  0.677  0.000  0.000
0.000  0.620  1.005  1.101  1.087  1.163  1.107  1.169  1.107  1.163  1.087  1.101  1.005  0.620  0.000
0.000  0.859  1.065  1.081  1.163  1.104  1.178  1.114  1.178  1.104  1.163  1.081  1.065  0.859  0.000
0.598  1.004  1.058  1.167  1.107  1.178  1.114  1.171  1.114  1.178  1.107  1.167  1.058  1.004  0.598
0.731  0.961  1.145  1.104  1.169  1.114  1.171  1.115  1.171  1.114  1.169  1.104  1.145  0.961  0.731
0.598  1.004  1.058  1.167  1.107  1.178  1.114  1.171  1.114  1.178  1.107  1.167  1.058  1.004  0.598
0.000  0.859  1.065  1.081  1.163  1.104  1.178  1.114  1.178  1.104  1.163  1.081  1.065  0.859  0.000
0.000  0.620  1.005  1.101  1.087  1.163  1.107  1.169  1.107  1.163  1.087  1.101  1.005  0.620  0.000
0.000  0.000  0.677  0.907  1.101  1.081  1.167  1.104  1.167  1.081  1.101  0.907  0.677  0.000  0.000
0.000  0.000  0.000  0.677  1.005  1.065  1.058  1.145  1.058  1.065  1.005  0.677  0.000  0.000  0.000
0.000  0.000  0.000  0.000  0.620  0.859  1.004  0.961  1.004  0.859  0.620  0.000  0.000  0.000  0.000
0.000  0.000  0.000  0.000  0.000  0.000  0.598  0.731  0.598  0.000  0.000  0.000  0.000  0.000  0.000
\end{lstlisting}
\end{minipage}
\begin{minipage}{\linewidth}
\begin{lstlisting}[caption={almaraz/cycle1/power2D\_Burn15100.txt}]
0.000  0.000  0.000  0.000  0.000  0.000  0.614  0.750  0.614  0.000  0.000  0.000  0.000  0.000  0.000
0.000  0.000  0.000  0.000  0.636  0.876  1.027  0.976  1.027  0.876  0.636  0.000  0.000  0.000  0.000
0.000  0.000  0.000  0.691  1.017  1.073  1.058  1.142  1.058  1.073  1.017  0.691  0.000  0.000  0.000
0.000  0.000  0.691  0.919  1.109  1.080  1.156  1.089  1.156  1.080  1.109  0.919  0.691  0.000  0.000
0.000  0.636  1.017  1.109  1.081  1.152  1.087  1.140  1.087  1.152  1.081  1.109  1.017  0.636  0.000
0.000  0.876  1.073  1.080  1.152  1.089  1.155  1.091  1.155  1.089  1.152  1.080  1.073  0.876  0.000
0.614  1.027  1.058  1.156  1.087  1.155  1.091  1.148  1.091  1.155  1.087  1.156  1.058  1.027  0.614
0.750  0.976  1.142  1.089  1.140  1.091  1.148  1.100  1.148  1.091  1.140  1.089  1.142  0.976  0.750
0.614  1.027  1.058  1.156  1.087  1.155  1.091  1.148  1.091  1.155  1.087  1.156  1.058  1.027  0.614
0.000  0.876  1.073  1.080  1.152  1.089  1.155  1.091  1.155  1.089  1.152  1.080  1.073  0.876  0.000
0.000  0.636  1.017  1.109  1.081  1.152  1.087  1.140  1.087  1.152  1.081  1.109  1.017  0.636  0.000
0.000  0.000  0.691  0.919  1.109  1.080  1.156  1.089  1.156  1.080  1.109  0.919  0.691  0.000  0.000
0.000  0.000  0.000  0.691  1.017  1.073  1.058  1.142  1.058  1.073  1.017  0.691  0.000  0.000  0.000
0.000  0.000  0.000  0.000  0.636  0.876  1.027  0.976  1.027  0.876  0.636  0.000  0.000  0.000  0.000
0.000  0.000  0.000  0.000  0.000  0.000  0.614  0.750  0.614  0.000  0.000  0.000  0.000  0.000  0.000
\end{lstlisting}
\end{minipage}
\subsubsection{Almaraz-2 Cycle 2}
\begin{minipage}{\linewidth}
\begin{lstlisting}[caption={almaraz/cycle2/fuelShuffling.txt}]
                                 0    0    0    0    0
                       0    0    0   -5   -5   -5    0    0    0
                  0    0   -5   -5   -5  H-7   -5   -5   -5    0    0
             0    0   -5  J-1  J-6  J-5 H-13  G-5  G-6  G-1   -5    0    0
        0    0   -5  N-4  K-5  E-2  G-2  H-5  J-2  L-2  F-5  C-4   -5    0    0
        0   -5  R-7  L-6  M-3  L-4 K-14  J-4 F-14  E-4  D-3  E-6  A-7   -5    0
   0    0   -5  K-7 P-11  M-5  N-5  K-3 H-15  F-3  C-5  D-5 B-11  F-7   -5    0    0
   0   -5   -5  L-7  P-9  B-6  N-6  L-3  G-4  E-3  C-6  P-6  B-9  E-7   -5   -5    0
   0   -5  J-8  C-8  L-8  M-9  A-8  M-7 F-12  D-9  R-8  D-7  E-8  N-8  G-8   -5    0
   0   -5   -5  L-9  P-7 B-10 N-10 L-13 J-12 E-13 C-10 P-10  B-7  E-9   -5   -5    0
   0    0   -5  K-9  P-5 M-11 N-11 K-13  H-1 F-13 C-11 D-11  B-5  F-9   -5    0    0
        0   -5  R-9 L-10 M-13 L-12  K-2 G-12  F-2 E-12 D-13 E-10  A-9   -5    0
        0    0   -5 N-12 K-11 E-14 G-14 H-11 J-14 L-14 F-11 C-12   -5    0    0
             0    0   -5 J-15 J-10 J-11  H-3 G-11 G-10 G-15   -5    0    0
                  0    0   -5   -5   -5  H-9   -5   -5   -5    0    0
                       0    0    0   -5   -5   -5    0    0    0
                                 0    0    0    0    0
\end{lstlisting}
\end{minipage}
\begin{minipage}{\linewidth}
\begin{lstlisting}[caption={almaraz/cycle2/boronExp.txt}]
0       1212
0       1212
178     774
212     771
745     708
1863    616
3000    499
4461    363
5540    258
6589    155
7617    64
8436    14
8826    4
9551    4
\end{lstlisting}
\end{minipage}
\begin{minipage}{\linewidth}
\begin{lstlisting}[caption={almaraz/cycle2/axOff.txt}]
0       58.1
212     3.7
1863    -1.9
4461    -3.3
6589    -4.3
8436    -1.8
\end{lstlisting}
\end{minipage}
\begin{minipage}{\linewidth}
\begin{lstlisting}[caption={almaraz/cycle2/power1D\_Burn0.txt}]
0.00000e+00 7.00000e-02
1.75440e+00 9.90000e-02
3.50880e+00 1.13000e-01
5.26310e+00 1.54000e-01
7.01750e+00 1.88000e-01
8.77190e+00 2.16000e-01
1.05263e+01 2.38000e-01
1.22806e+01 2.59000e-01
1.40350e+01 2.78000e-01
1.57894e+01 2.92000e-01
1.75438e+01 3.03000e-01
1.92982e+01 3.00000e-01
2.10525e+01 3.35000e-01
2.28069e+01 3.61000e-01
2.45613e+01 3.84000e-01
2.63157e+01 4.06000e-01
2.80700e+01 4.28000e-01
2.98244e+01 4.51000e-01
3.15788e+01 4.67000e-01
3.33332e+01 4.71000e-01
3.50876e+01 5.31000e-01
3.68420e+01 5.78000e-01
3.85964e+01 6.17000e-01
4.03507e+01 6.57000e-01
4.21051e+01 6.99000e-01
4.38595e+01 7.38000e-01
4.56139e+01 7.62000e-01
4.73682e+01 7.73000e-01
4.91226e+01 8.68000e-01
5.08770e+01 9.45000e-01
5.26314e+01 1.00400e+00
5.43857e+01 1.06500e+00
5.61401e+01 1.12400e+00
5.78945e+01 1.17800e+00
5.96489e+01 1.21800e+00
6.14033e+01 1.20400e+00
6.31577e+01 1.35800e+00
6.49120e+01 1.46200e+00
6.66664e+01 1.54400e+00
6.84208e+01 1.61700e+00
7.01752e+01 1.69000e+00
7.19296e+01 1.75300e+00
7.36839e+01 1.79200e+00
7.54383e+01 1.73100e+00
7.71927e+01 1.92000e+00
7.89471e+01 2.02900e+00
8.07015e+01 2.08800e+00
8.24558e+01 2.12700e+00
8.42102e+01 2.14200e+00
8.59646e+01 2.13400e+00
8.77190e+01 2.06600e+00
8.94734e+01 1.86900e+00
9.12277e+01 1.87800e+00
9.29821e+01 1.77600e+00
9.47365e+01 1.57500e+00
9.64909e+01 1.29800e+00
9.82453e+01 9.97000e-01
1.00000e+02 8.39000e-01
\end{lstlisting}
\end{minipage}
\begin{minipage}{\linewidth}
\begin{lstlisting}[caption={almaraz/cycle2/power1D\_Burn212.txt}]
0.        0.425
1.7544    0.468
3.5088    0.631
5.2631    0.756
7.0175    0.849
8.7719    0.917
10.5263   0.963
12.2806   0.992
14.035    1.008
15.7894   0.995
17.5438   0.927
19.2982   1.015
21.0525   1.04
22.8069   1.048
24.5613   1.05
26.3157   1.05
28.07     1.046
29.8244   1.029
31.5788   0.939
33.3332   1.03
35.0876   1.055
36.842    1.062
38.5964   1.065
40.3507   1.066
42.1051   1.063
43.8595   1.046
45.6139   0.959
47.3682   1.057
49.1226   1.084
50.877    1.093
52.6314   1.098
54.3857   1.102
56.1401   1.101
57.8945   1.086
59.6489   1.001
61.4033   1.099
63.1577   1.131
64.912    1.144
66.6664   1.153
68.4208   1.158
70.1752   1.161
71.9296   1.15
73.6839   1.059
75.4383   1.161
77.1927   1.193
78.9471   1.201
80.7015   1.201
82.4558   1.189
84.2102   1.166
85.9646   1.121
87.719    0.992
89.4734   1.014
91.2277   0.961
92.9821   0.863
94.7365   0.729
96.4909   0.571
98.2453   0.479
100.      0.395
\end{lstlisting}
\end{minipage}
\begin{minipage}{\linewidth}
\begin{lstlisting}[caption={almaraz/cycle2/power1D\_Burn8436.txt}]
0.        0.593
1.7544    0.574
3.5088    0.771
5.2631    0.907
7.0175    0.996
8.7719    1.054
10.5263   1.087
12.2806   1.104
14.035    1.105
15.7894   1.084
17.5438   0.972
19.2982   1.058
21.0525   1.092
22.8069   1.093
24.5613   1.09
26.3157   1.085
28.07     1.077
29.8244   1.055
31.5788   0.945
33.3332   1.045
35.0876   1.076
36.842    1.081
38.5964   1.081
40.3507   1.078
42.1051   1.071
43.8595   1.05
45.6139   0.948
47.3682   1.044
49.1226   1.075
50.877    1.082
52.6314   1.084
54.3857   1.082
56.1401   1.077
57.8945   1.051
59.6489   0.946
61.4033   1.052
63.1577   1.079
64.912    1.085
66.6664   1.087
68.4208   1.087
70.1752   1.083
71.9296   1.06
73.6839   0.959
75.4383   1.065
77.1927   1.097
78.9471   1.101
80.7015   1.101
82.4558   1.091
84.2102   1.074
85.9646   1.035
87.719    0.907
89.4734   0.947
91.2277   0.913
92.9821   0.824
94.7365   0.701
96.4909   0.57
98.2453   0.524
100.      0.479
\end{lstlisting}
\end{minipage}
\begin{minipage}{\linewidth}
\begin{lstlisting}[caption={almaraz/cycle2/power2D\_Burn0.txt}]
0.000  0.000  0.000  0.000  0.000  0.000  0.788  0.958  0.788  0.000  0.000  0.000  0.000  0.000  0.000
0.000  0.000  0.000  0.000  0.695  1.056  1.331  1.007  1.331  1.056  0.695  0.000  0.000  0.000  0.000
0.000  0.000  0.000  0.729  1.015  0.951  0.827  0.877  0.827  0.951  1.015  0.729  0.000  0.000  0.000
0.000  0.000  0.729  1.002  0.967  1.195  1.081  0.898  1.081  1.195  0.967  1.002  0.729  0.000  0.000
0.000  0.695  1.015  0.967  1.200  1.063  1.163  0.987  1.163  1.063  1.200  0.967  1.015  0.695  0.000
0.000  1.056  0.951  1.195  1.063  1.160  1.056  1.172  1.056  1.160  1.063  1.195  0.951  1.056  0.000
0.788  1.331  0.827  1.081  1.163  1.056  1.067  0.907  1.067  1.056  1.163  1.081  0.827  1.331  0.788
0.958  1.007  0.877  0.898  0.987  1.172  0.907  0.728  0.907  1.172  0.987  0.898  0.877  1.007  0.958
0.788  1.331  0.827  1.081  1.163  1.056  1.067  0.907  1.067  1.056  1.163  1.081  0.827  1.331  0.788
0.000  1.056  0.951  1.195  1.063  1.160  1.056  1.172  1.056  1.160  1.063  1.195  0.951  1.056  0.000
0.000  0.695  1.015  0.967  1.200  1.063  1.163  0.987  1.163  1.063  1.200  0.967  1.015  0.695  0.000
0.000  0.000  0.729  1.002  0.967  1.195  1.081  0.898  1.081  1.195  0.967  1.002  0.729  0.000  0.000
0.000  0.000  0.000  0.729  1.015  0.951  0.827  0.877  0.827  0.951  1.015  0.729  0.000  0.000  0.000
0.000  0.000  0.000  0.000  0.695  1.056  1.331  1.007  1.331  1.056  0.695  0.000  0.000  0.000  0.000
0.000  0.000  0.000  0.000  0.000  0.000  0.788  0.958  0.788  0.000  0.000  0.000  0.000  0.000  0.000
\end{lstlisting}
\end{minipage}
\begin{minipage}{\linewidth}
\begin{lstlisting}[caption={almaraz/cycle2/power2D\_Burn212.txt}]
0.000  0.000  0.000  0.000  0.000  0.000  0.766  0.925  0.766  0.000  0.000  0.000  0.000  0.000  0.000
0.000  0.000  0.000  0.000  0.695  1.020  1.255  0.985  1.255  1.020  0.695  0.000  0.000  0.000  0.000
0.000  0.000  0.000  0.741  1.007  0.954  0.850  0.904  0.850  0.954  1.007  0.741  0.000  0.000  0.000
0.000  0.000  0.741  1.000  0.973  1.176  1.090  0.928  1.090  1.176  0.973  1.000  0.741  0.000  0.000
0.000  0.695  1.007  0.973  1.193  1.069  1.170  1.010  1.170  1.069  1.193  0.973  1.007  0.695  0.000
0.000  1.020  0.954  1.176  1.069  1.161  1.080  1.185  1.080  1.161  1.069  1.176  0.954  1.020  0.000
0.766  1.255  0.850  1.090  1.170  1.080  1.109  0.958  1.109  1.080  1.170  1.090  0.850  1.255  0.766
0.925  0.985  0.904  0.928  1.010  1.185  0.958  0.789  0.958  1.185  1.010  0.928  0.904  0.985  0.925
0.766  1.255  0.850  1.090  1.170  1.080  1.109  0.958  1.109  1.080  1.170  1.090  0.850  1.255  0.766
0.000  1.020  0.954  1.176  1.069  1.161  1.080  1.185  1.080  1.161  1.069  1.176  0.954  1.020  0.000
0.000  0.695  1.007  0.973  1.193  1.069  1.170  1.010  1.170  1.069  1.193  0.973  1.007  0.695  0.000
0.000  0.000  0.741  1.000  0.973  1.176  1.090  0.928  1.090  1.176  0.973  1.000  0.741  0.000  0.000
0.000  0.000  0.000  0.741  1.007  0.954  0.850  0.904  0.850  0.954  1.007  0.741  0.000  0.000  0.000
0.000  0.000  0.000  0.000  0.695  1.020  1.255  0.985  1.255  1.020  0.695  0.000  0.000  0.000  0.000
0.000  0.000  0.000  0.000  0.000  0.000  0.766  0.925  0.766  0.000  0.000  0.000  0.000  0.000  0.000
\end{lstlisting}
\end{minipage}
\begin{minipage}{\linewidth}
\begin{lstlisting}[caption={almaraz/cycle2/power2D\_Burn1863.txt}]
0.000  0.000  0.000  0.000  0.000  0.000  0.794  0.953  0.794  0.000  0.000  0.000  0.000  0.000  0.000
0.000  0.000  0.000  0.000  0.703  1.033  1.271  1.011  1.271  1.033  0.703  0.000  0.000  0.000  0.000
0.000  0.000  0.000  0.754  1.016  0.960  0.863  0.922  0.863  0.960  1.016  0.754  0.000  0.000  0.000
0.000  0.000  0.754  1.003  0.974  1.170  1.085  0.935  1.085  1.170  0.974  1.003  0.754  0.000  0.000
0.000  0.703  1.016  0.974  1.174  1.056  1.132  0.990  1.132  1.056  1.174  0.974  1.016  0.703  0.000
0.000  1.033  0.960  1.170  1.056  1.145  1.048  1.156  1.048  1.145  1.056  1.170  0.960  1.033  0.000
0.794  1.271  0.863  1.085  1.132  1.048  1.086  0.949  1.086  1.048  1.132  1.085  0.863  1.271  0.794
0.953  1.011  0.922  0.935  0.990  1.156  0.949  0.798  0.949  1.156  0.990  0.935  0.922  1.011  0.953
0.794  1.271  0.863  1.085  1.132  1.048  1.086  0.949  1.086  1.048  1.132  1.085  0.863  1.271  0.794
0.000  1.033  0.960  1.170  1.056  1.145  1.048  1.156  1.048  1.145  1.056  1.170  0.960  1.033  0.000
0.000  0.703  1.016  0.974  1.174  1.056  1.132  0.990  1.132  1.056  1.174  0.974  1.016  0.703  0.000
0.000  0.000  0.754  1.003  0.974  1.170  1.085  0.935  1.085  1.170  0.974  1.003  0.754  0.000  0.000
0.000  0.000  0.000  0.754  1.016  0.960  0.863  0.922  0.863  0.960  1.016  0.754  0.000  0.000  0.000
0.000  0.000  0.000  0.000  0.703  1.033  1.271  1.011  1.271  1.033  0.703  0.000  0.000  0.000  0.000
0.000  0.000  0.000  0.000  0.000  0.000  0.794  0.953  0.794  0.000  0.000  0.000  0.000  0.000  0.000
\end{lstlisting}
\end{minipage}
\begin{minipage}{\linewidth}
\begin{lstlisting}[caption={almaraz/cycle2/power2D\_Burn4461.txt}]
0.000  0.000  0.000  0.000  0.000  0.000  0.781  0.935  0.781  0.000  0.000  0.000  0.000  0.000  0.000
0.000  0.000  0.000  0.000  0.718  1.019  1.249  0.994  1.249  1.019  0.718  0.000  0.000  0.000  0.000
0.000  0.000  0.000  0.770  1.016  0.957  0.876  0.936  0.876  0.957  1.016  0.770  0.000  0.000  0.000
0.000  0.000  0.770  1.009  0.974  1.160  1.093  0.950  1.093  1.160  0.974  1.009  0.770  0.000  0.000
0.000  0.718  1.016  0.974  1.162  1.049  1.141  1.001  1.141  1.049  1.162  0.974  1.016  0.718  0.000
0.000  1.019  0.957  1.160  1.049  1.135  1.053  1.155  1.053  1.135  1.049  1.160  0.957  1.019  0.000
0.781  1.249  0.876  1.093  1.141  1.053  1.097  0.960  1.097  1.053  1.141  1.093  0.876  1.249  0.781
0.935  0.994  0.936  0.950  1.001  1.155  0.960  0.823  0.960  1.155  1.001  0.950  0.936  0.994  0.935
0.781  1.249  0.876  1.093  1.141  1.053  1.097  0.960  1.097  1.053  1.141  1.093  0.876  1.249  0.781
0.000  1.019  0.957  1.160  1.049  1.135  1.053  1.155  1.053  1.135  1.049  1.160  0.957  1.019  0.000
0.000  0.718  1.016  0.974  1.162  1.049  1.141  1.001  1.141  1.049  1.162  0.974  1.016  0.718  0.000
0.000  0.000  0.770  1.009  0.974  1.160  1.093  0.950  1.093  1.160  0.974  1.009  0.770  0.000  0.000
0.000  0.000  0.000  0.770  1.016  0.957  0.876  0.936  0.876  0.957  1.016  0.770  0.000  0.000  0.000
0.000  0.000  0.000  0.000  0.718  1.019  1.249  0.994  1.249  1.019  0.718  0.000  0.000  0.000  0.000
0.000  0.000  0.000  0.000  0.000  0.000  0.781  0.935  0.781  0.000  0.000  0.000  0.000  0.000  0.000
\end{lstlisting}
\end{minipage}
\begin{minipage}{\linewidth}
\begin{lstlisting}[caption={almaraz/cycle2/power2D\_Burn6589.txt}]
0.000  0.000  0.000  0.000  0.000  0.000  0.764  0.913  0.764  0.000  0.000  0.000  0.000  0.000  0.000
0.000  0.000  0.000  0.000  0.724  1.007  1.212  0.978  1.212  1.007  0.724  0.000  0.000  0.000  0.000
0.000  0.000  0.000  0.781  1.021  0.962  0.885  0.945  0.885  0.962  1.021  0.781  0.000  0.000  0.000
0.000  0.000  0.781  1.016  0.981  1.160  1.095  0.960  1.095  1.160  0.981  1.016  0.781  0.000  0.000
0.000  0.724  1.021  0.981  1.160  1.050  1.142  1.008  1.142  1.050  1.160  0.981  1.021  0.724  0.000
0.000  1.007  0.962  1.160  1.050  1.134  1.058  1.160  1.058  1.134  1.050  1.160  0.962  1.007  0.000
0.764  1.212  0.885  1.095  1.142  1.058  1.108  0.976  1.108  1.058  1.142  1.095  0.885  1.212  0.764
0.913  0.978  0.945  0.960  1.008  1.160  0.976  0.848  0.976  1.160  1.008  0.960  0.945  0.978  0.913
0.764  1.212  0.885  1.095  1.142  1.058  1.108  0.976  1.108  1.058  1.142  1.095  0.885  1.212  0.764
0.000  1.007  0.962  1.160  1.050  1.134  1.058  1.160  1.058  1.134  1.050  1.160  0.962  1.007  0.000
0.000  0.724  1.021  0.981  1.160  1.050  1.142  1.008  1.142  1.050  1.160  0.981  1.021  0.724  0.000
0.000  0.000  0.781  1.016  0.981  1.160  1.095  0.960  1.095  1.160  0.981  1.016  0.781  0.000  0.000
0.000  0.000  0.000  0.781  1.021  0.962  0.885  0.945  0.885  0.962  1.021  0.781  0.000  0.000  0.000
0.000  0.000  0.000  0.000  0.724  1.007  1.212  0.978  1.212  1.007  0.724  0.000  0.000  0.000  0.000
0.000  0.000  0.000  0.000  0.000  0.000  0.764  0.913  0.764  0.000  0.000  0.000  0.000  0.000  0.000
\end{lstlisting}
\end{minipage}
\begin{minipage}{\linewidth}
\begin{lstlisting}[caption={almaraz/cycle2/power2D\_Burn8436.txt}]
0.000  0.000  0.000  0.000  0.000  0.000  0.756  0.901  0.756  0.000  0.000  0.000  0.000  0.000  0.000
0.000  0.000  0.000  0.000  0.731  1.003  1.192  0.968  1.192  1.003  0.731  0.000  0.000  0.000  0.000
0.000  0.000  0.000  0.792  1.029  0.967  0.886  0.946  0.886  0.967  1.029  0.792  0.000  0.000  0.000
0.000  0.000  0.792  1.024  0.986  1.160  1.097  0.969  1.097  1.160  0.986  1.024  0.792  0.000  0.000
0.000  0.731  1.029  0.986  1.159  1.046  1.138  1.013  1.138  1.046  1.159  0.986  1.029  0.731  0.000
0.000  1.003  0.967  1.160  1.046  1.128  1.053  1.162  1.053  1.128  1.046  1.160  0.967  1.003  0.000
0.756  1.192  0.886  1.097  1.138  1.053  1.107  0.984  1.107  1.053  1.138  1.097  0.886  1.192  0.756
0.901  0.968  0.946  0.969  1.013  1.162  0.984  0.861  0.984  1.162  1.013  0.969  0.946  0.968  0.901
0.756  1.192  0.886  1.097  1.138  1.053  1.107  0.984  1.107  1.053  1.138  1.097  0.886  1.192  0.756
0.000  1.003  0.967  1.160  1.046  1.128  1.053  1.162  1.053  1.128  1.046  1.160  0.967  1.003  0.000
0.000  0.731  1.029  0.986  1.159  1.046  1.138  1.013  1.138  1.046  1.159  0.986  1.029  0.731  0.000
0.000  0.000  0.792  1.024  0.986  1.160  1.097  0.969  1.097  1.160  0.986  1.024  0.792  0.000  0.000
0.000  0.000  0.000  0.792  1.029  0.967  0.886  0.946  0.886  0.967  1.029  0.792  0.000  0.000  0.000
0.000  0.000  0.000  0.000  0.731  1.003  1.192  0.968  1.192  1.003  0.731  0.000  0.000  0.000  0.000
0.000  0.000  0.000  0.000  0.000  0.000  0.756  0.901  0.756  0.000  0.000  0.000  0.000  0.000  0.000
\end{lstlisting}
\end{minipage}
\subsubsection{Fessenheim-2 Cycle 1}
\begin{minipage}{\linewidth}
\begin{lstlisting}[caption={fessenheim/cycle1/boronExp.txt}]
0       1325
280     1025
500     983
1173    898
2748    860
4000    805
5363    730
6500    660
7755    572
9000    485
10844   352
13340   137
14979   19
\end{lstlisting}
\end{minipage}
\begin{minipage}{\linewidth}
\begin{lstlisting}[caption={fessenheim/cycle1/power2D\_Burn7755.txt}]
0.000  0.000  0.000  0.000  0.000  0.000  0.000  0.6265 0.000  0.000  0.000  0.000  0.000  0.000  0.000
0.000  0.000  0.000  0.000  0.000  0.000  0.000  0.000  0.000  0.7559 0.000  0.000  0.000  0.000  0.000
0.000  0.000  0.000  0.5921 0.000  0.000  1.1917 1.0332 0.000  0.000  0.000  0.5701 0.000  0.000  0.000
0.000  0.000  0.000  0.000  0.9804 0.000  0.000  1.2639 0.000  1.2332 0.000  0.000  0.000  0.000  0.000
0.000  0.000  0.8040 0.000  1.2380 0.000  1.2940 0.000  0.000  0.000  1.2428 0.9861 0.000  0.5435 0.000
0.000  0.000  0.000  0.000  1.0743 0.000  0.000  1.3113 0.000  1.2933 0.000  0.000  0.000  0.000  0.000
0.000  0.000  1.1537 0.000  0.000  0.000  1.3133 0.000  1.3127 0.000  0.000  1.0690 0.000  0.8050 0.000
0.6275 0.000  1.0580 0.000  1.0886 0.000  0.000  0.000  0.000  1.0389 0.000  0.000  1.0243 1.0617 0.000
0.000  0.000  0.000  0.000  1.2932 0.000  0.000  0.000  1.3139 1.1104 0.000  0.000  0.000  0.000  0.5210
0.000  0.000  0.9659 0.000  0.000  0.000  1.1130 0.000  0.000  0.000  0.000  1.2325 0.000  0.7512 0.000
0.000  0.000  0.000  0.000  1.2417 0.000  0.000  1.0876 0.000  1.0864 1.2342 0.000  0.000  0.000  0.000
0.000  0.000  0.5981 0.000  0.000  0.000  1.0766 0.000  0.000  0.000  0.000  1.0149 0.5910 0.000  0.000
0.000  0.000  0.000  0.000  0.000  0.000  0.000  1.0279 0.000  0.9683 0.000  0.000  0.000  0.000  0.000
0.000  0.000  0.000  0.000  0.5477 0.000  0.000  0.000  0.8083 0.000  0.000  0.000  0.000  0.000  0.000
0.000  0.000  0.000  0.000  0.000  0.000  0.5258 0.000  0.000  0.000  0.000  0.000  0.000  0.000  0.000
\end{lstlisting}
\end{minipage}

\subsubsection{Turkey-Point-3 Cycle 1}
\begin{minipage}{\linewidth}
\begin{lstlisting}[caption={tp3/cycle1/boronExp.txt}]
0       1184
350     882
3000    747
6600    570
9000    405
12800   101
\end{lstlisting}
\end{minipage}

\subsubsection{Turkey-Point-3 Cycle 2}

\begin{minipage}{\linewidth}
\begin{lstlisting}[caption={tp3/cycle2/boronExp.txt}]
0       1184
350     882
6600    570
12800   101
\end{lstlisting}
\end{minipage}

\subsubsection{Turkey-Point-3 Cycle 3}
\begin{minipage}{\linewidth}
\begin{lstlisting}[caption={tp3/cycle3/boronExp.txt}]
0       1170
900     712
4150    423
6300    251
7730    95
\end{lstlisting}
\end{minipage}


\newpage
\cfoot{References}
\printbibliography

@phdthesis{Salino,
	author = {Vivian Salino},
	title = {Incertitudes et ajustements de données nucléaires au moyen de méthodes déterministes, probabilistes et de mesures effectuées sur des réacteurs à eau sous pression.},
	url= "https://publications.polymtl.ca/10545/",
	institution = {Polytechnique Montréal, PolyPublie},
	year = 2022
}

@article{TMI,
title = {Reactor core physics design and operating data for Cycles 1 and 2 of TMI Unit 1 PWR Power Plant. Final report},
author = {{Babcock \& Wilcox}},
doi = {10.2172/5082638},
url = {https://www.osti.gov/biblio/5082638},
place = {United States},
year = {1980},
month = {8}
}

@article{Zion,
title = {Reactor core physics design and operating data for Cycles 1 and 2 of the Zion Unit 2 PWR power plant. Final report},
author = {{Albert J. Impink, Jr.} and {B. Alan Guthrie III}},
doi = {10.2172/5303509},
url = {https://www.osti.gov/biblio/5303509}, journal = {},
place = {United States},
year = {1979},
month = {12}
}

@article{TurkeyPoint,
title = {Reactor core physics design and operating data for Cycles 1, 2, and 3 of the Turkey Point No. 3 PWR power plant. Final report},
author = {{Nuclear Associates International corporation (NAI)}},
url = {https://www.osti.gov/biblio/6803027}, journal = {},
place = {United States},
year = {1978},
month = {7}
}

@article{Surry,
title = {Reactor Core Physics Design and Operating Data for Cycles 1, 2, and 3 of Surry Unit 1 PWR Power Plant},
author = {R. W. Carlson},
url = {https://repository.gatech.edu/entities/publication/c8dbbf0e-e00d-4f24-b0dd-a0e503495b50/full}, journal = {},
place = {United States},
year = {1979},
month = {3}
}

@phdthesis{Hassini,
	author = {Ahmed Hassini},
	title = {Établissement de bibliothèques à contre-réactions pour le suivi du 1er cycle du réacteur Fessenheim.},
	institution = {Paris-Sud University},
	year = 1982
}

@phdthesis{Panek,
	author = {Henri Panek},
	title = {Qualification of the system Neptune: interpretation of critical experiments - core calculations of a power reactor - first seting-up of a new calculation chain based on the codes Apollo and Tortise: tests on a PWR core.},
	institution = {Paris-Sud University},
	year = 1979
}

@book{IAEA,
  title={In-Core Fuel Management Code Package Validation for PWRs},
  series={TECDOC Series},
  number={815},
  year={},
  url={https://www.iaea.org/publications/5461/in-core-fuel-management-code-package-validation-for-pwrs},
  publisher={International Atomic Energy Agency},
  address={Vienna}
}

@phdthesis{Enaam,
	author = {Melle Kamha Enaam},
	title = {Contribution à l'élaboration et à la qualification d'un schéma de calcul pour la gestion des reacteurs PWR, à l'aide du système NEPTUNE - Suivi du réacteur Fessenheim-2.},
	institution = {Paris-Sud University},
	year = 1981
}

@article{Aliberti,
title = {Nuclear data sensitivity, uncertainty and target accuracy assessment for future nuclear systems},
journal = {Annals of Nuclear Energy},
volume = {33},
number = {8},
pages = {700-733},
year = {2006},
issn = {0306-4549},
doi = {https://doi.org/10.1016/j.anucene.2006.02.003},
url = {https://www.sciencedirect.com/science/article/pii/S0306454906000296},
author = {G. Aliberti and G. Palmiotti and M. Salvatores and T.K. Kim and T.A. Taiwo and M. Anitescu and I. Kodeli and E. Sartori and J.C. Bosq and J. Tommasi}
}

@misc{ICSBEPI,
    url = {https://www.oecd-nea.org/jcms/pl_24498/international-criticality-safety-benchmark-evaluation-project-icsbep},
    note = {International Criticality Safety Benchmark Evaluation Project (ICSBEPI)}
}

@book{PARCS_1_Inputs,
  title={PARCS - VOLUME I: INPUT MANUAL,},
  series={NRC - v3.3.1 Release},
  author = {Thomas Downar and Andrew Ward and Yunlin Xu and Volkan Seker and Nathanael Hudson and Douglas Barber and Lance Larsen and Boyan Neykov and Glenn Roth},
  year={2018},
  url={https://nuram.engin.umich.edu/software/parcs/},
  publisher={{University of Michigan} and {U.S. Nuclear Regulatory Commission} and {Information Systems Laboratories Inc.}}
}

@misc{IRPhE,
    url = {https://www.oecd-nea.org/jcms/pl_20279/international-handbook-of-evaluated-reactor-physics-benchmark-experiments-irphe},
    note = {International Handbook of Evaluated Reactor Physics Benchmark Experiments (IRPhE)}
}

@book{Dragon,
    title = {DRAGON lattice code user guide},
    author = {G. Marleau and A. Hébert and R. Roy},
    url = {http://merlin.polymtl.ca/version5.htm},
    publisher = {École Polytechnique de Montreal}
}

@book{gitSalino,
    title = {Git repository with the work of Dr. Vivian Salino's PhD thesis (accessed on August 4, 2024)},
    author = {Vivian Salino},
    url = {https://github.com/IRSN/SalinoPhD},
    publisher= {IRSN and Polytechnique Montréal}
}

@book{JEFF,
    title = {Joint Evaluated Fission and Fusion (JEFF) Test Nuclear Data Library},
    url = {https://www.oecd-nea.org/dbdata/jeff/},
    date = {2022},
    month = {10},
    publisher = {Nuclear Energy Agency (NEA)}
}

@book{ENDF,
    title = {ENDF/B-VIII.0 Evaluated Nuclear Data Library},
    url = {https://www.nndc.bnl.gov/endf-b8.0/},
    author = {{Cross Section Evaluation Working Group (CSEWG)}},
    date = {2018},
    month = {2},
    publisher = {{National Nuclear Data Center (NNDC)}}
}

@book{dataIAEA,
    title = {ENDF database.},
    url = {https://www-nds.iaea.org/exfor/endf.htm},
    author = {{International Atomic Energy Agency (IAEA)}},
    date = {2024},
    month = {05}
}

@article{bullshit,
    title = {ChatGPT is bullshit},
    author = {{Hicks, M.T}. and {Humphries, J.} and {Slater, J.}},
    journal = {Ethics Inf Technol},
    year = 2024,
    doi = {https://doi.org/10.1007/s10676-024-09775-5}
}

@article{PyNJOY,
author = {Salino, V and Hébert, Alain},
year = {2023},
pages = {},
title = {PyNjoy2016: an Open Source System for Producing Cross Sections Libraries for DRAGON5 and SERPENT2},
url = {https://github.com/irsn/pynjoy2016}
}

@book{PNR2,
  title={Physics of Nuclear Reactor II},
  author={Konstantin Mikityuk},
  year={2022},
  publisher={{ETHZ lecture}}
}

@article{SHEM295,
author = {Alain Hébert},
title = {Development of the Subgroup Projection Method for Resonance Self-Shielding Calculations},
journal = {Nuclear Science and Engineering},
volume = {162},
number = {1},
pages = {56--75},
year = {2009},
publisher = {Taylor \& Francis},
doi = {10.13182/NSE162-56}
}

@book{DragonUserGuide,
  title={A USER GUIDE FOR DRAGON VERSION5},
  author={G. Marleau, A. Hébert and R. Roy},
  year={2024},
  url ={http://merlin.polymtl.ca/version5.htm},
  publisher={{Institut de génie nucléaire, Polytechnique Montréal}}
}

@book{CLE2000UserGuide,
  title={THE CLE-2000 TOOL-BOX},
  author={Robert ROY},
  year={1999},
  url ={http://merlin.polymtl.ca/version5.htm},
  publisher={{Institut de génie nucléaire, École Polytechnique de Montréal}}
}

@book{GenPMAXS_manual,
  title={GenPMAXS - v6.2, Code for Generating the PARCS Cross Section Interface File PMAXS (ALPHA RELEASE)},
  author={{A. Ward} and {Y. Xu} and {T. Downar}},
  year={2016},
  month={7},
  url={http://nuram.engin.umich.edu/software/genpmaxs/},
  publisher={{University of Michigan}}
}

@book{NRC_PARCS,
  title={Computer Codes (accessed on August 4, 2024)},
  url={https://www.nrc.gov/about-nrc/regulatory/research/safetycodes.html},
  publisher={{U.S. Nuclear Regulatory Commission (NRC)}}
}

@book{gitORNL,
  title={SCALE Reactor Physics  Validation (accessed on August 4, 2024)},
  author={Ugur Mertyurek},
  url={https://code.ornl.gov/scale/analysis/2023-polaris_parcs-validation},
  year={2023},
  month={10},
  publisher={{Oak Ridge National Laboratory (ORNL)}}
}

@book{ORNL_TP3,
  title={Benchmark Calculation for Turkey Point Unit 3 Cycles 1–3 Using the SCALE 6.3/Polaris–PARCS v3.4.2 Code Package (accessed on August 4, 2024)},
  author={Byoung-kyu Jeon and Kang Seog Kim and William A. Wieselquist},
  url={https://code.ornl.gov/scale/analysis/2023-polaris_parcs-validation},
  year={2023},
  month={8},
  publisher={{Oak Ridge National Laboratory (ORNL)}}
}

@book{ORNL_Surry,
  title={Benchmark Calculation for Surry Unit 1 Cycles 1–3 Using the SCALE 6.3/Polaris– PARCS v3.4.2 Code Package},
  author={Byoung-kyu Jeon and Kang Seog Kim and William A. Wieselquist},
  url={https://code.ornl.gov/scale/analysis/2023-polaris_parcs-validation},
  year={2023},
  month={8},
  publisher={{Oak Ridge National Laboratory (ORNL)}}
}

@book{morey_glassProperties,
  title={The Properties of Glass (2nd edition)},
  author={{Morey, G.W.}},
  year={1954},
  publisher={Reinhold Publishing Corporation}
}

@article{hursinIntegralParamCASMO,
	author = {{Hursin, Mathieu} and {Rochman, Dimitri} and {Vasiliev, Alexander} and {Ferroukhi, Hakim} and {Pautz, Andreas}},
	title = {Impact of various source of covariance on integral parameters uncertainty during depletion calculations with CASMO-5},
	DOI= "10.1051/epjconf/202124709005",
	url= "https://doi.org/10.1051/epjconf/202124709005",
	journal = {EPJ Web Conf.},
	year = 2021,
	volume = 247,
	pages = "09005"
}

@article{whyNotMC,
	author = {Martin, William R. },
	title = {Challenges and prospects for whole-core Monte Carlo analysis},
	DOI= "10.5516/NET.01.2012.502",
	url= "https://doi.org/10.5516/NET.01.2012.502",
	journal = {Nuclear Engineering and Technology, Korean Nuclear Society},
	year = 2012,
	volume = 33,
	pages = "151-160"
}

@article{hursinTowardMC_FullCore,
title = {Towards establishment of an efficient approach for validation of PWR full core Monte Carlo simulations at hot zero power conditions},
journal = {Progress in Nuclear Energy},
volume = {172},
pages = {105203},
year = {2024},
issn = {0149-1970},
doi = {https://doi.org/10.1016/j.pnucene.2024.105203},
url = {https://www.sciencedirect.com/science/article/pii/S0149197024001537},
author = {L. Berry and A. Vasiliev and M. Hursin and D. Rochman and M. Frankl and H. Ferroukhi}
}

@ARTICLE{Kaiwen_MC_complex,

AUTHOR={Li, Kaiwen  and An, Nan  and Luo, Hao  and Huang, Shanfang  and Wang, Kan },

TITLE={A better hash method for high-fidelity Monte Carlo simulations on nuclear reactors},

JOURNAL={Frontiers in Energy Research},

VOLUME={11},

YEAR={2023},

URL={https://www.frontiersin.org/journals/energy-research/articles/10.3389/fenrg.2023.1161861},

DOI={10.3389/fenrg.2023.1161861},

ISSN={2296-598X}
}

@article{Serpent_LatticeCalculations,
title = {Lattice Calculations and Power Distribution for Nigeria Research Reactor-1 (NIRR-1) using Serpent Code},
journal = {Communication in Physical Sciences},
volume = {10},
pages = {99-107},
year = {2023},
url = {https://www.journalcps.com/index.php/volumes/article/view/427},
author = {Abubakar Aliyu Umar and Aminu Ismaila  and Khaidzir Hamza }
}

@article{hursin_M_C_2023,
title = {Improved evaluated data in the resonance region by combining energy dependent cross section and depletion calculations: example U-238+n},
journal = {{The International Conference on Mathematics and Computational Methods Applied to Nuclear Science and Engineering}},
year = {2023},
author = {M. Hursin and D. Rochman and S. Kopecky and Peter Schillebeeckx}
}

@article{hursin_physor_2024,
title = {Improved evaluated data in the resonance region by combining energy dependent cross section and depletion calculations: application to Pu-239+n, Pu-240+n and Pu-241+n},
journal = {{International Conference on Physics of Reactors (PHYSOR)}},
year = {2024},
author = {M. Hursin and D. Rochman and S. Kopecky and S. van der Marck}
}

@article{JANIS,
title = {JANIS 4: An Improved Version of the NEA Java-based Nuclear Data Information System},
author = {N. Soppera and M. Bossant and E. Dupont},
journal={Nuclear Data Sheets},
volume={120},
year={2014},
pages={294-296},
url={http://dx.doi.org/10.1016/j.nds.2014.07.071},
doi={10.1016/j.nds.2014.07.071}
}

@book{POLARIS,
  title={University of Michigan, Nuclear Reactor Analysis and Methods, Softwares (accessed on August 10, 2024)},
  url={https://nuram.engin.umich.edu/software/},
  publisher={{developped by the Oak Ridge National Laboratory (ORNL) for the U.S. Nuclear Regulatory Commission (NRC)}}
}

@article{gridSpacerImpact,
title = {Investigation of the Effect of Spacer Grids Modeling on Reactivity and Power Distribution in PWR Fuel Assembly},
journal = {{Journal of Nuclear and Particle Physics}},
year = {2016},
author = {M. Hursin and D. Rochman and S. Kopecky and Peter Schillebeeckx},
doi={10.5923/j.jnpp.20160603.01},
url={http://article.sapub.org/10.5923.j.jnpp.20160603.01.html}
}

@book{img2table,
    title = {img2table 1.2.11 (accessed on August 11, 2024)},
    author = {Xavier Canton},
    url = {https://pypi.org/project/img2table/},
    publisher = {Python Library}
}

@book{PaddleOCR,
    title = {PaddleOCR 2.8.1 (accessed on August 11, 2024)},
    author = {{Collaborative project}},
    url = {https://github.com/PaddlePaddle/PaddleOCR},
    publisher = {Python Library}
}

@book{pandasPython,
    title = {Pandas 2.2.2 (accessed on August 11, 2024)},
    author = {{Collaborative project}},
    url = {https://pandas.pydata.org/},
    publisher = {Python Library}
}

@misc{chatgpt4o,
  author       = {OpenAI},
  title        = {ChatGPT-4o: Language Model},
  year         = {2024},
  url          = {https://www.openai.com/chatgpt},
  note         = {Accessed: August 11, 2024}
}

@misc{docker,
  author       = {Docker, Inc.},
  title        = {Docker: Empowering App Development},
  year         = {2024},
  url          = {https://www.docker.com},
  note         = {Accessed: August 11, 2024}
}

@article{SerpentFullCoreValidation1,
title = {Validation of Serpent-SUBCHANFLOW-TRANSURANUS pin-by-pin burnup calculations using experimental data from the Temelín II VVER-1000 reactor},
journal = {Nuclear Engineering and Technology},
volume = {53},
number = {10},
pages = {3133-3150},
year = {2021},
issn = {1738-5733},
doi = {https://doi.org/10.1016/j.net.2021.04.023},
url = {https://www.sciencedirect.com/science/article/pii/S1738573321002424},
author = {Manuel García and Radim Vočka and Riku Tuominen and Andre Gommlich and Jaakko Leppänen and Ville Valtavirta and Uwe Imke and Diego Ferraro and Paul {Van Uffelen} and Lukáš Milisdörfer and Victor Sanchez-Espinoza}
}

@article{SerpentFullCoreValidation2,
title = {Validation of Serpent-SUBCHANFLOW-TRANSURANUS pin-by-pin burnup calculations using experimental data from a Pre-Konvoi PWR reactor},
journal = {Nuclear Engineering and Design},
volume = {379},
pages = {111173},
year = {2021},
issn = {0029-5493},
doi = {https://doi.org/10.1016/j.nucengdes.2021.111173},
url = {https://www.sciencedirect.com/science/article/pii/S0029549321001254},
author = {Manuel García and Yurii Bilodid and Joaquín {Basualdo Perello} and Riku Tuominen and Andre Gommlich and Jaakko Leppänen and Ville Valtavirta and Uwe Imke and Diego Ferraro and Paul {Van Uffelen} and Marcus Seidl and Victor Sanchez-Espinoza}
}

@article{hursinSalinoFessenheimCycle1,
title = {Towards benchmarking nuclear data on reactor models},
journal = {JEFF Meeting},
year = {2023},
url = {https://www.oecd-nea.org/dbdata/nds_jefdoc/jefdoc-2236.pdf},
author = {Mathieu Hursin and Vivian Salino and Julien Taforeau and Dimitri Rochman},
}

@article{NJOY2016,
author = {Macfarlane, R. and Muir, Douglas},
year = {1994},
month = {10},
pages = {},
title = {The NJOY nuclear data processing system},
doi = {10.2172/10115999}
}

@article{nTRACERdirectTransport,
title = {Practical numerical reactor employing direct whole core neutron transport and subchannel thermal/hydraulic solvers},
journal = {Annals of Nuclear Energy},
volume = {62},
pages = {357-374},
year = {2013},
issn = {0306-4549},
doi = {https://doi.org/10.1016/j.anucene.2013.06.031},
url = {https://www.sciencedirect.com/science/article/pii/S0306454913003344},
author = {Yeon Sang Jung and Cheon Bo Shim and Chang Hyun Lim and Han Gyu Joo},
}

@article{MPACTdirectTransport,
author = {Brendan Kochunas and Benjamin Collins and Shane Stimpson and {al}},
title = {VERA Core Simulator Methodology for Pressurized Water Reactor Cycle Depletion},
journal = {Nuclear Science and Engineering},
volume = {185},
number = {1},
pages = {217--231},
year = {2017},
publisher = {Taylor \& Francis},
doi = {10.13182/NSE16-39},
URL = {https://doi.org/10.13182/NSE16-39},
eprint = {https://doi.org/10.13182/NSE16-39}
}

@article{JEFF_UncertaintyXS,
author = {{N. Slosse} and {C. R. Schneidesch}},
title = {Testing JEFF3.3 for whole core calculations on plant data using WIMS/PANTHER route},
journal = {JEFF Stakeholder Workshop},
year = {2020}
}

@misc{Taforeau,
    author = {J. Taforeau},
    note = {Personnal communication}
}

@article{MEUNIER2025,
title = {Modeling, verification and validation of multiple PWR depletion cycles with DRAGON and PARCS},
journal = {Annals of Nuclear Energy},
volume = {219},
pages = {111427},
year = {2025},
issn = {0306-4549},
doi = {https://doi.org/10.1016/j.anucene.2025.111427},
url = {https://www.sciencedirect.com/science/article/pii/S0306454925002440},
author = {B. Meunier and M. Hursin}
}

@misc{confParisNuc2024,
      title={Towards benchmarking nuclear data on reactor models}, 
      howpublished={JEFF Nuclear Data Week, Paris},
      year={2024},
      month={Nov},
      author={{Mathieu Hursin} and {Benjamin Meunier} and {al.}}
      %url={https://www.oecd-nea.org/dbdata/jeff/jeffdoc.html}
}

\end{document}